%% file: PRF2026_channel_wnl_arxiv_final.tex
\documentclass[aps,prf,superscriptaddress,longbibliography,amsmath,amssymb,dvipsnames]{revtex4-2}

\usepackage{amssymb,latexsym,color,amsmath,pifont,graphicx,booktabs}
\usepackage{tikz}
\usetikzlibrary{arrows,shapes,trees,calc,positioning,decorations.pathreplacing}
\usepackage{subfigure}  
\usepackage[colorlinks=true, allcolors=blue]{hyperref}
\usepackage{mathrsfs}
\usepackage{scalefnt}
\usepackage{comment}
\usepackage{xcolor,varwidth}
\usepackage{cleveref}
\usepackage[normalem]{ulem}

\input{commands_ozaslan}
\input{notation_ozaslan}

\newcommand{\mri}{\mathrm{i}}
\newcommand{\mre}{\mathrm{e}}

\newcommand{\bu}{\mathbf{u}}
\newcommand{\bv}{\mathbf{v}}
\newcommand{\bd}{\mathbf{d}}
\newcommand{\bx}{\mathbf{x}}
\newcommand{\bk}{\mathbf{k}}
\newcommand{\bq}{\mathbf{q}}

\newcommand{\bdelta}{\mbox{\boldmath{$\delta$}}}
\newcommand{\bvartheta}{\mbox{\boldmath{$\vartheta$}}}

\newcommand{\bom}{\omega}
\newcommand{\bkx}{k_x}
\newcommand{\bkz}{k_z}
\newcommand{\bc}{c}
\newcommand{\cS}{\mathcal{S}}
\newcommand{\bbN}{\mathbb{N}}

\definecolor{dblue}{rgb}{0,0,0.5}
\definecolor{dred}{rgb}{0.6,0.2,0}
\definecolor{dg}{RGB}{0, 100, 0}

\definecolor{LayerOne}{RGB}{0,76,153}  
\definecolor{LayerTwo}{RGB}{166,28,0}  

\newcommand{\tc}{\textcolor}
\newcommand{\cO}{\mathcal{O}}

\newcommand{\cG}{\mathcal{G}}

\newcommand{\cAk}{\mathcal{A}_{\mathbf{k}}}
\newcommand{\cBk}{\mathcal{B}_{\mathbf{k}}}
\newcommand{\cCk}{\mathcal{C}_{\mathbf{k}}}
\newcommand{\cHk}{\mathcal{H}_{\mathbf{k}}}
\newcommand{\cGk}{\mathcal{G}_{\mathbf{k}}}

\newtheorem{theorem}{Theorem}

\newtheorem{remark}{Remark}

\begin{document}

\tikzstyle{block} = [draw, fill=blue!20, rectangle, very thick,
    minimum height=3em, minimum width=6em]
\tikzstyle{bigblock} = [draw, fill=blue!20, rectangle, very thick,
    minimum height=4em, minimum width=8em]
\tikzstyle{input} = [coordinate]
\tikzstyle{output} = [coordinate]

\tikzstyle{vertex1}=[draw, fill=orange!75, circle, very thick, minimum size=33pt, inner sep=0pt]
\tikzstyle{vertex}=[draw, fill=orange!75, circle, very thick, minimum size=30pt, inner sep=0pt]
\tikzstyle{GrayVertex}=[circle, fill=gray!20, minimum size=30pt, inner sep=0pt]
\tikzstyle{BigVertex}=[draw, fill=orange!75, circle, very thick, minimum size=210pt, inner sep=0pt]
\tikzstyle{BigWhiteVertex}=[draw, fill=white, circle, very thick, minimum size=210pt, inner sep=0pt]
\tikzstyle{selected vertex} = [vertex, fill=red!24]
\tikzstyle{edge} = [draw, black, line width=3pt, -]
\tikzstyle{weak edge} = [draw, gray, line width=3pt, dashed]
\tikzstyle{weight} = [font=\small]
\tikzstyle{selected edge} = [draw, line width=4pt, -]
\tikzstyle{ignored edge} = [draw, line width=5pt, -, black!20]

\tikzstyle{na} = [baseline=-.5ex]
\tikzstyle{opblock} = [draw, rectangle, very thick, minimum height=3em,
    minimum width=6em, fill=white, text=black]
\definecolor{customRed}{RGB}{255,155,155}

\title{\LARGE \bf From oblique-wave forcing to streak reinforcement: 
	\\
A perturbation-based frequency-response framework}

\author{Du\v{s}an Bo\v{z}i\'c}
\email[E-mail:]{dbozic@usc.edu}
\altaffiliation{Ming Hsieh Department of Electrical and Computer Engineering, University of Southern California, Los Angeles, CA 90089, USA}
\author{Anubhav Dwivedi}
\email[E-mail:]{anubhavd91@gmail.com}
\altaffiliation{Department of Aerospace Engineering and Mechanics, University of Minnesota, Minneapolis, MN 55455, USA}
\author{Mihailo R.\ Jovanovi\'c}
\email[E-mail:]{mihailo@usc.edu}
\altaffiliation{Ming Hsieh Department of Electrical and Computer Engineering, University of Southern California, Los Angeles, CA 90089, USA}

\date{\today}
             
	\begin{abstract}
	We develop a perturbation-based frequency-response framework for analyzing amplification mechanisms that are central to subcritical routes to transition in wall-bounded shear flows. By systematically expanding the input--output dynamics of fluctuations about the laminar base flow with respect to forcing amplitude, we establish a rigorous correspondence between linear resolvent analysis and higher-order nonlinear interactions.   
	
	At second order, quadratic interactions of unsteady oblique waves generate steady streamwise streaks via the lift-up mechanism. We demonstrate that the spatial structure of these streaks is captured by the second output singular function of the streamwise-constant resolvent operator. At higher orders, nonlinear coupling between oblique waves and induced streaks acts as structured forcing of the laminar linearized dynamics, yielding additional streak components whose relative phase governs reinforcement or attenuation of the leading-order streak response.
	
	Our analysis identifies a critical forcing amplitude marking the breakdown of the weakly nonlinear regime, beyond which direct numerical simulations exhibit sustained unsteadiness. We show that this breakdown coincides with the onset of secondary instability, revealing that the nonlinear interactions responsible for streak formation also drive the modal growth central to classical transition theory. The resulting framework provides a mechanistically transparent and computationally efficient description of transition that unifies non-modal amplification, streak formation, and modal instability within a single formulation derived directly from the Navier--Stokes equations.			
	\end{abstract}

\maketitle	

	\vspace*{-6ex}
\section{Introduction}

	\vspace*{-1ex}
Transition to turbulence in wall-bounded shear flows is a foundational problem in fluid mechanics with significant implications for aerodynamic drag and transport efficiency. Despite over a century of inquiry, predicting the onset of turbulence across diverse flow configurations remains a formidable challenge. Building on classical ideas of hydrodynamic stability~\cite{drarei81}, the prevailing approach to transition has relied on modal decompositions of the Navier--Stokes (NS) equations linearized about a laminar base flow. 

For sufficiently large Reynolds numbers, this analysis predicts exponentially growing two-dimensional Tollmien--Schlichting (TS) waves in Blasius boundary-layer and pressure-driven channel (Poiseuille) flows~\cite{ors71,schhen01}. However, it fails to identify modal instabilities in shear-driven (Couette) and pipe flows~\cite{schhen01}, yielding predictions that contradict experimental and numerical observations of both transitional Reynolds numbers and observed flow structures. Secondary-instability analyses of TS waves improve these predictions through the K- and H-type transition scenarios~\cite{her88}, yet they remain unable to account for the bypass transition commonly observed in realistic flows with surface imperfections or free-stream turbulence~\cite{braschhen04}. Experiments and simulations instead show that three-dimensional perturbations in the form of streamwise vortices generate streaks that break down and trigger transition at amplitudes far below those required for TS-wave instability~\cite{RSBH1998}.
	
These observations motivated the development of non-modal frameworks that emphasize transient and non-normal amplification mechanisms governing the early stages of subcritical transition~\cite{butfar92,trefethen1993hydrodynamic,jovbamJFM05,schmid2007nonmodal,jovARFM21}. A defining feature of such transition is the formation of streamwise streaks---elongated regions of alternating high- and low-speed fluid. The ubiquity of these structures in wall-bounded shear flows can be traced to the strong non-normality of the linearized NS operator, which enables large transient growth and strong amplification of exogenous disturbances through the lift-up mechanism~\cite{ellpal75,lan75}. Non-modal linear theory captures this process by identifying the most energetic infinitesimal fluctuations via singular value decomposition of the state-transition and resolvent operators, corresponding to time- and frequency-domain representations. However, these analyses remain inherently qualitative: they do not specify their range of validity with respect to disturbance amplitude, nor do they explain the eventual breakdown of streaks into turbulence. Nonlinear non-modal stability and harmonic balance analyses partially address these limitations by employing gradient-based optimization to identify minimal-energy perturbations that trigger transition~\cite{ker18} or to evaluate finite-amplitude input--output amplification~\cite{rigsipcol21}, but both approaches entail substantial computational cost and offer limited physical interpretability.

While non-modal analyses of the linearized NS equations emphasize steady streamwise-constant vortical disturbances, direct numerical simulations (DNS) show that oblique waves can provide an especially effective pathway for triggering transition~\cite{schmid1992new,RSBH1998}. Beyond the onset of transition, nonlinear interactions between streaks and oblique waves also underpin the self-sustaining process that maintains turbulence in wall-bounded flows~\cite{hamkimwal95,wal97}. Exponentially growing three-dimensional modes of the streak-modified base flow excite such disturbances, which through nonlinear coupling regenerate streaks and close the self-sustaining cycle. Unsteady oblique waves are thus recognized as a robust route to transition across a wide range of flows, including low-speed incompressible~\citep{berlin1999numerical,rigsipcol21}, compressible~\citep{chang_malik_1994,mayer2011direct}, and high-speed boundary layers~\citep{ma2005receptivity,sivasubramanian_fasel_2015,hader_fasel_2019}, as well as separated flows involving shock-wave/boundary-layer interactions~\cite{dwisidjovJFM22}.

To bridge the gap between linear and fully nonlinear descriptions, weakly nonlinear analysis offers a tractable and physically transparent framework. By accounting for nonlinear interactions among small fluctuations about the laminar base flow, this approach successfully describes streak generation in Poiseuille and boundary-layer flows using only a second-order expansion~\cite{PonzianiWNLPois,brandt2002weakly,dwisidjovJFM22}. Related formulations have employed multiple-scale asymptotic expansions to correct the amplitude of the most amplified resolvent modes in strongly non-normal flows~\cite{ducbougal22}. Rather than incorporating induced streaks into a predefined modified base flow, as is common in conventional secondary-stability analyses~\cite{RSBH1998,andersson2001breakdown,schbralanhen04}, the present work builds on the weakly nonlinear framework of~\cite{brandt2002weakly} by systematically extending the perturbation expansion of fluctuations about the laminar base flow beyond second order with respect to the amplitude of external forcing. This formulation retains the advantages of expansion about the laminar base flow while enabling a frequency-response characterization of the nonlinear interactions responsible for streak generation and their subsequent influence on transition dynamics. In our recent study~\cite{dwisidjovJFM22}, we employed a weakly nonlinear resolvent framework to identify the mechanisms triggering oblique transition in a separated hypersonic flow and validated its predictions against direct numerical simulations. Here, we adopt a canonical channel-flow configuration to extend this framework to higher perturbation orders, enabling systematic examination of nonlinear interactions and the coupling between streak amplification and modal instability.

Recent extensions of the resolvent framework have addressed flows with time- or spatially-periodic base states~\cite{padottorow20,padrow22,jovfarAUT08,farjovbam08}. In these settings, the natural modes are no longer purely harmonic but instead take the form of Bloch waves~\cite{odekel64}, leading to coupling between frequencies separated by the fundamental frequency of the base flow~\cite{wer91,farjovbam08}. Lifting-based formulations recast linear time-periodic dynamics as equivalent time-invariant systems~\cite{jovARFM21}, enabling analysis of phenomena such as vortex pairing~\cite{padrow22} as well as the impact of wall oscillations on transition and drag reduction~\cite{jovPOF08,moajovJFM12}. Analogous ideas extend to spatially-periodic base flows, providing a framework for assessing fundamental limitations of sensor-less flow control strategies such as streamwise traveling waves and riblets~\cite{moajovJFM10,liemoajovJFM10,chaluh20,ranzarjovJFM21}. For spatially-periodic flows, generalized quasilinear approximations further enable scale separation via spectral filtering, retaining large-scale interactions while linearizing small-scale dynamics~\cite{FarIoaJimConLozNik16}, with applications in both geophysical and canonical flows~\cite{marchitob16,HerCarQiaHwa22}. In contrast, the present framework retains a steady laminar base state and treats departures from it as emergent outputs of the Navier--Stokes equations, thereby preserving a direct link between classical resolvent analysis and the nonlinear interactions that it mediates.

In this paper, we introduce a perturbation-based frequency-response framework for analyzing amplification mechanisms that underpin subcritical routes to transition in wall-bounded shear flows. Rather than relying on stability analysis of a fixed streak-modified base flow, we examine the hierarchy of nonlinear interactions that emerge when the input--output dynamics of fluctuations about the laminar base flow are expanded with respect to the amplitude of external forcing. This formulation unifies linear resolvent analysis with weakly nonlinear theory, enabling systematic quantification of energy transfer across successive orders of nonlinearity and revealing how streamwise streaks arise and evolve through repeated coupling between oblique-wave disturbances and streaky flow components.

At second order, we focus on three-dimensional, unsteady, single-harmonic disturbances and show that quadratic interactions of the resulting oblique waves generate steady streamwise streaks via the lift-up mechanism. Embedding frequency-response analysis within this weakly nonlinear framework enables identification of the spatial structure of the most amplified flow fluctuations and clarifies the physical origin of dominant amplification pathways. In particular, the second output singular function of the streamwise-constant resolvent operator emerges as central to streak formation through nonlinear oblique-wave interactions, yielding excellent agreement with DNS.

At higher orders, nonlinear interactions between oblique waves and streaks generate additional streak components whose spatial structure is also captured by the second output singular function of the streamwise-constant resolvent operator. These higher-order contributions can be either {\em aligned\/} or {\em anti-aligned\/} with the second-order response, corresponding to reinforcement or attenuation of the leading-order streaks. Over a broad region of spectral parameter space, reinforcement persists across successive perturbation orders, leading to sustained amplification of streak energy that can ultimately trigger transition. Our analysis further identifies a critical forcing amplitude beyond which the perturbation series diverges, marking the breakdown of the weakly nonlinear regime. Direct numerical simulations corroborate these predictions, exhibiting sustained unsteadiness once this threshold is exceeded.

By employing convergence-acceleration techniques based on the Shanks transformation~\cite{shanks55,vandyke64}, we extend the practical applicability of the perturbation framework to disturbance amplitudes approaching this critical threshold. The resulting model captures both the quantitative evolution of streak energy and the accompanying mean-flow modification, providing a unified and computationally efficient description of subcritical transition dynamics. Collectively, these results establish higher-order perturbation analysis of the frequency response as a self-consistent, physics-based framework for characterizing the mechanisms governing streak reinforcement and transition onset.

We further demonstrate that the breakdown of the weakly nonlinear expansion coincides with the emergence of secondary instability. Unlike classical secondary-stability analysis---which assumes a prescribed streak amplitude to construct a distorted base flow---our approach determines streak-induced distortions directly through higher-order perturbation analysis of the frequency response, thereby linking non-modal amplification pathways to the modal instabilities that arise at larger amplitudes.

\vspace*{1ex}
\noindent {\bf Contributions and preview of key results.}
The main contributions of this work are threefold. First, we develop a perturbation-based frequency-response framework that unifies linear resolvent analysis with weakly nonlinear theory, providing a mechanistically transparent description of how non-modal amplification evolves into finite-amplitude streaks. Second, we show that higher-order nonlinear interactions generate self-similar streak corrections that reinforce or attenuate the leading-order response, with phase alignment between the interacting unsteady oblique fluctuations governing both streak-shape persistence and energy growth. We also identify a critical forcing amplitude marking the breakdown of the weakly nonlinear regime.
Third, we demonstrate that this breakdown coincides with the onset of secondary instability, thereby establishing a direct link between resolvent-driven non-modal amplification and classical modal transition theory without imposing an {\em a priori\/} streak-modified base flow.

To complement this high-level summary, we next provide a brief technical preview of the main results.

\vspace*{-1ex}
\begin{itemize}

\item A perturbation expansion of the frequency response with respect to forcing amplitude transforms {\em nonlinear fluctuation dynamics\/} into {\em a coupled hierarchy of linear systems.\/} At each perturbation order, the dynamics are governed by the {\em laminar resolvent operator\/}, with nonlinear interactions among lower-order responses driving higher-order evolution and providing a predictive description of subcritical amplification.
\vspace*{-1ex}

\item At second order, quadratic interactions of unsteady oblique waves generate steady streamwise streaks via the lift-up mechanism. The wall-normal structure of the resulting streaks is accurately captured by {\em the second output singular function of the streamwise-constant resolvent operator.}
\vspace*{-1ex}

\item At higher orders, nonlinear coupling between oblique waves and induced streaks acts as structured forcing of the laminar linearized dynamics, yielding additional streak components with persistent wall-normal structure whose phase governs their reinforcing or attenuating effect.
\vspace*{-1ex}

\item A critical forcing amplitude is identified that marks the breakdown of the perturbation expansion, beyond which DNS exhibits sustained unsteadiness. This breakdown coincides with the onset of secondary instability of the streak-distorted flow, thereby linking resolvent-driven non-modal amplification to classical modal transition theory without imposing an {\em a priori\/} streak-modified base flow.
 \vspace*{-1ex}

\item All theoretical predictions of the perturbation-based framework are validated against a harmonic-balance-based nonlinear optimization framework and against DNS.
\end{itemize}
		
	  \vspace*{-1ex}
\noindent {\bf Scope and limitations.}
The present framework is intentionally developed within a weakly nonlinear regime, in which flow responses can be organized systematically through a perturbation expansion about the laminar base flow. Accordingly, it does not aim to describe fully developed turbulence or strongly nonlinear saturation dynamics. Instead, the focus is on elucidating the amplification mechanisms that govern the emergence and early evolution of streaks, and on identifying the amplitude scales at which weakly nonlinear descriptions cease to be predictive. The critical forcing amplitude $\epsilon_{\mathrm{cr}}$ identified here should therefore be interpreted not as a sharp bifurcation threshold, but as a practical indicator of the transition from weakly nonlinear amplification to regimes in which additional physical mechanisms---such as modal instability and sustained unsteadiness---become relevant. While the analysis is presented for canonical channel flow with externally imposed disturbances, the underlying methodology is general and readily extendable to other wall-bounded shear flows, alternative forcing mechanisms, and pre-transitional settings in which nonmodal amplification, nonlinear interactions, and base-flow modification play a central role.
	
The paper is organized as follows.
Section~\ref{sec.setupWNL} develops the perturbation framework and reformulates the fluctuation dynamics around the laminar base flow as a coupled interconnection of identical linear systems, highlighting key properties of the frequency-response (resolvent) operators.
Section~\ref{sec.freq-NS} presents a weakly nonlinear frequency-response analysis and shows that the second output singular function of the streamwise-constant resolvent operator captures the spatial structure and dominant spanwise scale of streaks generated by small-amplitude oblique forcing.
Section~\ref{sec.higher-order-anal} extends the analysis beyond second order, examines the role of higher-order nonlinear interactions in streak reinforcement and attenuation, and establishes quantitative agreement between the accelerated perturbation expansion and direct numerical simulations.
Section~\ref{sec.sec-instab} validates the framework via secondary-stability analysis, demonstrating that the breakdown of the perturbation expansion coincides with the onset of modal instability of the streak-distorted base flow.
Finally, Section~\ref{eq.conclusion} summarizes the main findings and outlines directions for future work.

 \vspace*{-3ex}
\section{Dynamics of flow fluctuations}
	\label{sec.setupWNL}
	
\begin{figure}
\centering
\includegraphics[width = 0.5\textwidth]{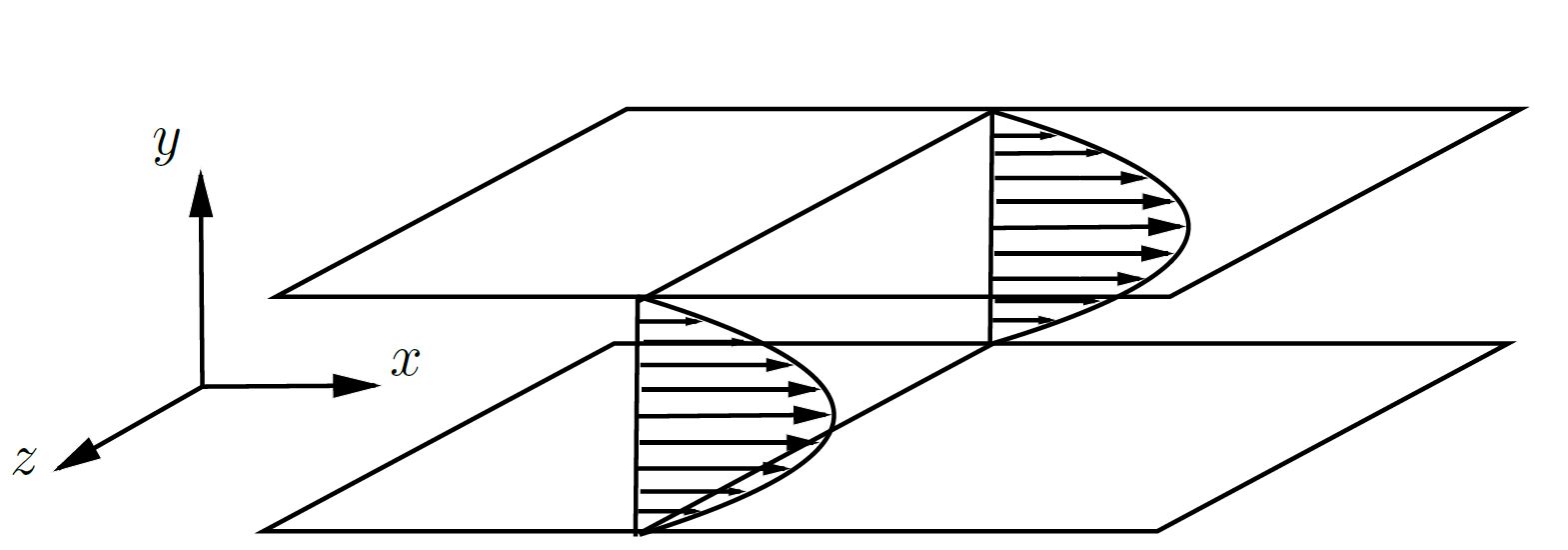}
\caption{Pressure driven flow between two parallel infinitely long plates with parabolic laminar velocity profile.}
    \label{fig:channel_sketch}
    \end{figure}

	\vspace*{-1ex}
	Our objective is to analyze the response of velocity fluctuations to exogenous disturbances in the channel flow of an incompressible Newtonian fluid; see Fig.~\ref{fig:channel_sketch} for the flow geometry. In this section, we first present a dynamical description of these fluctuations. For small-amplitude disturbances, we then employ perturbation analysis to recast the Navier--Stokes equations into a coupled interconnection of identical linear systems, each driven by the nonlinear modulation of velocity fluctuations from lower perturbation orders.
			
	The evolution of velocity and pressure fluctuations, $(\bu, p)$, around a laminar base flow $(\bar{\bu}, \bar{p})$ is governed by the incompressible NS equations,
\beq
\begin{aligned}
	\partial_t \bu 
	& 
	\; = \;
	\frac{1}{Re} \, \Delta \bu
	\, - \,
	( \bar{\bu} \cdot \nabla ) \bu 
	\, - \,
	( \bu \cdot \nabla ) \bar{\bu}
	\, - \,
	( \bu \cdot \nabla ) \bu
	\, - \,
	\nabla p
	\, + \,
	\bd 
	\\[0.1cm]
	0 
	& \; = \; 
	\nabla \cdot \bu
\end{aligned}
	\label{eq.NS}
\eeq
where $\partial_t$ denotes the time derivative, $\nabla$ the gradient, and $\Delta \DefinedAs \nabla \cdot \nabla$ the Laplacian. The velocity and forcing fluctuations in the three spatial directions $\bx \DefinedAs (x, y, z)$ are $\bu \DefinedAs (u, v, w)$ and $\bd \DefinedAs (d_u, d_v, d_w)$, respectively. The steady-state solution of the NS equations is $\bar{\bu} \DefinedAs (U(y), 0, 0)$, with $U(y) = 1 - y^2$ for pressure-driven Poiseuille flow.

All quantities have been non-dimensionalized with the following scales: spatial coordinates by the channel half-width $\delta$, velocities by the maximum laminar velocity $\bar{U}$, time by the inertial scale $\delta / \bar{U}$, and pressure by $\rho \bar{U}^2$, where $\rho$ denotes the fluid density. This scaling introduces the Reynolds number, $Re \DefinedAs \bar{U} \delta / \nu$, with $\nu$ representing the kinematic viscosity.

For small-amplitude inputs of magnitude $\epsilon$,
	\begin{subequations}
	\label{eq:distpert}
	\beq
	\bd (\bx, t)
	\; = \;
	\epsilon \bd^{(1)}(\bx, t)
	\label{eq.d-eps}
	\eeq
a perturbation analysis can be employed to study the externally forced NS equations~\eqref{eq.NS} and to determine corrections to the steady base flow $\bar{\bu}$. Expanding the velocity and pressure fluctuations $(\bu, p)$ in powers of~$\epsilon$ yields
	\beq
	\ba{rcl}
	\bu (\bx, t)
	& = &
	\displaystyle{ \sum_{n \, = \, 1}^{\infty} }
	\epsilon^n \bu^{(n)} (\bx, t) 
	\; = \;
	\epsilon \bu^{(1)} (\bx, t) 
	\; + \; 
	\epsilon^2 \bu^{(2)} (\bx, t) 
	\; + \; 
	\ldots
	\\[0.25cm]
	p (\bx, t)
	& = &
	\displaystyle{ \sum_{n \, = \, 1}^{\infty} }
	\epsilon^n p^{(n)} (\bx, t) 
	\; = \;
	\epsilon p^{(1)} (\bx, t) 
	\; + \; 
	\epsilon^2 p^{(2)} (\bx, t) 
	\; + \; 
	\ldots
	\ea
	\label{eq.u-eps}
	\eeq
	\end{subequations}
so that, for each order $n \geq 1$, the quantities $\bu^{(n)} (\bx, t)$ and $p^{(n)} (\bx, t)$ satisfy
	\begin{subequations} 
	\label{eq.NSn-dn}	
	\beq
	\begin{aligned}
	\partial_t \bu^{(n)} 
	& \; = \;
	\frac{1}{Re} \, \Delta \bu^{(n)}
	\, - \,
	( \bar{\bu} \cdot \nabla ) \bu^{(n)}
	\, - \,
	( \bu^{(n)} \! \cdot \nabla ) \bar{\bu}
	\, - \,
	\nabla p^{(n)}
	\, + \,
	\bd^{(n)}
	\\[0.1cm]
	0 
	& \; = \;
	\nabla \cdot \bu^{(n)}
	\end{aligned}
	\label{eq.NSn}
	\eeq
where $\bd^{(1)}$ represents the exogenous input defined in~\eqref{eq.d-eps}. For $n \geq 2$, the effective input $\bd^{(n)}$ accounts for the quadratic interactions among lower-order terms,
	\beq
	\bd^{(n)}
	\; \DefinedAs \;
	\, - \,
	\sum_{r\,=\,1}^{n-1}
	( \bu^{(r)} \! \cdot \nabla ) \bu^{(n-r)},
	~~
	n \, \geq \, 2.
	\label{eq.d_n}
	\eeq	
	\end{subequations}

Equations~\eqref{eq.NSn}-\eqref{eq.d_n} are obtained by substituting~\eqref{eq.d-eps} and~\eqref{eq.u-eps} into the NS equations~\eqref{eq.NS} and neglecting $\mathcal{O}(\epsilon^{n+1})$ terms. These equations are parameterized by the Reynolds number $Re$ and the laminar velocity profile $\bar{\bu}$, and apart from the small-amplitude assumption~\eqref{eq.d-eps}, no additional structural restrictions are imposed on the input $\bd(\bx, t)$. The precise meaning of ``small amplitude'' is quantified in Section~\ref{sec.higher-order-anal}.

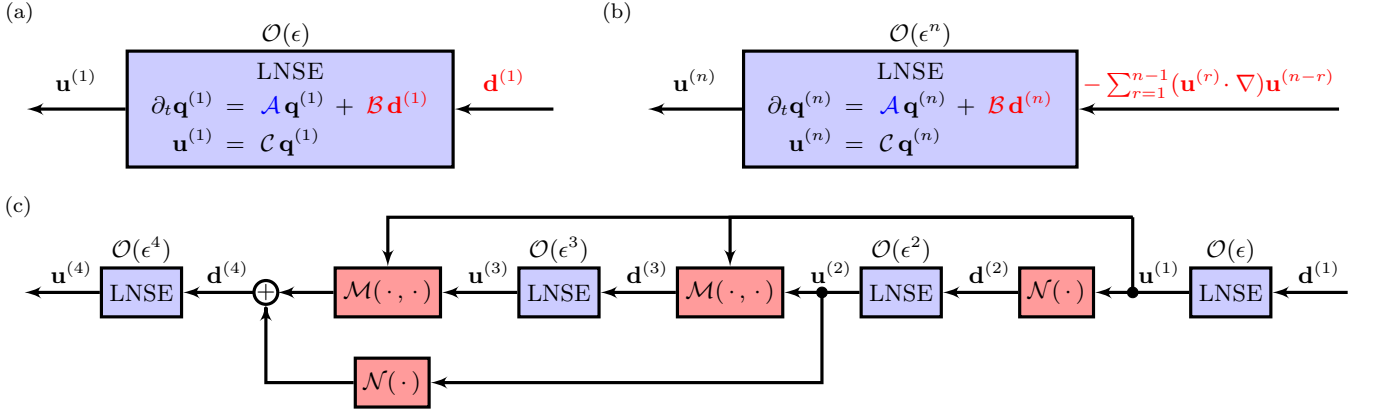
\begin{figure*}[t!]
  \centering
    \begin{tabular}{ll}
    \hspace{-0.1cm}
    \subfigure[]{\label{fig.Oeps-lnse}}
    &
    \hspace{0.2cm}
    \subfigure[]{\label{fig.Oepsn-lnse}}
    \\[-0.2cm]
	\hspace*{-0.45cm}
	\begin{tabular}{c}
	$\cO (\epsilon)$
	\\[0.cm]
	\input{figures/Figure2a}	
	\end{tabular}
	&
	\hspace*{-0.5cm}
	\begin{tabular}{c}
	\hspace*{-3.28cm}
	$\cO (\epsilon^n)$
	\\[0.cm]
	\input{figures/Figure2b}	
	\end{tabular}
  	\\
    \subfigure[]{\label{fig.bd-eps4}}
    \\[-0.2cm] & \hspace{-7.5cm}
    \input{figures/Figure2c}
\end{tabular}
	\caption{For small-amplitude exogenous inputs, $\bd = \epsilon \bd^{(1)}$, perturbation analysis transforms the Navier--Stokes equations~\eqref{eq.NS} into a coupled hierarchy of linearized systems around the equilibrium profile $\bar{\bu}$. The dynamics of velocity fluctuations $\bu^{(n)}$ at $\mathcal{O}(\epsilon^n)$ are driven by:~(a) the exogenous input $\bd^{(1)}$ for $n = 1$; and~(b) the nonlinear cross-interaction $\bd^{(n)} \DefinedAs - \sum_{r = 1}^{n - 1} (\bu^{(r)} \! \cdot \nabla)\bu^{(n - r)}$ of lower-order responses $\bu^{(r)}$ for $r = 1, \dots, n - 1$, when $n \geq 2$.~(c) Block diagram illustrating the coupling structure, i.e., the propagation of disturbances, up to $\mathcal{O}(\epsilon^4)$; operators $\cN_1$ and $\cN_2$, which generate quadratic self- and cross-interactions of lower-order responses, are defined in~\eqref{eq.NM}.}
 	\label{fig.bd-wn-psi_n}
\end{figure*}

To eliminate the pressure term and obtain an evolution model suitable for input--output analysis, we employ the standard wall-normal velocity/vorticity representation $(v, \eta) = (v, \partial_z u - \partial_x w)$~\cite{schhen01},	
	\begin{subequations}
	\beq
	\ba{rcl}
	\partial_t  \bq^{(n)} (\bx, t)
	& = &
	[ \cA \bq^{(n)} ( \, \cdot \, , t) ] (\bx)
	\; + \; 
	[ \cB \bd^{(n)} ( \, \cdot \, , t) ] (\bx)
	\\[0.1cm]
	\bu^{(n)} (\bx, t)
	& = &
	[ \cC \bq^{(n)} ( \, \cdot \, , t) ] (\bx)
	\ea
	\label{eq.ss-On}
	\eeq
where $\bq^{(n)}$ denotes the $\mathcal{O}(\epsilon^n)$ term in the power-series expansion of the state $\bq \DefinedAs (v, \eta)$,
	\beq
	\bq (\bx, t)
	\; = \; 
	\sum_{n = 1}^{\infty}
	\epsilon^n \bq^{(n)} (\bx, t).
	\eeq
Here, $\cA$ is the Orr-Sommerfeld/Squire operator obtained by linearizing the NS equations~\eqref{eq.NS} about the laminar base flow $(\bar{\bu}, \bar{p})$; $\cB$ specifies how the input $\bd^{(n)}$ enters the $\mathcal{O}(\epsilon^n)$ dynamics; and $\cC$ maps the state $\bq^{(n)}$ to the corresponding velocity field~\cite{jovbamJFM05}. The no-slip and no-penetration conditions on velocity fluctuations imply
	\beq
 	v^{(n)} (x,y= \pm 1,z,t) 
	\; = \; 
	\partial_y v^{(n)} (x,y= \pm 1,z,t) 
	\; = \;
	\eta^{(n)} (x,y= \pm 1,z,t) 
	\; = \; 
	0.
	\eeq
 
It is worth noting that, for all $n \geq 1$, the same spatial operators $\cA$, $\cB$, and $\cC$ appear in~\eqref{eq.ss-On}. Consequently, while the standard linearized model (corresponding to the $\mathcal{O}(\epsilon)$ dynamics) is driven by the exogenous input $\bd^{(1)}(\bx, t)$, the higher-order dynamics at $\mathcal{O}(\epsilon^n)$ for $n \geq 2$ are driven by the nonlinear modulation~\eqref{eq.d_n} of lower-order responses $\bu^{(r)}$ with $r = 1, \dots, n - 1$; see Fig.~\ref{fig.bd-wn-psi_n} for an illustration.
  
By introducing the nonlinear operators $\cN_1$ and $\cN_2$, which generate quadratic self- and cross-interactions between velocity fluctuations,
	\beq
	\cN_1 (\bu)
	\; \DefinedAs \;
	- 
	( \bu \cdot \nabla ) \bu,
	~~
	\cN_2 (\bu, \bv)
	\; \DefinedAs \;
	- 
	( \bu \cdot \nabla ) \bv
	\, - \,
	( \bv \cdot \nabla ) \bu
	\label{eq.NM}
	\eeq	
the input $\bd^{(n)}$ to~\eqref{eq.NSn} for $n \geq 2$ can be expressed as
	\beq
	\bd^{(n)}
	\; = \,
	\left\{
	\ba{rl}
	\cN_1(\bu^{(n/2)})
	\; + 
	\displaystyle{ \sum_{r\,=\,1}^{(n - 2)/2} }
	\cN_2 (\bu^{(r)}, \bu^{(n-r)}),
	& 
	~
	n~\mbox{even}
	\\[0.45cm]
	\displaystyle{ \sum_{r\,=\,1}^{(n-1)/2} }
	\cN_2 (\bu^{(r)}, \bu^{(n-r)}), 
	&
	~
	n~\mbox{odd}
	\ea
	\right.
	\label{eq.dnMN}
	\eeq		
	\end{subequations}
and the resulting propagation of disturbances in the Navier--Stokes equations under small-amplitude exogenous forcing can be represented by the block diagram shown in Fig.~\ref{fig.bd-eps4}.

For streamwise-constant fluctuations, $\partial_x(\cdot) = 0$, the $\mathcal{O}(\epsilon^n)$ evolution model can be equivalently expressed in terms of the $(y,z)$-plane stream function $\psi$ and the streamwise velocity $u$,
	\beq 
	\ba{rcl}
	\partial_t \psi^{(n)}
	& = &
	\dfrac{1}{Re} \, \Delta^{-1} \Delta^2 \psi^{(n)}
	\; + \;
	d_{\psi}^{(n)}
	\\[0.2cm]
	\partial_t u^{(n)}
	& = &
	- U'(y) \partial_z
	\psi^{(n)}
	\; + \;
	\dfrac{1}{Re} \, \Delta u^{(n)}
	\; + \;
	d_u^{(n)}
	\ea
	\label{eq.ss-kx0-epsn}
	\eeq
where $\Delta = \partial_{yy} + \partial_{zz}$ is the two-dimensional Laplacian with homogeneous Dirichlet boundary conditions in $y$, and $\Delta^2 = \partial_{yyyy} + 2 \partial_{yy}\partial_{zz} + \partial_{zzzz}$ is its biharmonic counterpart with homogeneous Dirichlet and Neumann boundary conditions. The wall-normal and spanwise velocity fluctuations are given by $(v, w) = (\partial_z \psi, -\partial_y \psi)$, and $U'(y) = \mathrm{d}U(y)/\mathrm{d}y$ denotes the wall-normal derivative of the mean velocity. The forcing term $d_{\psi}^{(n)}$ is defined as
	\beq
	d_{\psi}^{(n)}
	\; = \;
	\Delta^{-1}
	\big(
	\partial_z d_v^{(n)} 
	\, - \,
	\partial_y d_w^{(n)} 
	\big).
	\eeq
This formulation will be used to investigate the evolution of streamwise streaks in flows driven by unsteady oblique disturbances.

	\vspace*{-4ex}
\subsection{Frequency responses of the NS equations linearized about the laminar base flow}
\label{sec.FrRes}

	\vspace*{-1ex}
	
Since the perturbation analysis transforms the NS equations into a coupled interconnection of identical linear systems, it is essential to understand the response of linearized dynamics to exogenous disturbances. This problem has been extensively studied (e.g., see~\cite{jovbamJFM05,schmid2007nonmodal,jovARFM21}); here, we briefly summarize the main aspects relevant to the frequency-response analysis of channel flow and present representative results for Poiseuille flow at $Re = 2000$.	

Owing to homogeneity in the wall-parallel directions $(x, z)$ and in time $t$, the Fourier transform in $(x, z, t)$ converts system~\eqref{eq.ss-On} into a form parameterized by the wave-number vector $\bk \DefinedAs (k_x, k_z)$ and the temporal frequency $\omega$. This decomposition represents all signals in~\eqref{eq.ss-On} as Fourier modes in $(x, z, t)$, e.g.,
	\begin{subequations}
	\begin{eqnarray}
	\bd (x,y,z,t) 
	& \; = ~ &
	{\bd}_{\bk} (y, \omega)
	\,
	\mre^{\mri (k_x x + k_z z + \omega t)}
	\label{eq.d-ft}
	\\[0.1cm]
	\bu (x,y,z,t) 
	& \; = ~ &
	{\bu}_{\bk} (y, \omega)
	\,
	\mre^{\mri (k_x x + k_z z + \omega t)}
	\label{eq.u-ft}
	\end{eqnarray}
	\end{subequations}	
where, for notational simplicity, we suppress the dependence on the perturbation-expansion order~$n$. The Fourier transforms of the input $\bd (\bx, t)$ and the output $\bu (\bx, t)$ in~\eqref{eq.ss-On} are related by
	\beq
	{\bu}_{\bk} ( y , \omega) 
	\; = \;
	[ \cHk (\omega)  {\bd}_{\bk} ( \, \cdot \, , \omega)] (y)
	\label{eq.input--output}
	\eeq
where the frequency-response operator	
	\beq
	 \cHk (\omega) 
	 \; = \;
	 \cCk (\mri \omega I \, - \, \cAk)^{-1} \cBk
	\label{eq.Hfr}	
	\eeq
characterizes the amplitude and phase of the response of the stable system~\eqref{eq.ss-On} to periodic inputs~\eqref{eq.d-ft} across spatial wavenumbers~$\bk$ and temporal frequencies~$\omega$; see Appendix~\ref{app.Operators} for definitions of the corresponding Fourier symbols $(\cAk, \cBk, \cCk)$.

The singular-value decomposition (SVD) of $\cHk(\omega)$ yields orthonormal bases for the input and output spaces and quantifies their respective input--output gains,
	\beq
	{\bu}_\bk (y,\omega)
	\; = \;
	\sum_{j \, = \, 1}^{\infty}
	\sigma_{\bk,j} (\omega)
	\inner{{\bdelta}_{\bk,j} ( \, \cdot \, , \omega)}{{\bd}_\bk ( \, \cdot \, , \omega)}
	{\bvartheta}_{\bk,j} (y,\omega).
		\label{eq.svd-adjoint}
	\eeq 	
The right and left singular functions, ${\bdelta}_{\bk,j}$ and ${\bvartheta}_{\bk,j}$, provide orthonormal bases for the input and output spaces and are referred to as the input and output singular functions (or modes), respectively.
Each singular value $\sigma_{\bk,j} (\omega)$ quantifies the amplification of ${\bdelta}_{\bk,j}$ by the linearized dynamics, and $\inner{\, \cdot \,}{\, \cdot \,}$ denotes the $L_2$ inner product in~$y$.
By sorting $\sigma_{\bk,j} (\omega)$ in descending order with respect to~$j$, the principal singular value
	\beq
	\sigma_{\bk,1} (\omega) 
	\; \DefinedAs \; 
	\sigma_{\max} (\cHk (\omega))
	\eeq 	
identifies the largest amplification at a given~$(\bk,\omega)$, and the supremum over~$\omega$ defines the $H_\infty$ norm of the system~\eqref{eq.ss-On}~\cite{zhodoyglo96},
	\beq
		\| \cHk \|_\infty^2
		\; = \; 
		\sup_{\omega} \; \sigma_{\max}^2 (\cHk (\omega)).
	\label{eq.Hinf}
	\eeq
For ${\bd}_\bk (y, \omega) = {\bdelta}_{\bk,j} (y,\omega)$ in~\eqref{eq.d-ft}, the steady-state response of the stable system~\eqref{eq.ss-On} is given by~\eqref{eq.u-ft}, where ${\bu}_\bk$ aligns with the $j$th output singular function ${\bvartheta}_{\bk,j}$,	
	\beq
	{\bu}_\bk (y, \omega)
	\; = \; 
	 \sigma_{\bk,j} (\omega) {\bvartheta}_{\bk,j} (y,\omega)
	 \eeq
and its energy is	
	\beq
	\| {\bu}_\bk (\omega) \|_2^2 
	\; \DefinedAs \;
	\sigma_{\bk,j}^2 (\omega)
	\eeq
with 
	$ \| \cdot \|_2$ 
denoting the $L_2$ norm in~$y$.

We employ ultraspherical and spectral-integration schemes~\cite{harkumjovJCP21} to compute the $H_{\infty}$ norm of the linearized NS equations in pressure-driven channel flow at $Re = 2000$. The method developed in~\cite{harkumjovJCP21} leverages the Hamiltonian formulation of the linearized system~\cite{boyd89} and enables accurate computation of the $H_{\infty}$ norm without explicitly evaluating $\sigma_{\bk,1} (\omega)$ across different frequencies~$\omega$.

\begin{figure}[t]
 \centering
  \begin{tabular}{ccccc}
    \subfigure[]{\label{fig:Hinf_a}}
    & $\log_{10} \| \cHk \|_\infty^2$ 
    &\subfigure[]{\label{fig.sigma-max-omega}}&
    \\[-.26cm]\hspace{0.1cm}
    \begin{tabular}{c}
      \vspace{.25cm}
      \hspace{-0.15cm}
      \normalsize{\rotatebox{90}{$k_x$}}
    \end{tabular}
    &
    \begin{tabular}{c}
      \includegraphics[width=0.3\textwidth]{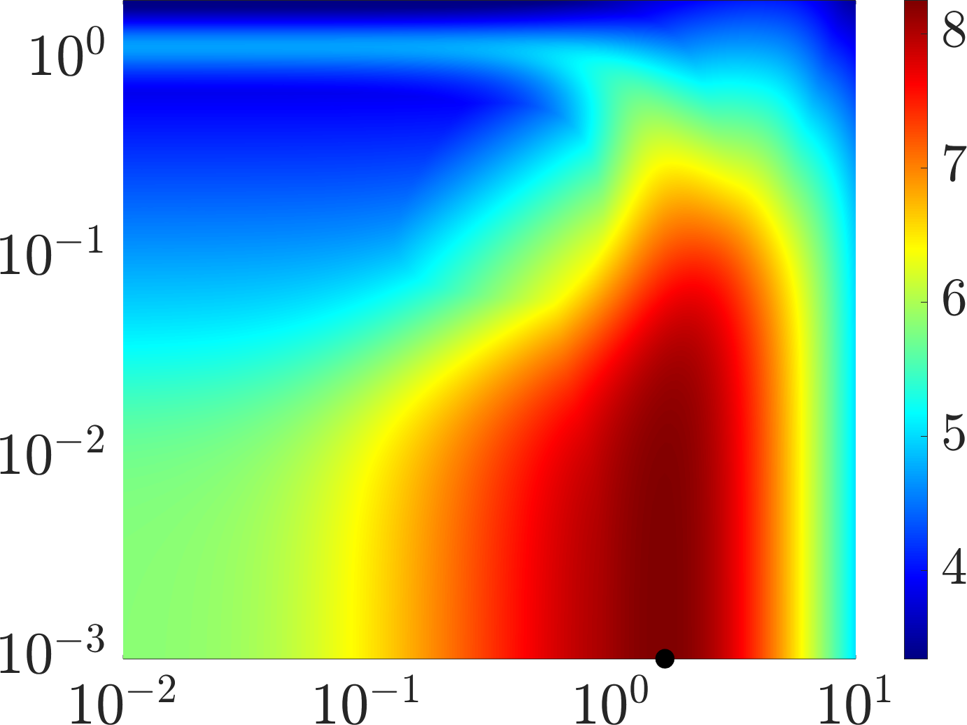}
      \\[-0.cm] {$k_z$}
    \end{tabular}
    &
    \begin{tabular}{c}
        \vspace{.25cm}
        \normalsize{\rotatebox{90}{$\sigma_{\max}^2( \cHk (\omega) )$}}
      \end{tabular}
      &
      \begin{tabular}{c}
        \includegraphics[width=0.313\textwidth]{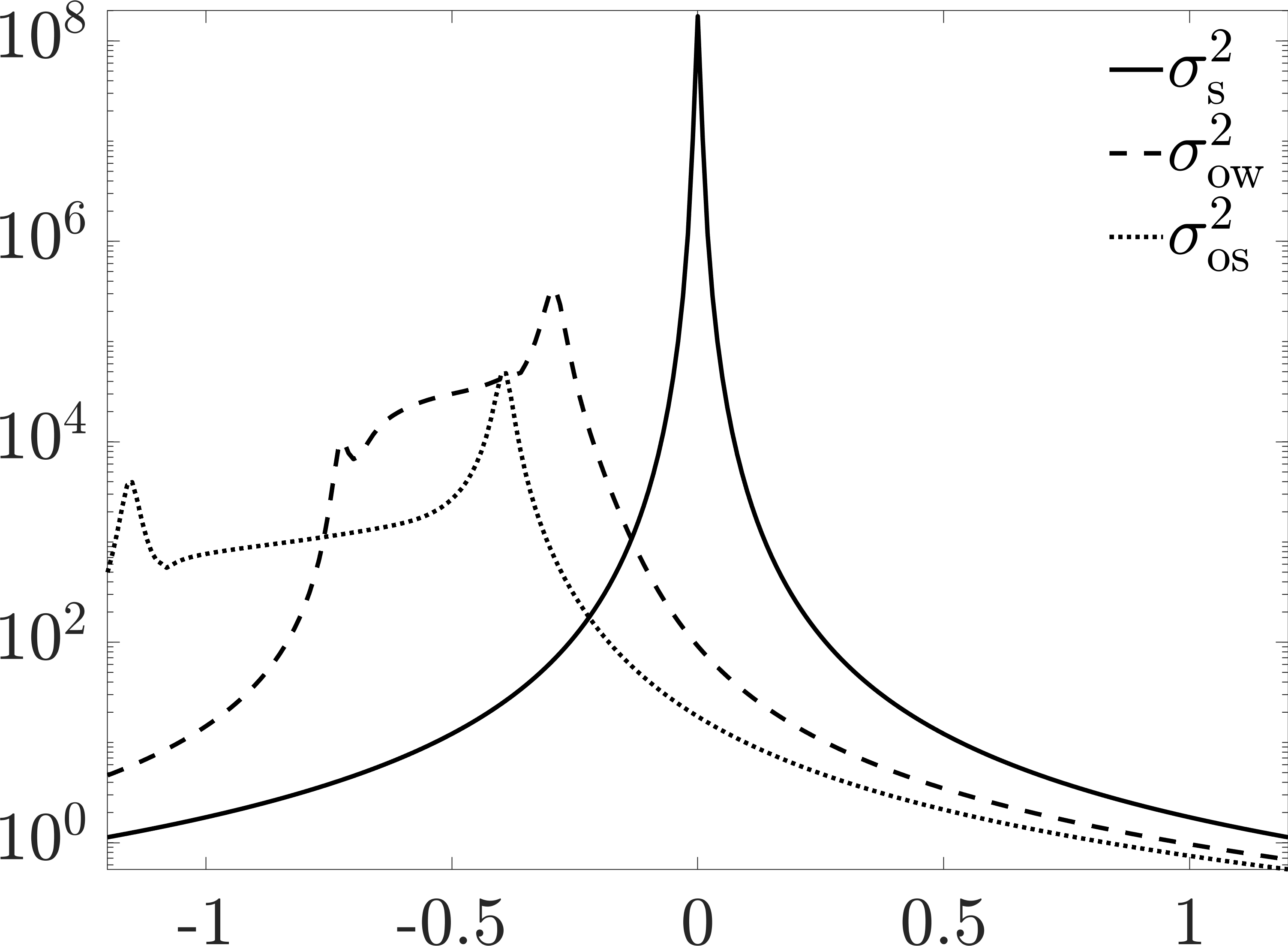}
        \\[-0.cm]
        \hspace{0.4cm}
        {$\omega$}
      \end{tabular}
  \end{tabular}
  \vspace{-0.1cm}
 \caption{(a) Square of the $H_{\infty}$ norm for Poiseuille flow with $Re = 2000$.
The black dot marks the pair $(k_x, k_z)$ corresponding to the largest amplification.
(b) Dependence of the principal singular value on temporal frequency~$\omega$ for streamwise streaks with $(k_x, k_z) = (0, 1.65)$ (solid), oblique waves with $(k_x, k_z) = (0.74, 1.14)$ (dashed), and two-dimensional OS modes with $(k_x, k_z) = (1.17, 0)$ (dotted).
While the strongest amplification of streamwise streaks occurs for steady disturbances ($\omega = 0$), oblique waves and OS modes exhibit pronounced amplification for unsteady disturbances with negative~$\omega$.}
\label{fig:Hinf}
\end{figure}

Figure~\ref{fig:Hinf_a} shows the dependence of the square of the $H_{\infty}$ norm on the wavenumbers $(k_x, k_z)$. The dark-red region indicates strong amplification of streamwise-elongated flow structures with a characteristic spanwise length scale of order~$\mathcal{O}(1)$. The largest amplification occurs at $(k_x, k_z) = (0, 1.65)$, and steady disturbances ($\omega = 0$) experience the greatest gain; see Figure~\ref{fig.sigma-max-omega}. These most energetic responses of the linearized NS equations take the form of streamwise velocity streaks, excited by vortical disturbances via the lift-up mechanism~\cite{schmid2007nonmodal,jovARFM21}.

Figure~\ref{fig.sigma-max-omega} depicts the dependence of the principal singular value of the frequency-response operator on the temporal frequency~$\omega$ for three representative cases: streamwise-constant fluctuations (streaks) at $(k_x, k_z) = (0, 1.65)$; three-dimensional fluctuations (oblique waves) with phase-speed vectors inclined obliquely to the base flow at $(k_x, k_z) = (0.74, 1.14)$; and two-dimensional Orr-Sommerfeld (OS) modes at $(k_x, k_z) = (1.17, 0)$. The OS modes correspond to underdamped two-dimensional eigenmodes of the linearized NS equations that become unstable when the Reynolds number exceeds the critical value $Re_c = 5772$.

	\vspace*{-2ex}
\section{Weakly nonlinear frequency response analysis}	
	\label{sec.freq-NS}
	\vspace*{-1ex}
    
Input--output analysis of the linearized Navier--Stokes equations identifies steady streamwise streaks as the most amplified flow structures.
In contrast, direct numerical simulations have shown that oblique waves provide a more effective pathway for triggering transition~\cite{schmid1992new,RSBH1998}.
Similar observations have been reported in wind-tunnel experiments and numerical studies across a range of shear flows, including low-speed boundary layers~\cite{BrandtWNL,berlin1999nonlinear}, compressible high-speed boundary layers~\cite{chang_malik_1994}, and separated flows involving shock-wave/boundary-layer interactions~\cite{dwisidjovJFM22}.
Recent investigations employing structured uncertainty representations that capture the nonlinear terms of the NS equations and that are based on structured singular values~\cite{liugay21,musbhaseihem23}, have likewise identified oblique waves as the most energetic structures in channel flow~\cite{liugay21}.

Building on these observations, and extending the weakly nonlinear framework of~\cite{PonzianiWNLPois,brandt2002weakly}, in this section we develop a second-order frequency-response perturbation analysis to examine fluctuations governed by the NS equations.
Our analysis focuses on three-dimensional, unsteady, single-harmonic disturbances and the resulting nonlinear interactions among their responses.
In particular, we show that quadratic interactions of unsteady oblique waves give rise to steady streamwise streaks.
By embedding linear frequency-response analysis within this weakly nonlinear framework, we identify the spatial structure of the most energetic flow fluctuations and elucidate the underlying amplification mechanisms.
We conclude this section by comparing model predictions with results from direct numerical simulations.

	\vspace*{-2ex}
	\subsection{Response to three-dimensional unsteady single-harmonic inputs}
	\label{sec.response-single-harm}
	
	\vspace*{-1ex}
We now confine our attention to small-amplitude harmonic inputs~\eqref{eq.d-eps} to the Navier--Stokes equations~\eqref{eq.NS}, characterized by temporal frequency~$\omega$ and wall-parallel wavenumbers $\bk = (k_x, k_z)$,	
	\beq
	\ba{rcl}
	\bd^{(1)} (\bx,t) 
	& = &
	\Re 
	\,
	\big\{
	\big( \bd_{(\bkx,\bkz)} ^{(1)} (y,\bom) \, \mre^{\mri ( \bkx x + \bom t )} 
	~ + ~
	\bd_{(-\bkx,\bkz)} ^{(1)} (y,-\bom) \, \mre^{-\mri ( \bkx x + \bom t )} 
	\big)
	\,\mre^{\mri \bkz z}
	\big\}
	\ea
	\label{eq.dow}
	\eeq
where $\Re\{\cdot\}$ denotes the real part. As shown in Figure~\ref{fig.sigma-max-omega}, 3D disturbances experience strong amplification when $k_x$ and $\omega$ have opposite signs; we therefore focus on oblique disturbances that travel downstream.
Setting $\omega = - k_x c$, where $c$ is the phase speed, removes explicit time dependence in~\eqref{eq.dow} via the Galilean transformation 
	$
    \bar{x} \DefinedAs x - \bc t,
    $
    \begin{subequations}
	\beq
	\ba{rcl}
	\bd^{(1)}(\bar{x},y,z) 
	& = &
	\Re 
	\,
	\big\{
	\big(
	\bd_{(\bkx,\bkz)} ^{(1)} (y,c) \, \mre^{\mri \bkx \bar{x}} 
	~ + ~
	\bd_{(-\bkx,\bkz)} ^{(1)} (y,c) \, \mre^{- \mri \bkx \bar{x}} 
	\big)
	\, \mre^{\mri \bkz z}
	\big\}
	\ea
	\label{eq.dow-new}
	\eeq
where
	\beq
	\bd_{(\pm \bkx,\bkz)} ^{(1)} (y,c)
	\; \DefinedAs \;
	\bd_{(\pm \bkx,\bkz)} ^{(1)} (y, \mp k_x c).
	\eeq
Spatial derivatives remain unchanged under this transformation, while the time derivative becomes    
	$
    \partial_{t}
    -
    \bc \, \partial_{\bar{x}},
    $
introducing an additional convective term $c \partial_{\bar{x}}\bu$ into the NS equations~\eqref{eq.NS}~\cite{moajovJFM10},
	\beq
\begin{aligned}
	\partial_t \bu 
	& 
	\; = \;
	\frac{1}{Re} \, \Delta \bu
	\, + \,
	c \partial_{\bar{x}} \bu
	\, - \,
	( \bar{\bu} \cdot \nabla ) \bu 
	\, - \,
	( \bu \cdot \nabla ) \bar{\bu}
	\, - \,
	( \bu \cdot \nabla ) \bu
	\, - \,
	\nabla p
	\, + \,
	\epsilon \, \bd^{(1)} 
	\\[0.1cm]
	0 
	& \; = \; 
	\nabla \cdot \bu.
	\end{aligned}
	\label{eq.NS-c}
	\eeq
	\end{subequations}
In this moving frame, the perturbation analysis quantifies the steady response of the transformed system~\eqref{eq.NS-c} to the time-independent input $\bd^{(1)}(\bar{x}, y, z)$ given by~\eqref{eq.dow-new}. For steady fluctuations, the time derivative $\partial_t \bu$ in~\eqref{eq.NS-c} is set to zero, and the frequency-response operator $\cGk(c)$ associated with the linearization of~\eqref{eq.NS-c} about $\bar{\bu}$ is defined as
	\beq
	\cGk (c) 
	\; \DefinedAs \;
	\cHk (-c k_x)
	\; = \;
	- \cCk (\mri c k_x I \, + \, \cAk)^{-1} \cBk.
	\eeq

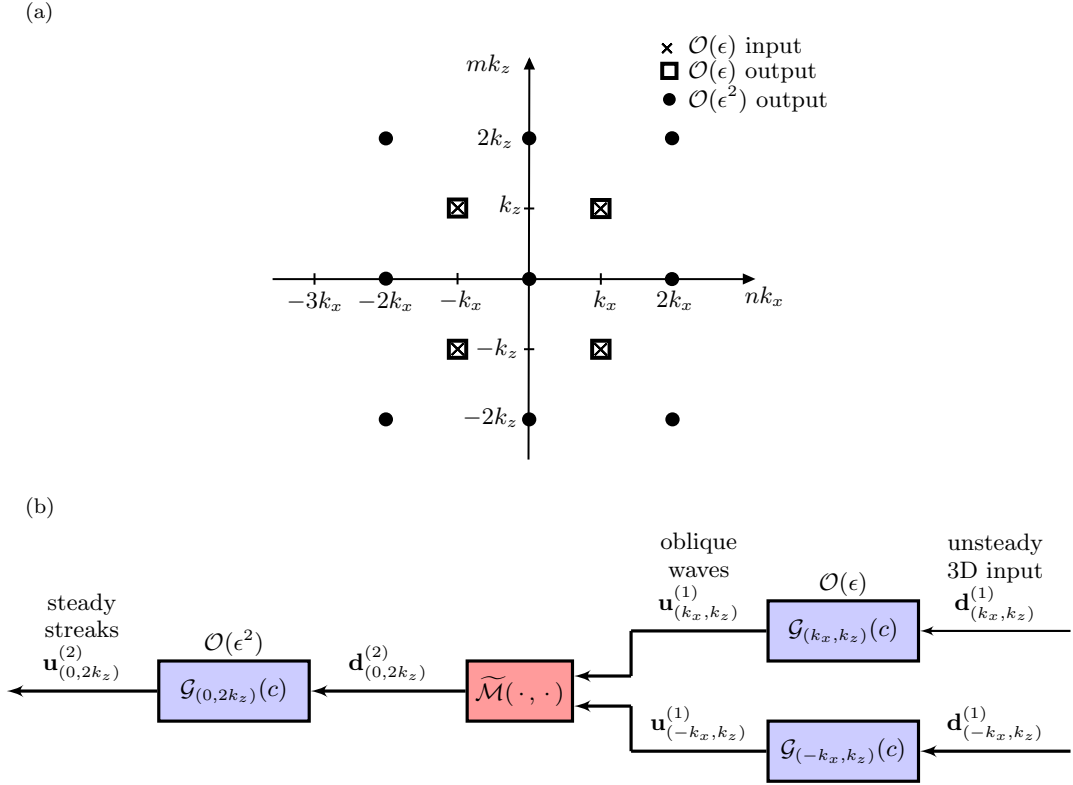
\begin{figure*}[t]
  \centering
  \begin{tabular}{cc}
    \subfigure[]{\label{fig.freq-graph}}
    &
    \\[-.13cm] 
    &
    \hspace{-0.5cm}
    \begin{tabular}{c}
      \vspace{.25cm}
        \input{figures/Figure4a}
    \end{tabular}
    \\[-0.13cm]
    \subfigure[]{\label{diag.OWstr}} 
    &
    \\[-0.13cm]
    &
    \hspace{-0.8cm}
    \begin{tabular}{l}
     \input{figures/Figure4b}
    \end{tabular}  
  \end{tabular}
  \vspace{-0.1cm}
  \caption{(a) Spatial wavenumbers that arise in the $\mathcal{O}(\epsilon^2)$ expansion of the response of the NS equations to small-amplitude three-dimensional disturbances. Input wavenumbers are marked by crosses, the $\mathcal{O}(\epsilon)$ response by squares, and nine $\mathcal{O}(\epsilon^2)$ terms by black dots. (b) Small-amplitude harmonic input~\eqref{eq.dow} to the NS equations~\eqref{eq.NS}, with wavenumbers $(\pm k_x, k_z)$ and temporal frequency $\omega = \mp k_x c$, excites $\mathcal{O}(\epsilon)$ oblique waves whose quadratic interactions generate $\mathcal{O}(\epsilon^2)$ steady vortical forcing and thereby streamwise streaks. The operator $\widetilde{\cN}_2$ denotes the Fourier-transformed cross-interaction nonlinear operator $\cN_2$ given in~\eqref{eq.NM}, $\widetilde{\cN}_2(\bu_{\bk}, \bu_{\bk'}) \DefinedAs -(\bu_{\bk} \cdot \nabla_{\bk'}) \bu_{\bk'} - (\bu_{\bk'} \cdot \nabla_{\bk}) \bu_{\bk}$, where $\nabla_{\bk} \DefinedAs [\, \mri k_x \; \partial_y \; \mri k_z ]^T$. The operators $\cG_{(\pm k_x, k_z)}(c)$ and $\cG_{(0, 2k_z)}(c)$, associated with the linearized NS equations at $(\pm k_x, k_z, \mp k_x c)$ and $(0, 2 k_z, 0)$, completely characterize the resulting responses.}
      \label{fig.Together}
\end{figure*}

Figure~\ref{fig.freq-graph} identifies all wavenumber pairs $(k_x, k_z)$ that appear in the $\mathcal{O}(\epsilon^2)$ expansion of the response to the $\mathcal{O}(\epsilon)$ input defined by~\eqref{eq.dow-new}. The resulting velocity field can be expressed as
	\beq
	\ba{rcl}
	\bu(\bar{x},y,z)
	& = &
	\overbrace{\big( 
	\bar{\bu} (y) 
	\; + \; 
	\tc{red}{\epsilon^2} \, 
	\bu_{(0,0)}^{(2)} (y,\bc)
	\big)}^{\ba{c} \mbox{\bf modified laminar flow} \ea} 
	+ ~\;	
	\tc{red}{\epsilon}
	\,
	\overbrace{\Re \,
	\big\{
	(
	\bu_{(\bkx,\bkz)}^{(1)} (y,\bc) \, \mre^{\mri \bkx \bar{x}}
	\; + \; 
	\bu_{(-\bkx,\bkz)}^{(1)} (y,\bc) \, \mre^{-\mri \bkx \bar{x}}
	)
	\,
	\mre^{\mri \bkz z} \big\}
    }^{\ba{c} \cO (\epsilon)~\mbox{\bf oblique waves $\bu_{\mathrm{ow}}^{(1)}$} \ea}
	~~ +
	\\[0.35cm]
	& &
	\tc{red}{\epsilon^2} 
	\underbrace{ 
	\Re
	\,
	\big\{
	{\bu}_{(0,2 \bkz)}^{(2)} (y,\bc) 
	\, 
	\mre^{2 \mri \bkz z}
	\big\}
	}_{\ba{c} \mbox{\tc{dred}{\bf $\cO (\epsilon^2)$ streaks $\bu_{\mathrm{s}}^{(2)}$}} \ea}
	~\, + ~\,
	\tc{red}{\epsilon^2} 
	\underbrace{
	\Re \,
	\big\{
	(
	\bu_{(2\bkx,2\bkz)}^{(2)} (y,\bc) \, \mre^{\mri 2 \bkx \bar{x}}
	\; + \; 
	\bu_{(-2\bkx,2\bkz)}^{(2)} (y,\bc) \, \mre^{-\mri 2 \bkx \bar{x}}
	)
	\, 
	\mre^{2 \mri k_z z}
	\big\}
    }_{\ba{c} \cO (\epsilon^2)~\mbox{\bf oblique waves $\bu_{\mathrm{ow}}^{(2)}$} \ea} 
	~~ +
	\\[0.95cm]
	& &
	\tc{red}{\epsilon^2} 
	\underbrace{
	\Re
	\,
	\big\{
	\bu_{(2\bkx,0)}^{(2)} (y,\bc) 
	\, \mre^{\mri 2 \bkx \bar{x}}
	\big\}
	}_{\mbox{$\cO (\epsilon^2)$~{\bf 2D modes $\bu_{\mathrm{2d}}^{(2)}$}}}
	~\, + ~\,
	{\cal O} (\epsilon^3).
	\ea
	\label{eq.u-eps2}
	\eeq
Here, $\bu_{(\pm k_x, k_z)}^{(1)}$ represents the $\mathcal{O}(\epsilon)$ oblique-wave response;
$\bu_{(0,0)}^{(2)}$ denotes the $\mathcal{O}(\epsilon^2)$ correction to the laminar base flow;
$\bu_{(0, 2 k_z)}^{(2)}$ corresponds to $\mathcal{O}(\epsilon^2)$ streamwise streaks;
$\bu_{(2 k_x, 0)}^{(2)}$ captures two-dimensional modes;
and $\bu_{(\pm 2 k_x, 2 k_z)}^{(2)}$ represents second-order oblique waves.	

Evaluation of the frequency response operator 
	$
	\cGk (c) 
	\DefinedAs
	\cHk (-c k_x)
	$
at selected values of $(k_x, k_z, \omega = -c k_x)$ yields the $\mathcal{O}(\epsilon)$ and $\mathcal{O}(\epsilon^2)$ contributions in~\eqref{eq.u-eps2},
	\beq
	\bu_{\bk_n}^{(n)} (y,c)
	\, = \,
	[
	\cG_{\bk_n} \! (c)
	\,
	\bd_{\bk_n}^{(n)} ( \, \cdot \, , c)
	] (y)
	\non
	\eeq  
where $\bk_1 = (\pm k_x, k_z)$ and $\bk_2 \in \{(0,0), (0, 2 k_z), (2 k_x, 0), (\pm 2 k_x, 2 k_z)\}$. The forcing terms $\bd_{\bk_2}^{(2)}$ arise from the nonlinear modulation of the response $\bu^{(1)}$ of the linearized NS equations to the external input $\bd_{\bk_1}^{(1)}$. Higher-order terms in the perturbation expansion can be obtained in an
analogous manner; see Appendix~\ref{app.terms-eval} \mbox{for details on the
$\mathcal{O}(\epsilon^4)$ computations.}
    
    	The block diagram in Figure~\ref{diag.OWstr} illustrates the mechanism by which unsteady oblique disturbances generate second-order steady streamwise streaks: small-amplitude three-dimensional inputs excite $\mathcal{O}(\epsilon)$ oblique waves through the linearized dynamics, and quadratic interactions of these unsteady waves induce $\mathcal{O}(\epsilon^2)$ steady streaks.
	    
	Finally, our formulation can be interpreted as a perturbation-analysis-based extension of classical harmonic balance methods, which collect terms at identical frequencies to obtain algebraic equations for the steady-state response~\cite{rigsipcol21}.
By employing perturbation analysis, we reveal how distinct response components interact and highlight the relative importance of each frequency via its amplitude scaling.
This framework thus provides a computationally efficient means for characterizing the response of the NS equations to small-amplitude harmonic three-dimensional inputs.

	\input{tables/table_eps2}

	\vspace*{-3ex}
\subsection{Second-order response to the principal oblique disturbance}
\label{sec:princWNL}

	\vspace*{-1ex}
We next examine the response of the Navier--Stokes equations to the small-amplitude input~\eqref{eq.dow-new} defined by the principal input singular function
	$
	{\bdelta}_{\bk,1} (y,c)
	$ 
of the frequency-response operator
	$
	\cG_{\bk} (c)
	$
associated with the oblique waves. Thus, the first-order input is chosen as
	\beq
	\bd_{\bk}^{(1)} (y,c)
	\; = \;
	{\bdelta}_{\bk,1} (y, c).
	\label{eq.dow-princ}
	\eeq
The corresponding $\mathcal{O}(\epsilon)$ output ${\bu}_{\bk}^{(1)}(y, c)$ aligns with the principal output singular function 
	$
	{\bvartheta}_{\bk,1} (y,c),
	$ 
	\beq
	\ba{rcl}
	{\bu}_{\bk}^{(1)} (y,c)
	& = &
	[ \cGk (c)  {\bdelta}_{\bk,1} ( \, \cdot \, , c) ] (y)
	\\[0.1cm]
	& = &
	\sigma_{\bk,1} (c) {\bvartheta}_{\bk,1} (y,c)
	\ea
	\label{eq.uhat-eps1}
	\eeq
and its energy is determined by the corresponding singular value,
   	$
	\| \bu_\bk^{(1)} (c) \|_2^2 
	=
	\sigma_{\bk,1}^2 (c).
	$
Owing to the channel symmetries in the streamwise and spanwise directions, the disturbances and their corresponding responses---capturing interactions between oblique waves and the induced streaks---can be summarized as shown in Table~\ref{tab.ow-ss}.
In particular, the $\mathcal{O}(\epsilon^2)$ streaks $\bu_{\mathrm{s}}^{(2)}(y, z)$ arise as the steady response of the streamwise-constant model~\eqref{eq.ss-kx0-epsn} to the input ($d_{\psi}^{(2)}, \,d_{u}^{(2)}$) generated by the nonlinear modulation of the $\mathcal{O}(\epsilon)$ oblique waves $\bu^{(1)}(\bar{x}, y, z)$ given in Table~\ref{tab.ow-ss}:	
	\beq
	\bu_{\mathrm{s}}^{(2)} (y,z)
	\; = \;
	\left[
	\ba{c}
	u_{\mathrm{s}, 2 k_z}^{(2)} (y) 	
	\cos 2 k_z z
	\\[0.15cm]
	v_{\mathrm{s}, 2 k_z}^{(2)} (y) 	
	\cos 2 k_z z
	\\[0.15cm]
	w_{\mathrm{s}, 2 k_z}^{(2)} (y) 	
	\sin 2 k_z z
	\ea
	\right].
	\eeq	
Since the streamwise velocity component captures the dominant energetic content of the second-order streaks induced by the quadratic interactions of oblique waves, our analysis focuses on $u_{\mathrm{s}, 2 k_z}^{(2)}$.
	
	To identify the wavenumbers and wave speed that maximize the energy of the $\mathcal{O}(\epsilon^2)$ streaks, we first examine the dependence of the energy of $u_{\mathrm{s}, 2k_z}^{(2)}$ on the doubled spanwise wavenumber $k_{z0} \DefinedAs 2k_z$. We then compare these results with predictions from the streamwise-constant linearized NS equations and with outcomes obtained using the harmonic-balance-based optimization procedure of~\cite{rigsipcol21}.
	
	We perform computations over a broad range of $(k_x, k_z, c)$ to identify parameter values that yield the largest amplification of the $\mathcal{O}(\epsilon^2)$ steady streaks in Poiseuille flow at $Re = 2000$. The wall-normal direction is discretized using $71$ Chebyshev collocation points, resulting in a finite-dimensional representation of the governing operators. Uniform grids consisting of $96$ points in $k_x \in [0.1,2]$, $196$ points in $k_z \in [0.1,4]$, and $601$ points in $c \in [0,1.2]$ are employed. The parameter ranges and grid resolutions are selected through a coarse-to-fine refinement procedure: the ranges are first expanded until further extension does not affect the amplification near the boundaries, after which the grid spacing is refined until both the magnitude and the location of the maximum gain converge to plotting accuracy.
	
\begin{figure*}[h]
  \centering
  \begin{tabular}{cccccc}
   \subfigure[]{\label{fig.kx0-compare-a}}
    &&
    \hspace{-0.1cm}
    \subfigure[]{\label{fig.kx0-compare-b}}
    &&
    \hspace{-0.2cm}
    \subfigure[]{\label{fig.kx0-compare-c}}
    &
    \\[-.19cm]
    \begin{tabular}{c}
      \vspace{.25cm}
      \hspace{-0.5cm}
      \normalsize{\rotatebox{90}{}}
    \end{tabular}
    &
    \begin{tabular}{c}
      \includegraphics[width=0.3\textwidth]{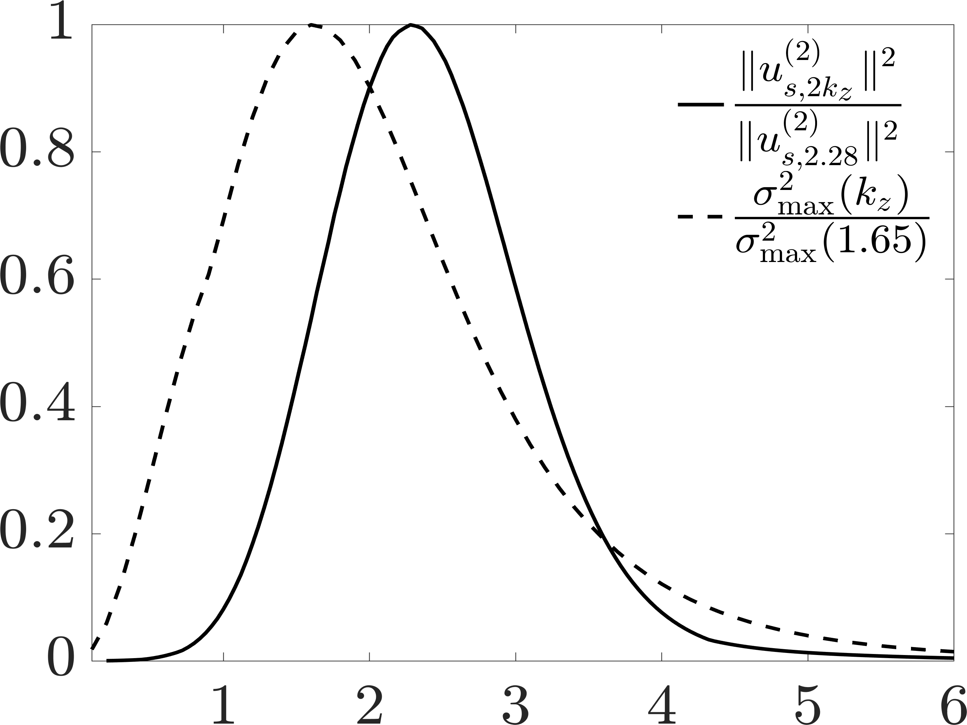}
      \\[-0.1cm]
      \hspace{0.1cm}
      {$k_{z}$}
    \end{tabular}
    &
    \begin{tabular}{c}
        \vspace{.75cm}
        \hspace{0.1cm}
        \normalsize{\rotatebox{90}{$y$}}
      \end{tabular}
      &
      \hspace{-0.37cm}
      \begin{tabular}{c}
      \includegraphics[width=0.3\textwidth]{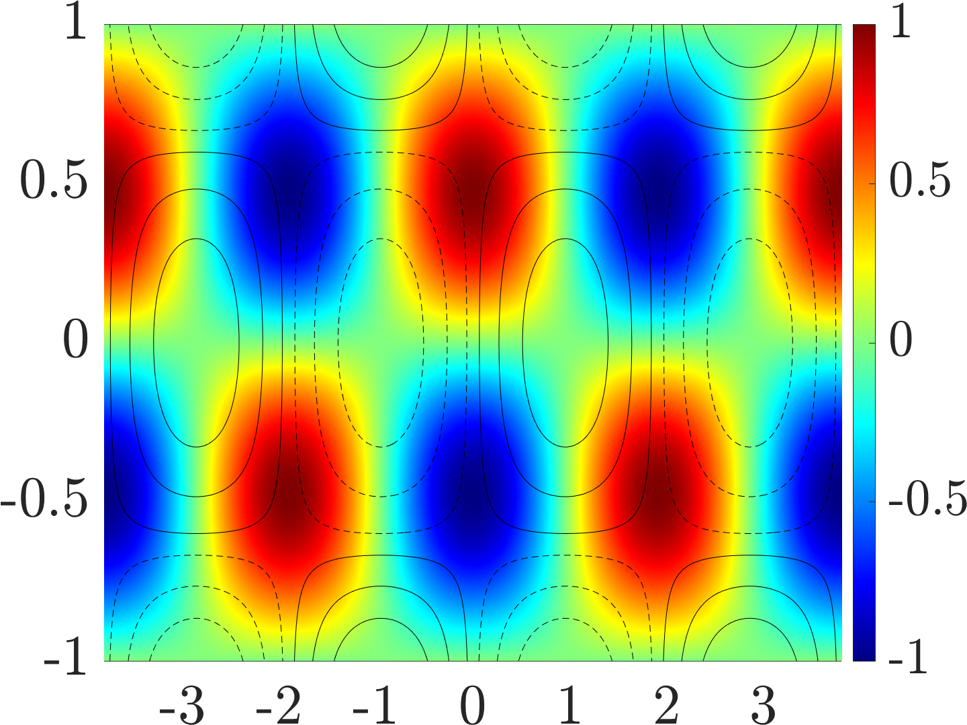}
        \\[-0.1cm] 
        \hspace{-0.28cm}
        {$z$}
      \end{tabular}
    &
    \begin{tabular}{c}
      \vspace{.75cm}
      \normalsize{\rotatebox{90}{$y$}}
    \end{tabular}
    &
    \hspace{-0.38cm}
    \begin{tabular}{c}
      \includegraphics[width=0.3\textwidth]{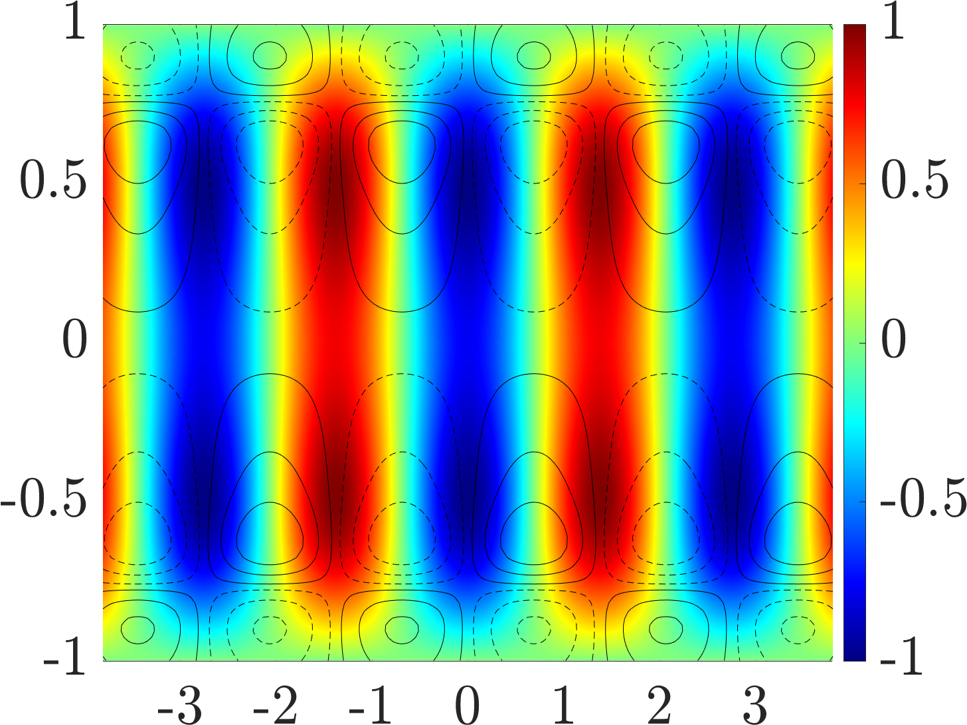}
      \\[-0.1cm]  
      \hspace{-0.28cm}
      {$z$}
    \end{tabular}  
  \end{tabular}
  \vspace{-0.1cm}
   \caption{(a) Energy amplification as a function of the spanwise wavenumber $k_z$ obtained from the primary linearized (dashed) and perturbation (solid) analyses. The dashed curve shows $\sigma_{\max}^2(\cH_{(0,k_z)}(0))$, while the solid curve displays the $\mathcal{O}(\epsilon^2)$ steady streak response induced by oblique waves with $(k_x,c)$ chosen to maximize streak amplification. Both curves are normalized to unit peak value; absolute magnitudes depend on the choice of $\epsilon$. (b,c) Color contours show the most energetic streamwise velocity fluctuations, while black contours indicate streamwise vortices. The spanwise wavelengths are $2\pi/1.65$ in (b) (primary linearized analysis) and $2\pi/2.28$ in (c) (perturbation analysis).}
   \label{fig.kx0-compare}
\end{figure*}

	The largest amplification of the $\mathcal{O}(\epsilon^2)$ streaks, $u_{\mathrm{s}, 2k_z}^{(2)}$, induced by forcing the NS equations with the principal oblique disturbance~\eqref{eq.dow-princ}, occurs at $k_{z0} = 2.28$.
Hence, the most energetic $\mathcal{O}(\epsilon^2)$ streak response is achieved for $(k_x, k_{z0}, \omega) = (0, 2.28, 0)$, driven by oblique waves with $(k_x, k_z, c) = (0.74, 1.14, 0.396)$.
In contrast, the dominant streaks predicted by the linearized $\mathcal{O}(\epsilon)$ analysis correspond to $(k_x, k_z, \omega) = (0, 1.65, 0)$; see Fig.~\ref{fig.kx0-compare-a}.
The two peaks attain comparable energy when
$\epsilon\sigma_\mathrm{max}^2(1.65) 
\approx
\epsilon^2 \|u_{\mathrm{s},2k_z}^{(2)}\|^2$
which occurs at $\epsilon \approx 10^{-10}$.
This suggests that weakly nonlinear effects may become significant even at very small disturbance amplitudes.
The spanwise wavelength of the streaks predicted by the perturbation framework is consistent with the observations of~\cite{RSBH1998}, which identified streaks with $k_z \in [2, 2.5]$ as having the lowest amplitude threshold for transition to turbulence at comparable Reynolds numbers.

Figures~\ref{fig.kx0-compare-b} and~\ref{fig.kx0-compare-c} compare the spatial structures of the most energetic streamwise velocity fluctuations obtained from the primary linear resolvent analysis and the $\mathcal{O}(\epsilon^2)$ perturbation analysis.
In both cases, the streamwise velocity fluctuations carrying the largest energy are steady, streamwise constant, and purely harmonic in $z$.
Their wall-normal profiles are determined, respectively, by:
(i) the principal output singular function of the frequency-response operator for $(k_x, k_z, \omega) = (0, 1.65, 0)$, shown in Fig.~\ref{fig.kx0-compare-b}; and
(ii) the $\mathcal{O}(\epsilon^2)$ response to the $\mathcal{O}(\epsilon)$ oblique-wave input~\eqref{eq.dow-princ} with $(k_x, k_z, c) = (0.74, 1.14, 0.396)$, shown in Fig.~\ref{fig.kx0-compare-c}.

In both regimes, pairs of counter-rotating streamwise vortices (contour lines) redistribute momentum in the $(y, z)$-plane, promoting the amplification of high- and low-speed streaks through the lift-up mechanism. However, the wall-normal symmetry of these vortices differs: in the weakly nonlinear analysis, the streaks are induced by antisymmetric vortical perturbations, consistent with the observations of~\cite{RSBH1998} for Poiseuille flow, where such antisymmetric structures exhibit the lowest amplitude threshold for transition.
 
\begin{figure*}[h]
  \centering
  \begin{tabular}{cccccc}
  \vspace{-0.1cm}
  \hspace{-1cm}
   \subfigure[]{\label{fig:WNL_Pois_a}}
    &
    $\log_{10} \| u_{\mathrm{s}, 2\bkz}^{(2)} \|_2^2$
    &
    \hspace{-0.5cm}
    \subfigure[]{\label{fig:WNL_Pois_b}}
    \vspace{-0.1cm}
    &
    $\log_{10} \| u_{\mathrm{s}, 2\bkz}^{(2)} \|_2^2$
    &
    \hspace{-0.5cm}
    \subfigure[]{\label{fig:WNL_Pois_c}}
    \\[-.19cm]
    \begin{tabular}{c}
      \vspace{.45cm}
      \hspace{-0.8cm}
      \normalsize{\rotatebox{90}{$\bkx$}}
    \end{tabular}
    &
    \begin{tabular}{c}
    \hspace{-0.35cm}
      \includegraphics[width=0.3\textwidth]{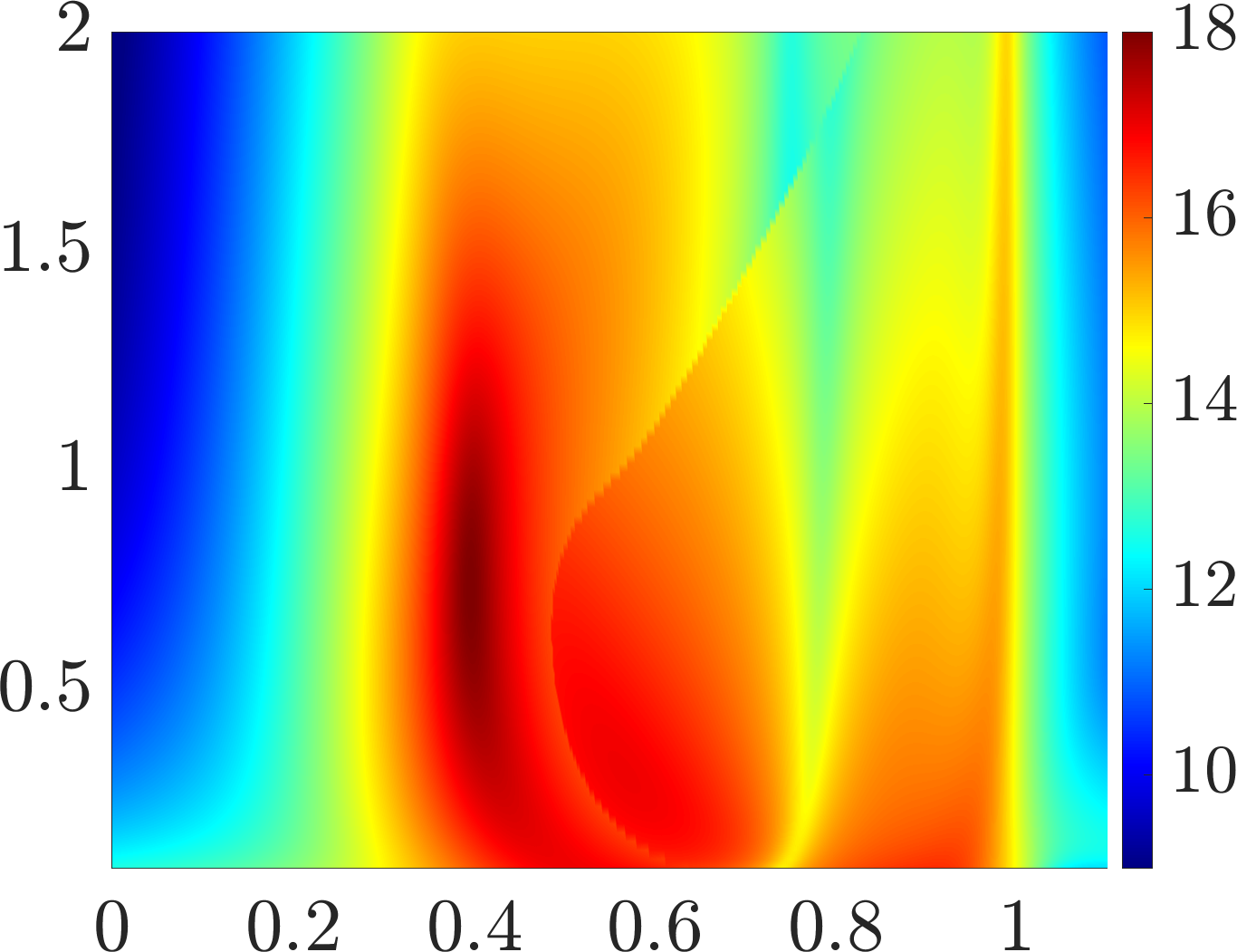}
      \\[-0.1cm] { $\bc$}
    \end{tabular}
    &
    \begin{tabular}{c}
      \vspace{.45cm}
      \hspace{-0.2cm}
      \normalsize{\rotatebox{90}{$\bkx$}}
    \end{tabular}
    &
    \begin{tabular}{c}
    \hspace{-0.35cm}
      \includegraphics[width=0.3\textwidth]{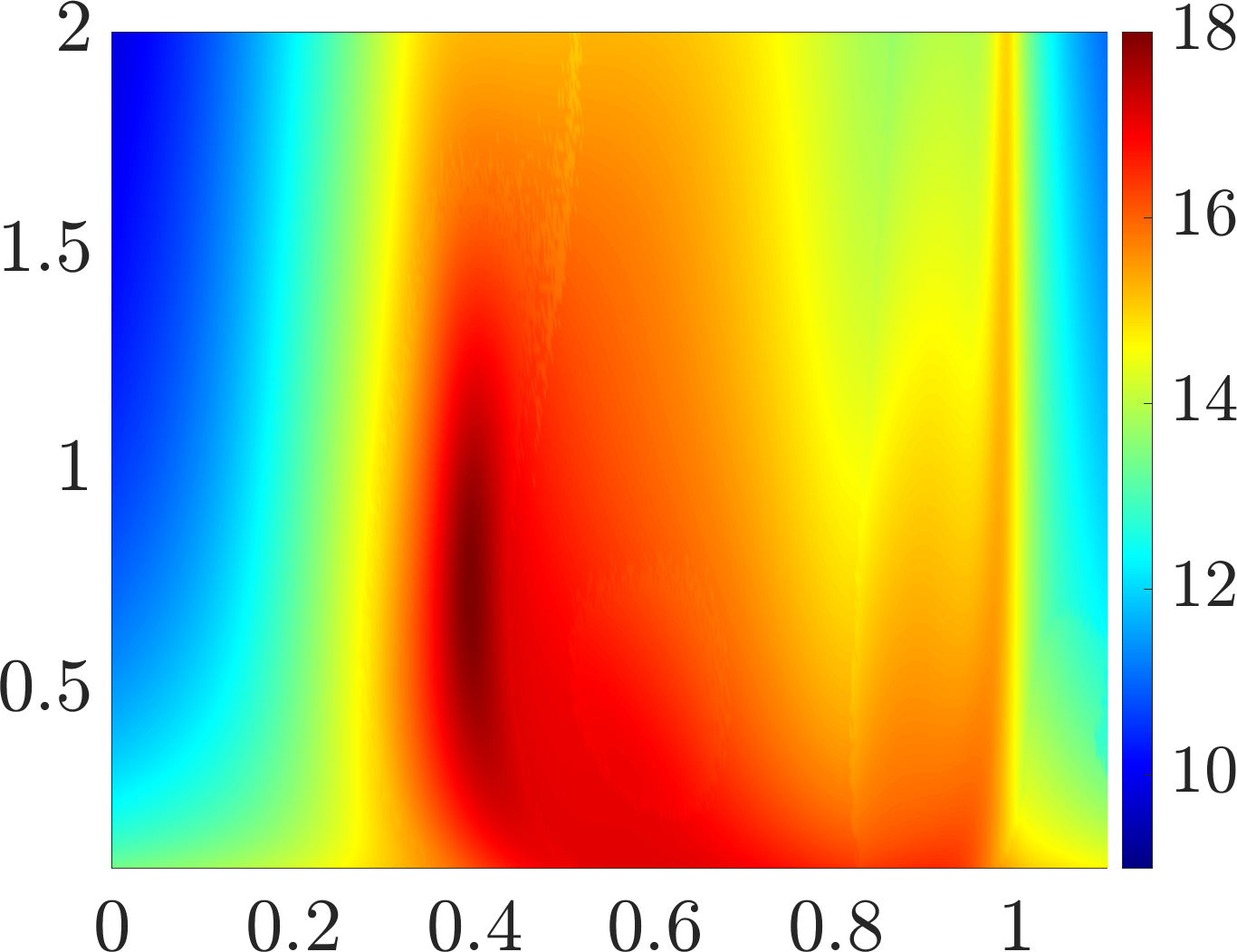}
      \\[-0.1cm]  {$\bc$}
    \end{tabular}
    &
    \begin{tabular}{c}
    \hspace{0.1cm}
      \vspace{0.6cm}
      \normalsize{\rotatebox{90}{$\displaystyle{\max_{k_x, \, c}} ~ \lVert u_{\mathrm{s}, 2\bkz}^{(2)}\rVert^2$}}
    \end{tabular}
    &
    \begin{tabular}{c}
      \includegraphics[width=0.3\textwidth]{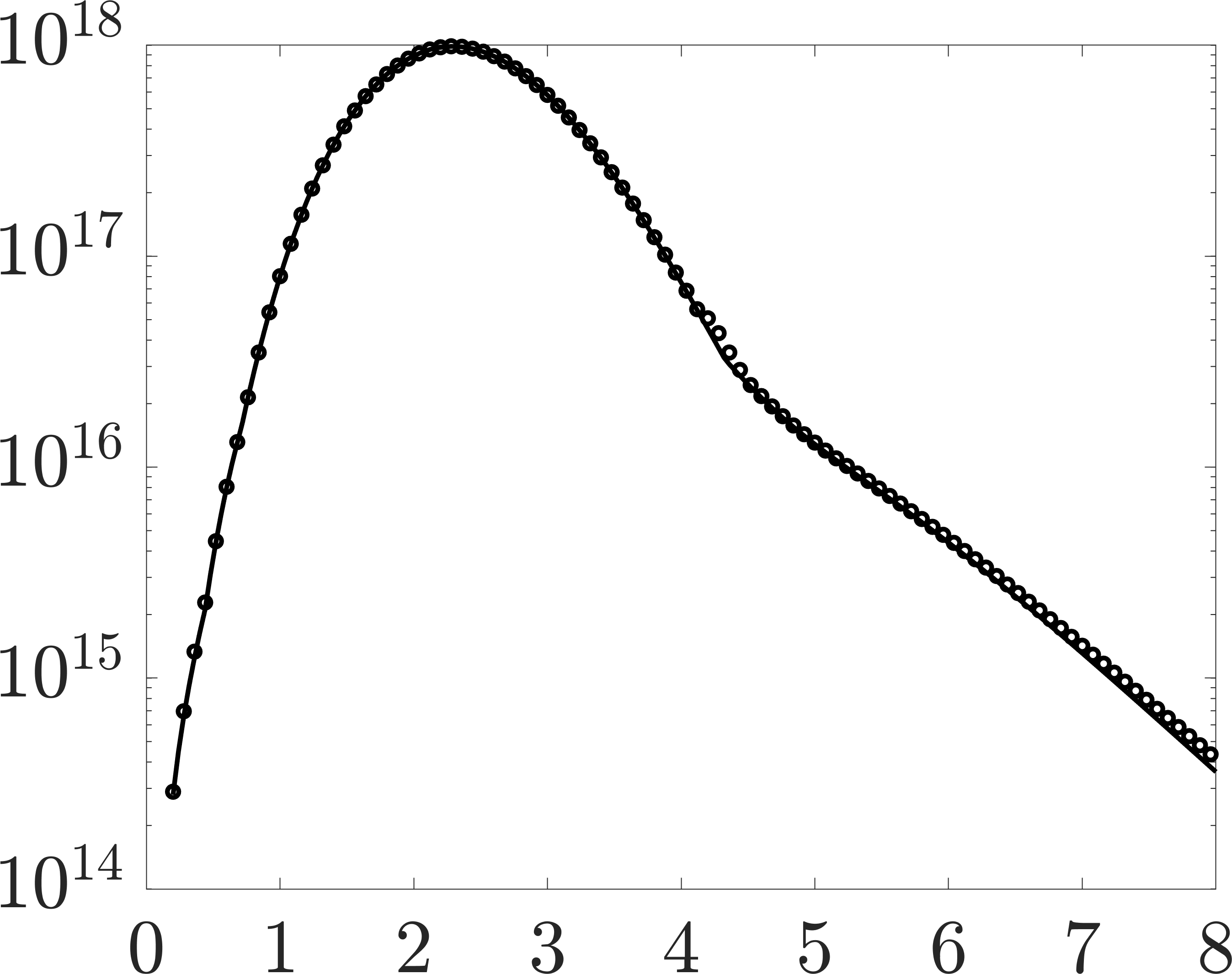}
      \\[-0.1cm]   {$k_{z0} \, = \, 2\bkz$}
    \end{tabular}
  \end{tabular}
  \vspace{-0.1cm} 
  \caption{(a,b) Energy amplification of $\mathcal{O}(\epsilon^2)$ steady streaks, $\| u_{\mathrm{s},2k_z}^{(2)} \|^2$, as a function of $(k_x,c)$ for the fixed spanwise wavenumber $k_{z0} \DefinedAs 2k_z = 2.28$. Panel~(a) shows results obtained using the $\mathcal{O}(\epsilon)$ principal oblique disturbance with $k_z = 1.14$, while panel~(b) shows those obtained using the nonlinear $\mathcal{O}(\epsilon^2)$ optimal disturbance.
(c) Largest amplification over $(k_x,c)$ as a function of $2k_z$; solid lines and circles denote results obtained using the disturbances specified in panels~(a) and~(b), respectively.}
  \label{fig:WNL_Pois}
\end{figure*}

 	To further elucidate the role of unsteady disturbances in the generation of steady streaks through quadratic interactions, we examine in Fig.~\ref{fig:WNL_Pois_a} the dependence of the streak energy on $k_x$ and $c$ for $k_z = 1.14$. The $\mathcal{O}(\epsilon^2)$ streamwise streaks exhibit significant energy, $\| u_{\mathrm{s}, 2k_z}^{(2)} \|^2$, over a broad range of streamwise wavenumbers, provided the wave speed is approximately $40 \%$ of the laminar centerline velocity.
The maximum amplification occurs at $c = 0.396$, although wave speeds slightly below this value also produce strong streak responses.
The apparent discontinuity near $c \approx 0.6$ arises from an interchange in dominance between the first and second singular functions.
While the singular values vary continuously (though non-differentiably) with $(k_x, k_z, c)$, a sudden change occurs in the corresponding singular functions, which modifies their mode shapes and, consequently, their projection onto the resulting streaks.

\noindent {\bf Comparison with nonlinear optimization.} To demonstrate that our perturbation-based approach can identify the characteristic length and time scales of the most amplified oblique disturbances without resorting to computationally intensive optimization procedures, we compare its predictions with those obtained by directly optimizing the nonlinear response to external disturbances with $k_z=1.14$ in Fig.~\ref{fig:WNL_Pois_b}. In this formulation, $\bd_{(k_x, k_z)}^{(1)}(y, c)$ in~\eqref{eq.dow-new} is treated as the optimization variable, and the kinetic energy of the $\mathcal{O}(\epsilon^2)$ streaks serves as the objective function; see Appendix~\ref{app.optim} for the problem statement.
The optimization is parameterized by $(k_x, k_z, c)$, using the same grid resolution employed in the perturbation analysis with the oblique-wave input~\eqref{eq.dow-princ}.
A gradient-based algorithm, similar to those used in~\cite{ker18,rigsipcol21}, is then applied to determine the optimal disturbance~\eqref{eq.dow-new} that maximizes the kinetic energy \mbox{of the $\mathcal{O}(\epsilon^2)$ streaks.} 

We observe close agreement between the results of the nonlinear optimization and the predictions of the perturbation-based framework. In particular, the $(k_x,c)$-dependence of the most energetic $\mathcal{O}(\epsilon^2)$ streak response obtained from the iterative search initialized with the $\mathcal{O}(\epsilon)$ principal input singular functions of $\cG_{\bk}(c)$ (Fig.~\ref{fig:WNL_Pois_b}) closely matches that predicted by the perturbation analysis (Fig.~\ref{fig:WNL_Pois_a}). Similar agreement is observed for external disturbances with other values of $k_z$.

This correspondence is further quantified in Fig.~\ref{fig:WNL_Pois_c}, which shows excellent agreement over a broad range of spanwise wavenumbers between the $\mathcal{O}(\epsilon^2)$ streak energy obtained from the optimization procedure and that resulting from forcing the NS equations with the $\cO(\epsilon)$ principal oblique disturbance~\eqref{eq.dow-princ}. The two curves are nearly indistinguishable, exhibiting a relative error of only $0.001$ at the peak. Moreover, both approaches identify the same parameters for maximum amplification, $(k_x, k_z, c) = (0.74, 1.14, 0.396)$. These results confirm that, within the perturbation-analysis framework, frequency-response analysis of the linearized NS equations successfully recovers the same wavenumber and frequency structure of the most amplified unsteady disturbance as the optimization-based approach that directly maximizes the $\mathcal{O}(\epsilon^2)$ streak energy, $\| u_{\mathrm{s},2k_z}^{(2)} \|^2$.

Additional comparison with the harmonic-balance-based optimization procedure of~\cite{rigsipcol21} is provided in Appendix~\ref{app.hbns}. The close correspondence between the $\mathcal{O}(\epsilon^2)$ streak energy produced by the linear optimal (principal) disturbance and that obtained from the weakly nonlinear locally optimal disturbance indicates that energy amplification is primarily governed by linear mechanisms, with nonlinear terms redistributing the amplified energy rather than generating it.

	\vspace*{-3ex}
\subsection{Spatial structure of steady streamwise streaks}
	\label{sec.ss-eps2-eps4}

	\vspace*{-1ex}
Perturbation analysis of the frequency response of the Navier--Stokes equations provides fundamental insight into the structural features of the resulting flow responses.
As discussed in Section~\ref{sec.FrRes}, the left and right singular functions of the resolvent operator form orthonormal bases for the output and input spaces, respectively, and thereby describe how external excitations are mapped to flow responses.
Here, we show that the second output singular function of the resolvent operator at $k_x = 0$ captures the spatial structure of the streamwise streaks generated by small-amplitude oblique inputs, and we provide an explanation for the emergence of this subdominant output singular function in the induced streak response.

\begin{figure}[h]
  \centering
  \begin{tabular}{cc}
    \begin{tabular}{c}
    \vspace{1.15cm}
        \normalsize{\rotatebox{90}{$y$}}
      \end{tabular}
      &
      \begin{tabular}{c}
    \includegraphics[width=0.3\textwidth]{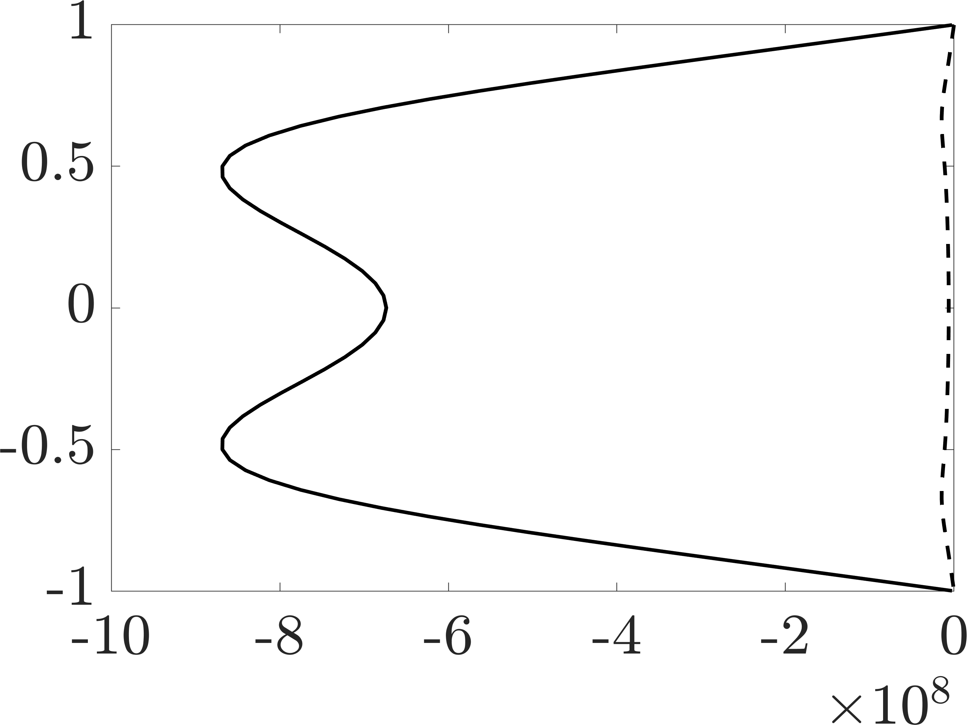}
        \\[-0.1cm] 
        {$u_{\mathrm{s}1, 2 k_z}^{(2)} (y)$, $u_{\mathrm{s}2, 2 k_z}^{(2)} (y)$}
      \end{tabular}
  \end{tabular}
  \caption{Wall-normal profiles of $u_{\mathrm{s}1, 2k_z}^{(2)} (y)$ and $u_{\mathrm{s}2, 2k_z}^{(2)} (y)$ resulting from forcing the steady, streamwise-constant model~\eqref{eq.ss-kx0-epsn} with $d_{\psi}^{(2)}$ (solid) and $d_{u}^{(2)}$ (dashed). The wall-normal and spanwise forcing component $d_{\psi}^{(2)}$ produces $u_{\mathrm{s}1, 2k_z}^{(2)}$, which accounts for $97.45 \%$ of the $\mathcal{O}(\epsilon^2)$ streak energy.}
    \label{fig.u(y)-streaks-eps2} 
\end{figure}

	\vspace*{1ex}
	\noindent {\bf Computational observations.}
We first examine the spatial structure of the $\mathcal{O}(\epsilon^{2})$ streaks that arise from nonlinear interactions of the most energetic $\mathcal{O}(\epsilon)$ oblique waves in Poiseuille flow with $Re = 2000$ and parameters $(k_x, k_z, c) = (0.74, 1.14, 0.396)$.
Figure~\ref{fig.u(y)-streaks-eps2} shows the wall-normal dependence of the components of	    
	\beq	
	u_{\mathrm{s},2k_z}^{(2)} (y) 
	\; = \;
	u_{\mathrm{s}1, 2 k_z}^{(2)} (y) 
	\; + \; 
	u_{\mathrm{s}2, 2 k_z}^{(2)} (y)
	\eeq
which respectively result from forcing the steady, streamwise-constant model~\eqref{eq.ss-kx0-epsn} with $d_{\psi}^{(2)}$ and $d_{u}^{(2)}$.
The wall-normal and spanwise forcing components, i.e., $d_{\psi}^{(2)}$, have a dominant influence on the $\mathcal{O}(\epsilon^{2})$ streaks, with $u_{\mathrm{s}1, 2k_z}^{(2)}$ containing $97.45 \%$ of the total energy.
	
The streak response $u_{\mathrm{s},2k_z}^{(2)}$ can be expressed using the SVD of the frequency response operator
\mbox{$\mathcal{G}_{2k_z} \DefinedAs \mathcal{H}_{(0,2k_z)} (0)$},
which corresponds to the steady, streamwise-constant linearized model~\eqref{eq.ss-kx0-epsn} with input
	\beq
	{\bd}_{2 k_z}^{(2)} (y)
	\; \DefinedAs \;
	\big[ \, {{d}}_{\psi,2 k_z}^{(2)} (y) ~\; {{d}}_{u,2 k_z}^{(2)} (y) \, \big]^T.
	\eeq
The decomposition reads	
	\beq
	u_{\mathrm{s},2 k_z}^{(2)} (y)
	\; = \;
	\big[
	\cG_{2\bkz}
	{\bd}_{2 k_z}^{(2)} ( \, \cdot \, )
	\big] (y)
	\; = \;
	\sum_j a_j {\bvartheta}_{2\bkz,j}(y)
	\label{eq:proj1}
	\eeq
where $\sigma_j$, ${\bdelta}_{2k_z,j}$, and ${\bvartheta}_{2k_z,j}$ denote the singular values and the corresponding input and output singular functions of $\mathcal{G}_{2k_z}$.The coefficient	
	\beq
	a_{j} 
	\; \DefinedAs \; 
	\sigma_{j}
	\langle 
	{\bdelta}_{2\bkz,j}, {\bd}_{2\bkz}^{(2)}
	\rangle
	\eeq
quantifies the contribution of the $j$th output singular function ${\bvartheta}_{2k_z,j}$ to the $\mathcal{O}(\epsilon^{2})$ steady streaks $u_{\mathrm{s},2k_z}^{(2)}$.

\begin{figure}[h]
  \centering
  \begin{tabular}{cccccc}
  \vspace{-0.1cm}
  \hspace{-.5cm}
   \subfigure[]{\label{fig.SingValsP}}
    & &
    \hspace{-0.25cm}
    \subfigure[]{\label{fig.SingProjP}}
    \vspace{-0.1cm}
    &
    &
    \hspace{-0.5cm}
    \subfigure[]{\label{fig.SpatProfP}}
    & 
    \\[-.cm]
    \begin{tabular}{c}
    \hspace{-0.2cm}
      \vspace{0.6cm}
      \normalsize{\rotatebox{90}{$\sigma_j$}}
    \end{tabular}
    &
    \begin{tabular}{c}
       \includegraphics[width=0.3\textwidth]{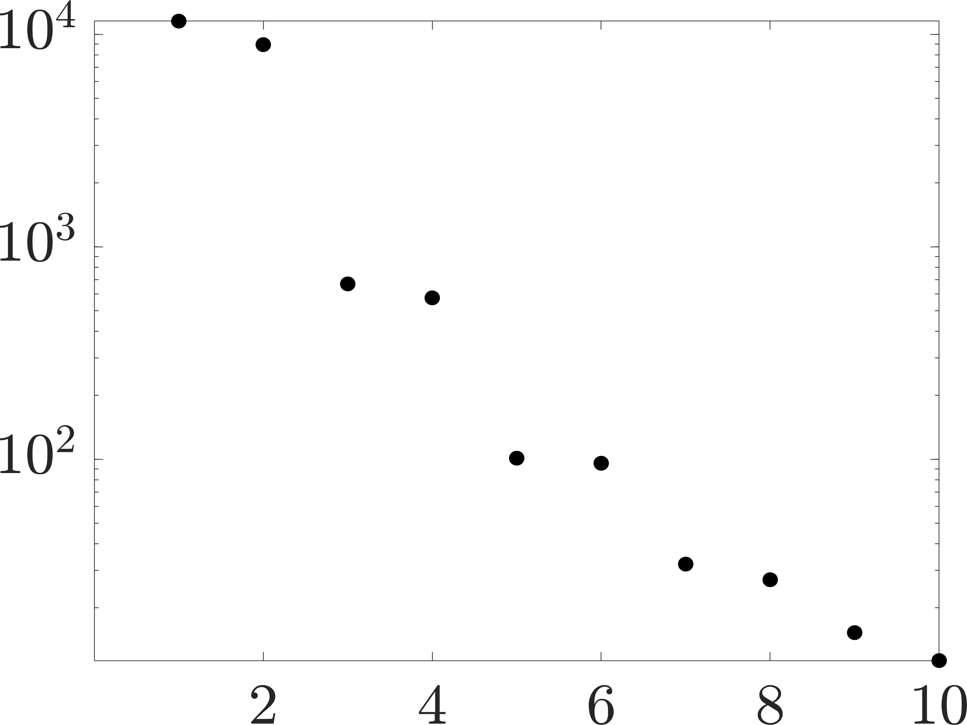}
      \\[-0.1cm] {index $j$}
    \end{tabular}
    &
    \begin{tabular}{c}
        \vspace{.55cm}
         \hspace{-0.25cm}
        \normalsize{\rotatebox{90}{$a_j^2/\lVert u_{\mathrm{s},2.28}^{(2)} \rVert^2$}}
      \end{tabular}
      &
      \begin{tabular}{c}
      \hspace{-0.4cm}
        \includegraphics[width=0.3\textwidth]{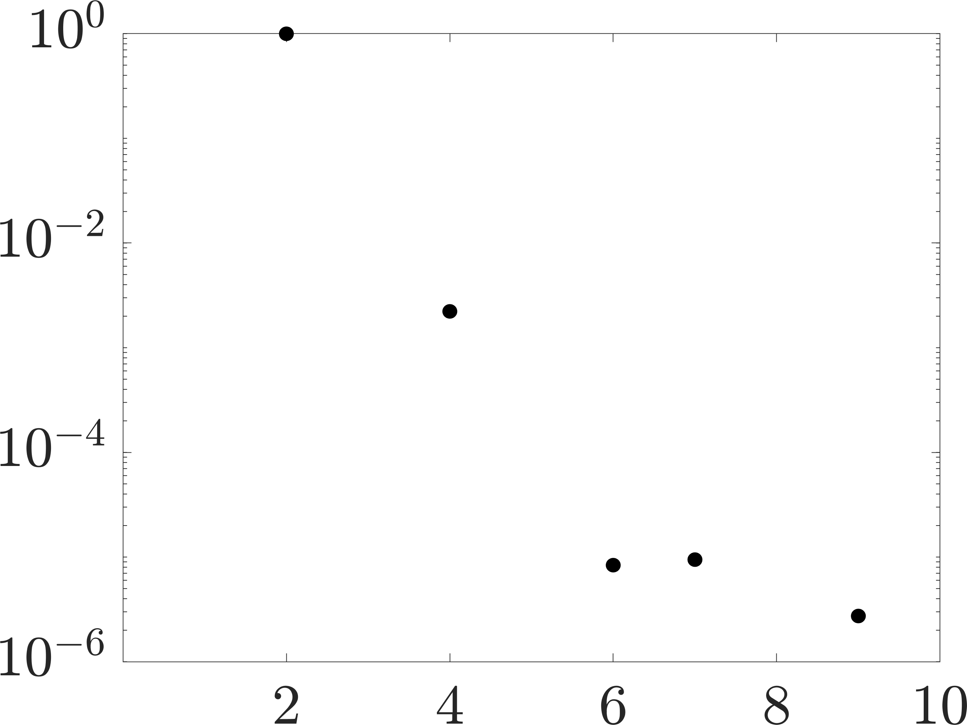}
        \\[-0.1cm] {index $j$}
      \end{tabular}
    &
    \begin{tabular}{c}
      \vspace{.7cm}
      \hspace{-0.35cm}
      \normalsize{\rotatebox{90}{$y$}}
    \end{tabular}
    &
    \begin{tabular}{c}
    \hspace{-0.35cm}
      \includegraphics[width=0.3\textwidth]{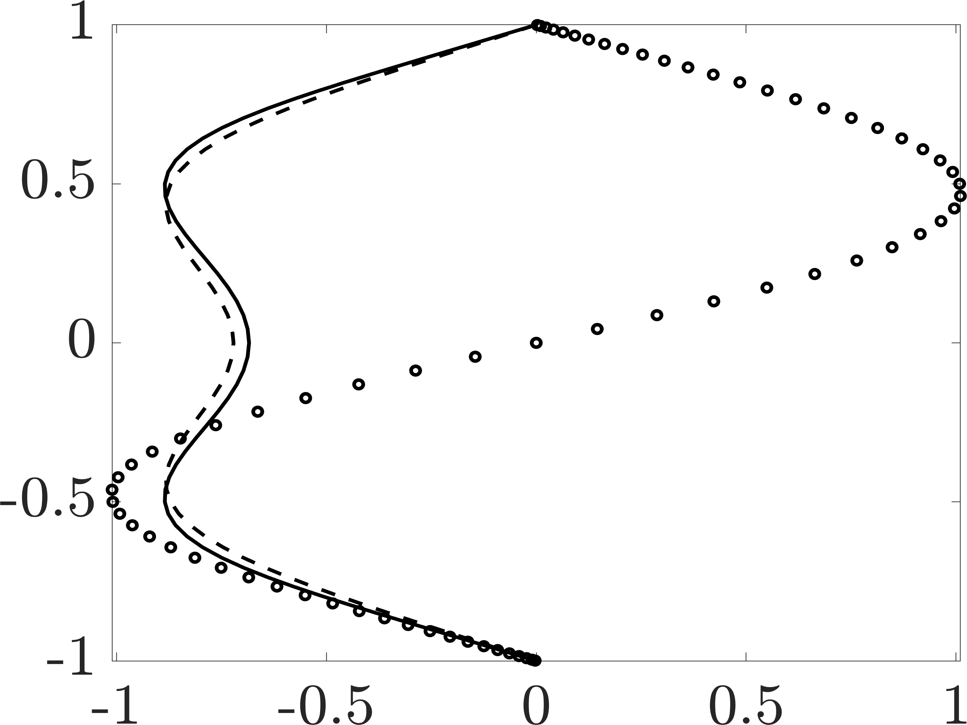}
      \\[-0.1cm]  
      \hspace{0.1cm}
      {$u_{\mathrm{s},2.28}^{(2)} (y)$, $\vartheta_{2.28,1} (y)$, $\vartheta_{2.28,2} (y)$}
    \end{tabular}  
  \end{tabular}
  \vspace{-0.1cm}  
  \caption{(a) Singular values of the frequency response operator $\cG_{2.28} = \cH_{(0,2.28)} (0)$ associated with the streamwise-constant linearized model~\eqref{eq.ss-kx0-epsn} in Poiseuille flow at $Re = 2000$.
(b) Contribution of the $j$th output mode ${\vartheta}_{2.28,j}$ to the energy of the $\mathcal{O}(\epsilon^{2})$ streaks $u_{\mathrm{s},2.28}^{(2)}$ induced by nonlinear interactions of $\mathcal{O}(\epsilon)$ oblique waves with $(k_x, k_z, c) = (0.74, 1.14, 0.396)$.
(c) Wall-normal profiles of $u_{\mathrm{s},2.28}^{(2)}(y)$ (solid), and the first (circles) and second (dashed) output singular functions of $\cG_{2.28}$; all curves are normalized to unit $L_2$ norm, e.g., $\| u_{\mathrm{s},2.28}^{(2)} \|_2^2 = 1$.}
    \label{fig.SingProj}
\end{figure}

Figure~\ref{fig.SingValsP} displays the ten largest singular values of
	$
	\cG_{2 k_z} = \cH_{(0,2 k_z)} (0)
	$
for $k_z = 1.14$. These values quantify the amplification of each input singular direction of the frequency response operator $\mathcal{H}_{(k_x,k_z)}(\omega)$ at the given $(k_x, k_z, \omega)$.
Although $\sigma_{1}$ and $\sigma_{2}$ are markedly larger than the remaining singular values, Fig.~\ref{fig.SingProjP} shows that the principal output singular function ${\vartheta}_{2k_z,1}$ of $\mathcal{G}_{2k_z}$ contributes negligibly to the $\mathcal{O}(\epsilon^2)$ streaks.
Instead, the dominant contribution to $u_{\mathrm{s},2k_z}^{(2)}$ originates from the second output mode ${\vartheta}_{2k_z,2}$ of $\mathcal{G}_{2k_z}$, which, at $k_z = 1.14$, accounts for $99.78 \%$ of the $\mathcal{O}(\epsilon^2)$ streak energy ($a_2^2 / \| u_{\mathrm{s},2.28}^{(2)} \|_2^2$).
The influence of all other output singular directions is negligible.

To verify that the second output singular function ${\vartheta}_{2k_z,2}$ of $\mathcal{G}_{2k_z}$ characterizes the $\mathcal{O}(\epsilon^2)$ steady streaks in Poiseuille flow with $Re = 2000$ and $k_z = 1.14$, we compare its spatial structure with that of $u_{\mathrm{s},2k_z}^{(2)}$ in Fig.~\ref{fig.SpatProfP}. This comparison reveals excellent agreement between the weakly nonlinear response and the second output singular function of $\mathcal{G}_{2.28}$, underscoring the key role of ${\vartheta}_{2k_z,2}$ in capturing the spatial structure of the most energetic steady response of the NS equations to small-amplitude three-dimensional harmonic forcing.

\begin{figure}[h]
  \centering
  \begin{tabular}{cc}
    \begin{tabular}{c}
    \vspace*{0.5cm}
        \normalsize{\rotatebox{90}{$a_2^2 / \| u_{\mathrm{s},2.28}^{(2)} \|_2^2$}}
      \end{tabular}
      &
      \begin{tabular}{c}
    \includegraphics[width=0.3\textwidth]{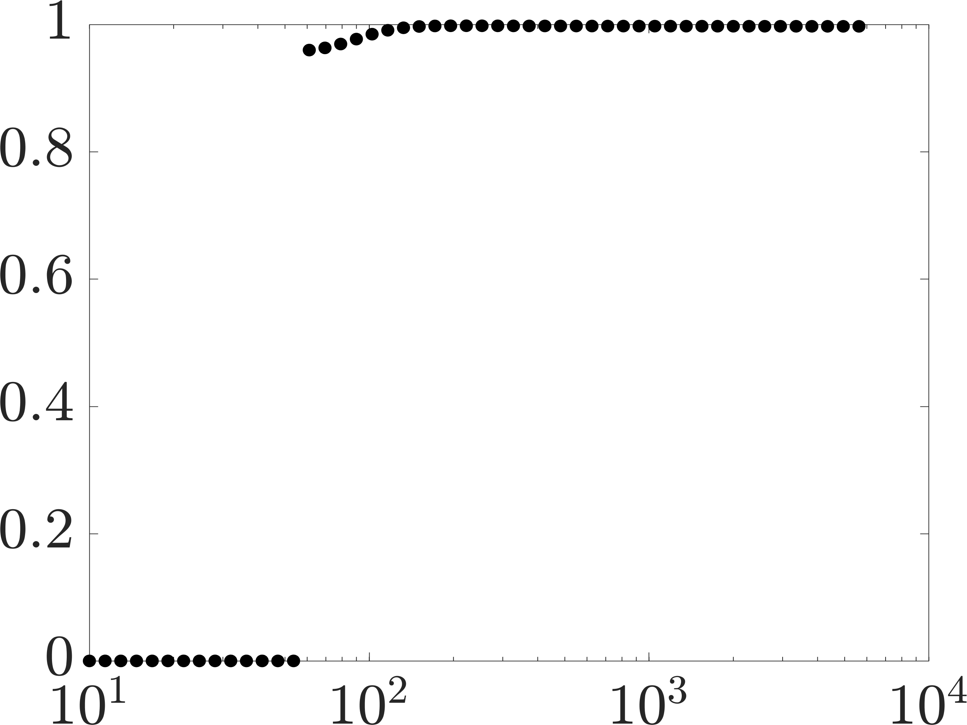}
        \\[-0.01cm] {$Re$}
      \end{tabular}
  \end{tabular}
  \caption{Contribution of the second output mode ${\vartheta}_{2.28,2}$ to the energy of the $\mathcal{O}(\epsilon^{2})$ steady streaks $u_{\mathrm{s},2.28}^{(2)}$ induced by nonlinear interactions of the principal disturbance, shown as a function of Reynolds number.}
    \label{PoisRe}
\end{figure}

Figure~\ref{PoisRe} shows the contribution of the second output singular function to the total energy of the $\mathcal{O}(\epsilon^{2})$ steady streamwise streaks as a function of the Reynolds number.
The results demonstrate that the second output singular function of the frequency response operator associated with the linearized, streamwise-constant model accurately captures the dominant features of the weakly nonlinear response of the NS equations across the entire range of transitional Reynolds numbers.

	\vspace*{0.1cm}
	\noindent{\bf Theoretical explanation.}
As discussed in Section~\ref{sec.setupWNL}, for fluctuations satisfying $\partial_x(\cdot)=0$, the dynamics may be formulated in terms of the $(y,z)$-plane stream function $\psi$ and the streamwise velocity component $u$. In this representation, the $\mathcal{O}(\epsilon^2)$ streamwise-constant dynamics governing streak generation are given by
\beq
\ba{rcl}
\partial_t \psi^{(2)}
& = &
\dfrac{1}{Re}\,\cL \psi^{(2)}
\; + \;
d_{\psi}^{(2)}
\\[0.2cm]
\partial_t u^{(2)}
& = &
\cC_p\,\psi^{(2)}
\; + \;
\dfrac{1}{Re}\,\cS u^{(2)}
\; + \;
d_u^{(2)}
\ea
\label{eq.ss-kx0-eps2}
\eeq
where $\cL=\Delta^{-1}\Delta^2$ is the streamwise-constant Orr--Sommerfeld operator with homogeneous Dirichlet and Neumann boundary conditions, $\cS=\Delta$ is the Squire operator with homogeneous Dirichlet boundary conditions, and $\cC_p=-U'(y)\partial_z$ denotes the lift-up coupling operator. The forcing terms $d_{\psi}^{(2)}$ and $d_u^{(2)}$ arise from quadratic interactions between the $\mathcal{O}(\epsilon)$ oblique waves. This formulation admits an explicit characterization of the steady response and its Reynolds-number scaling, summarized in Theorem~\ref{thm1}; see Appendix~\ref{app.stream-const} for the proof.

\begin{theorem}\label{thm1}
For the $\mathcal{O}(\epsilon)$ oblique-wave inputs listed in Table~\ref{tab.ow-ss}, consider the steady~{\rm{(}}$\partial_t=0${\rm{)}} $\mathcal{O}(\epsilon^2)$ streamwise-constant system~\eqref{eq.ss-kx0-eps2}. The corresponding steady solution contains a single nonzero spanwise harmonic at $2k_z$ and can be expressed as
	\begin{subequations}\label{eq.psi-u-eps2}
	\beq
	\ba{rcl}
	\psi_{\mathrm{s}}^{(2)}(y,z)
	& = &
	\psi_{{\mathrm{s}},2k_z}^{(2)}(y) \sin(2k_z z)
	\\[0.2cm]
	u_{\mathrm{s}}^{(2)}(y,z)
	& = &
	\left( 
	u_{{\mathrm{s}1},2k_z}^{(2)}(y) \, + \, u_{{\mathrm{s}2},2k_z}^{(2)}(y) 
	\right) 
	\cos(2k_z z)
	\ea
	\eeq
where
	\beq
	\ba{rcl}
	 {\psi}_{\mathrm{s}, 2 k_z}^{(2)} (y)
	 & = &
	 - Re \, \big[ \cL_{2 k_z}^{-1} {d}_{\psi, 2 k_z}^{(2)} (\cdot) \big] (y)
	 \ea
	 \label{eq.psis-us-eps2}
	\eeq
and the streak component admits the decomposition
	\beq
	\ba{rcl}
	 {u}_{\mathrm{s}1, 2k_z}^{(2)} (y)
	 & = &
	 Re^2
	 \big[  
	 \cS_{2 k_z}^{-1} \cC_{p, 2k_z} \cL_{2 k_z}^{-1} {d}_{\psi, 2 k_z}^{(2)} (\cdot) 
	 \big] (y)
	 \\[0.2cm]
	 {u}_{\mathrm{s}2, 2 k_z}^{(2)} (y)
	 & = &
	 - Re \, \big[ \cS_{2 k_z}^{-1} {d}_{u, 2 k_z}^{(2)} (\cdot) \big] (y)
	 \ea
	 \label{eq.us1-us2-eps2}
	\eeq
with $\cL_{2k_z}^{-1}$ and $\cS_{2k_z}^{-1}$ denoting the bounded inverse operators associated with~\eqref{eq.ss-kx0-eps2} under its boundary conditions.
\end{subequations}
\end{theorem}

\begin{remark}
The term $u_{\mathrm{s}1,2k_z}^{(2)}$ represents the contribution of wall-normal and spanwise forcing $d_{\psi}^{(2)}$ to streak formation through the lift-up mechanism, whereas $u_{\mathrm{s}2,2k_z}^{(2)}$ corresponds to the direct contribution of streamwise forcing $d_u^{(2)}$.
\end{remark}

\begin{figure}[t]
    \centering
   \input{figures/Figure10}
   \caption{Block diagram illustrating the coupling structure of the steady, streamwise-constant model governing the $\mathcal{O}(\epsilon^2)$ streamwise streaks.}
	\label{fig.bd-kx0-eps2}
\end{figure}
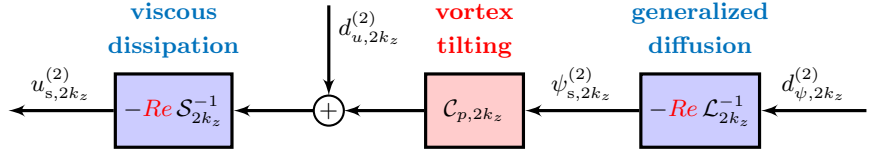

Theorem~\ref{thm1} follows directly from the one-way coupling between the Orr--Sommerfeld and Squire equations and from the Reynolds-number scaling of the associated linear operators. The block diagram in Fig.~\ref{fig.bd-kx0-eps2} schematically summarizes how the forcing components $d_{\psi,2k_z}^{(2)}$ and $d_{u,2k_z}^{(2)}$ influence the streamwise-constant flow fluctuations. Specifically, $d_{\psi,2k_z}^{(2)}$ drives both the stream function $\psi_{\mathrm{s},2k_z}^{(2)}$ and the streamwise velocity $u_{\mathrm{s},2k_z}^{(2)}$, whereas the streamwise forcing $d_{u,2k_z}^{(2)}$ affects only the latter and does not contribute to the evolution of $\psi_{\mathrm{s},2k_z}^{(2)}$.

The $\mathcal{O}(Re^2)$ amplification from wall-normal and spanwise forcing to streamwise streaks is a direct consequence of the lift-up mechanism, while the $\mathcal{O}(Re)$ contributions in~\eqref{eq.psis-us-eps2} and~\eqref{eq.us1-us2-eps2} reflect viscous dissipation. This separation of scalings highlights the dominance of lift-up in the streamwise-constant dynamics. Originating from the linearization of the convective terms in the Navier--Stokes equations, the lift-up mechanism constitutes the primary source of amplification in wall-bounded shear flows~\cite{schmid2007nonmodal,jovARFM21}.

Since the inputs $d_{\psi,2k_z}^{(2)}$ and $d_{u,2k_z}^{(2)}$ generated by the $\mathcal{O}(\epsilon)$ oblique waves in Poiseuille flow at $Re = 2000$ with $(k_x, k_z, c) = (0.74, 1.14, 0.396)$ have comparable energy magnitudes ($\| d_{\psi,2k_z}^{(2)} \| = 1.91 \times 10^5$ and $\| d_{u,2k_z}^{(2)} \| = 1.55 \times 10^5$), the Reynolds-number scaling identified in Theorem~\ref{thm1} implies that the contribution of $d_{\psi}^{(2)}$ dominates the resulting $\mathcal{O}(\epsilon^2)$ streak response. Accordingly, the dominant streak dynamics are governed by the wall-normal/spanwise forcing pathway associated with the lift-up mechanism. This explains the numerical observations reported earlier and provides a principled justification for neglecting $d_{u}^{(2)}$ in the subsequent analysis.

As shown in Appendix~\ref{app.GalilTrEven}, the most amplified $\mathcal{O}(\epsilon)$ oblique waves generate a forcing term $d_{\psi,2k_z}^{(2)}$ that is antisymmetric with respect to the wall-normal coordinate $y$, regardless of the symmetry properties of the oblique waves themselves. This follows from the theory of even operators (see Appendix~\ref{app.ParityOp} for definitions and properties). By the same symmetry arguments, each singular function of the operator $\cS_{2k_z}^{-1}\cC_{p,2k_z}\cL_{2k_z}^{-1}$ is either symmetric or antisymmetric in $y$. Consequently, $d_{\psi,2k_z}^{(2)}$ projects exclusively onto antisymmetric modes, corresponding to even-indexed singular functions, for all spanwise wavenumbers of interest, $2k_z \in [0.92, 2\pi]$. 

For $2k_z > 2\pi$, the spanwise wavelength becomes smaller than the channel half-height, and the singular values come in equal pairs~\cite{moashatromck13}, rendering the choice of symmetric or antisymmetric modes immaterial. The lower bound of this range can be estimated via a Galerkin projection onto the Orr--Sommerfeld and Squire eigenfunctions; see Appendix~\ref{app.Galerkin}. The dominance of the second output singular function follows from the fact that its associated singular value exceeds those of all other even-indexed modes by orders of magnitude; cf.\ Fig.~\ref{fig.SingValsP}. As a result, the emergence and dominance of a single streamwise-streak structure at $\mathcal{O}(\epsilon^2)$ is not dictated by parameter selection, but is enforced by symmetry constraints and resolvent amplification inherent to the linearized NS operator.

This analysis also clarifies why the energy contribution of the second singular function is not dominant at low Reynolds numbers; cf.\ Fig.~\ref{PoisRe}. In this regime, the influence of $d_{u,2k_z}^{(2)}$, which is not subject to symmetry constraints, becomes non-negligible, resulting in a finite projection onto the first principal mode.

Overall, our results demonstrate that the second output singular function of the operator $\cG_{2\bkz}$ accurately captures the spatial structure of the $\mathcal{O}(\epsilon^2)$ steady streaks. Consequently, a low-order approximation of the streak response should be constructed by projecting onto the second, rather than the first, singular direction. For completeness, Appendix~\ref{app.Galerkin} provides an analytical characterization of this response in terms of Orr--Sommerfeld and Squire eigenfunction expansions.

Taken together, these results firmly anchor the weakly nonlinear framework to the dominant resolvent modes of the linearized dynamics, furnishing a principled basis for the DNS validation that follows.
  
	\vspace*{-3ex}
	\subsection{Validation against direct numerical simulations}
\label{sec.dns}
\vspace{-1ex}

We employ direct numerical simulations (DNS) to validate the predictive capability of the perturbation-based framework. All computations are performed using a spectral scheme implemented in the open-source solver~\cite{channelflow}. Spatial derivatives are evaluated spectrally, with Fourier expansions in the homogeneous streamwise ($x$) and spanwise ($z$) directions and Chebyshev polynomials in the wall-normal ($y$) direction. Time integration is carried out using a third-order semi-implicit backward differentiation formula (SBDF3). A fixed pressure gradient is maintained in all simulations, which are conducted at $Re = 2000$ based on the channel half-height and centerline velocity. The computational domain spans $y \in [-1,1]$, with streamwise and spanwise lengths $L_x = 10\pi/3.7$ and $L_z = 10\pi/5.7$, respectively, and a grid resolution of $(N_x, N_y, N_z) = (48, 73, 48)$.

To assess the ability of the perturbation analysis to capture steady three-dimensional responses, we perform DNS in the presence of externally imposed oblique-wave forcing of the form
\beq
\ba{rcl}
\bd(\bx,t)
& = &
\epsilon^*
\Re
\big\{
\big(
\bd_{(\bkx,\bkz)}^{(1)}(y,\bom)\,\mre^{\mri(\bkx x + \bom t)}
~+~
\bd_{(-\bkx,\bkz)}^{(1)}(y,-\bom)\,\mre^{-\mri(\bkx x + \bom t)}
\big)
\,\mre^{\mri \bkz z}
\big\}
\ea
\label{eq.dow-dns}
\eeq
where $\epsilon^*$ denotes the forcing amplitude and $\bd_{(\pm\bkx,\bkz)}^{(1)}(y,\mp\bc\bkx)$ are chosen as the principal input modes obtained from resolvent analysis of the laminar channel flow at $Re = 2000$.

\begin{figure}[t]
  \centering
  \begin{tabular}{c@{\hspace{-0.04 cm}}c@{\hspace{-0.1 cm}}l@{\hspace{-0.04 cm}}c@{\hspace{-0.1 cm}}}
    \subfigure[]{\label{fig.Oeps2-scale-str}}
    &&
    \subfigure[]{\label{fig.Oeps2-scale-mean}}&
    \\[-.13cm]
    \begin{tabular}{c}
      \vspace{1cm}
      \normalsize{\rotatebox{90}{$y$}}
    \end{tabular}
    &
    \begin{tabular}{c}
      \includegraphics[width=0.332\textwidth]{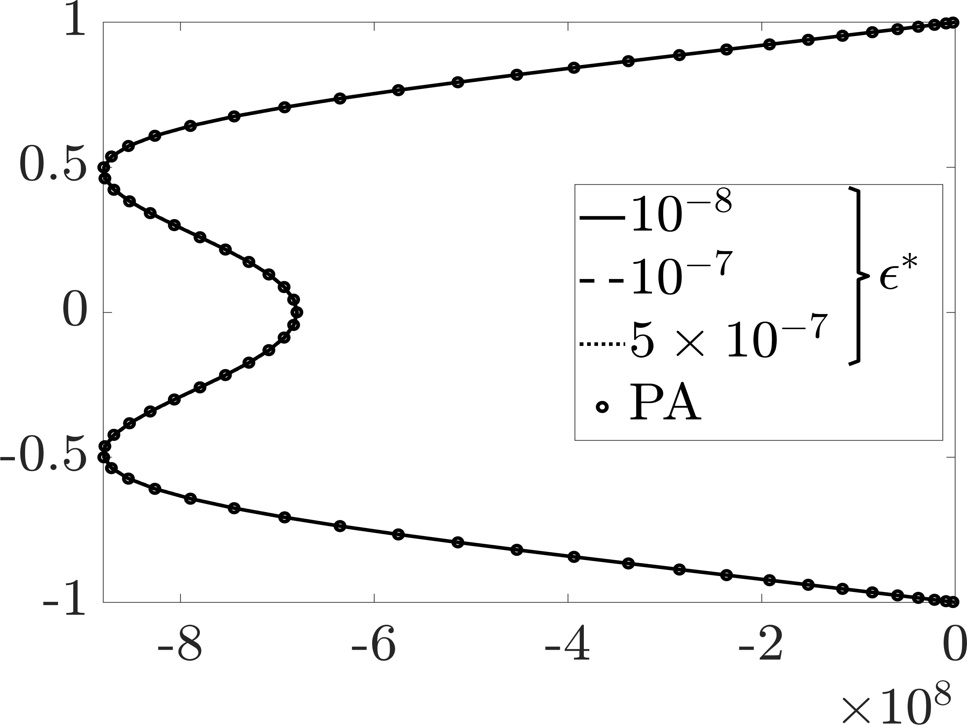}
      \\ [-0.3cm]
      \hspace{0.4cm}{$u_{\mathrm{s},2.28}(y)/\epsilon^{*2}$}
    \end{tabular}
      &
      \begin{tabular}{c}
        \includegraphics[width=0.332\textwidth]{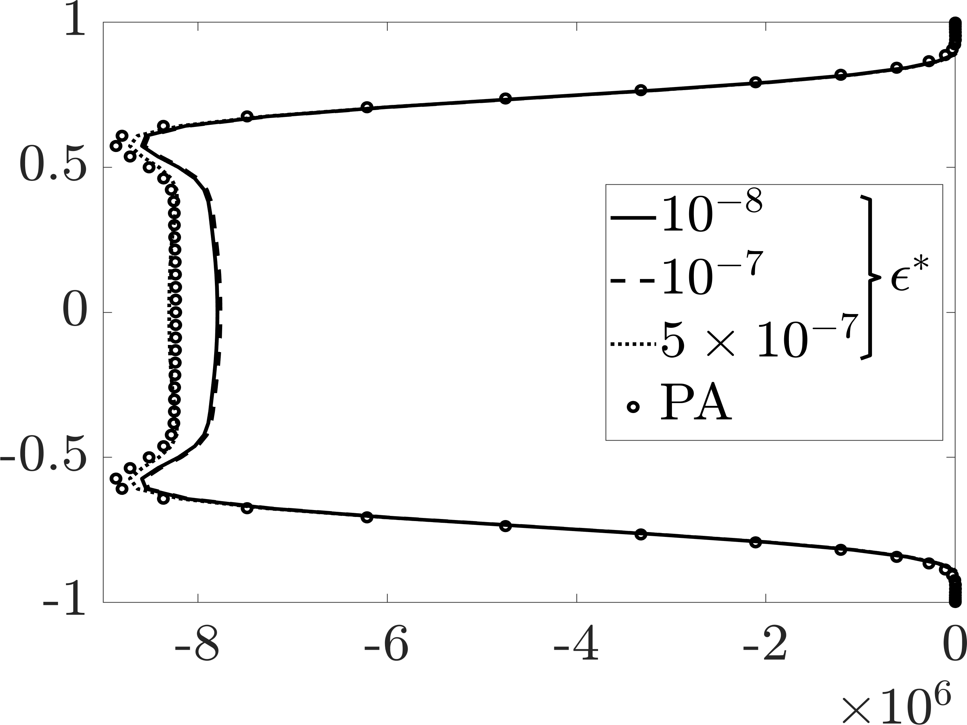}
        \\ [-0.3cm]
        \hspace{0.4cm}{$u_{(0,0)}(y)/\epsilon^{*2}$}
      \end{tabular}
  \end{tabular}
  \vspace{-0.1cm}
  \caption{DNS results for (a) the streak response and (b) the mean-flow deformation, both scaled with $\epsilon^{*2}$, for oblique forcing at $(\bkx, \bkz, \bc) = (0.74, 1.14, 0.396)$. The forcing amplitudes are indicated in the legend, and the results are compared with the $\mathcal{O}(\epsilon^2)$ perturbation-analysis (PA) prediction.}
    \label{fig.Oeps2-scale}
\end{figure}

Figure~\ref{fig.Oeps2-scale} compares the steady-state streak and mean-flow deformation obtained from DNS with predictions from the perturbation expansion for a range of forcing amplitudes and $(\bkx,\bkz,\bc) = (0.74,1.14,0.396)$. Across a broad range of $\epsilon^*$, the DNS results exhibit excellent agreement with the perturbation-based predictions. For sufficiently small forcing amplitudes, the DNS response displays a clear quadratic dependence on $\epsilon^*$, consistent with the second-order scaling. Accordingly, for $\epsilon^* \ll 1$, the leading-order streak component can be extracted as
\beq
u_{\mathrm{s},2\bkz}^{(2)}(y)
\;=\;
u_{\mathrm{s},2\bkz}(y) / \epsilon^{*2}
\eeq
where $u_{\mathrm{s},2\bkz}(y)$ denotes the steady DNS response.

\begin{figure}[t]
  \centering
  \begin{tabular}{c@{\hspace{-0.04 cm}}c@{\hspace{-0.1 cm}}l@{\hspace{-0.04 cm}}c@{\hspace{-0.1 cm}}}
    \subfigure[]{\label{fig.Oeps2-scale-str-over}}
    &&
    \subfigure[]{\label{fig.Oeps2-scale-mean-over}}&
    \\[-.13cm]
    \begin{tabular}{c}
      \vspace{1cm}
      \normalsize{\rotatebox{90}{$y$}}
    \end{tabular}
    &
    \begin{tabular}{c}
      \includegraphics[width=0.332\textwidth]{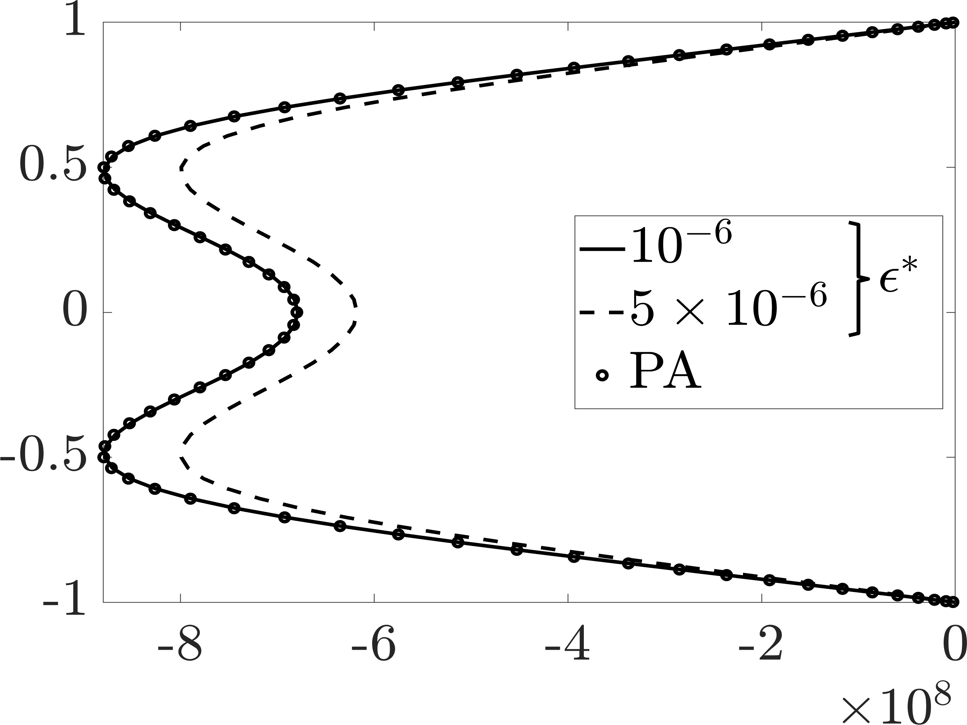}
      \\ [-0.3cm]
      \hspace{0.4cm}{$u_{\mathrm{s},2.28}(y)/\epsilon^{*2}$}
    \end{tabular}
      &
      \begin{tabular}{c}
        \includegraphics[width=0.332\textwidth]{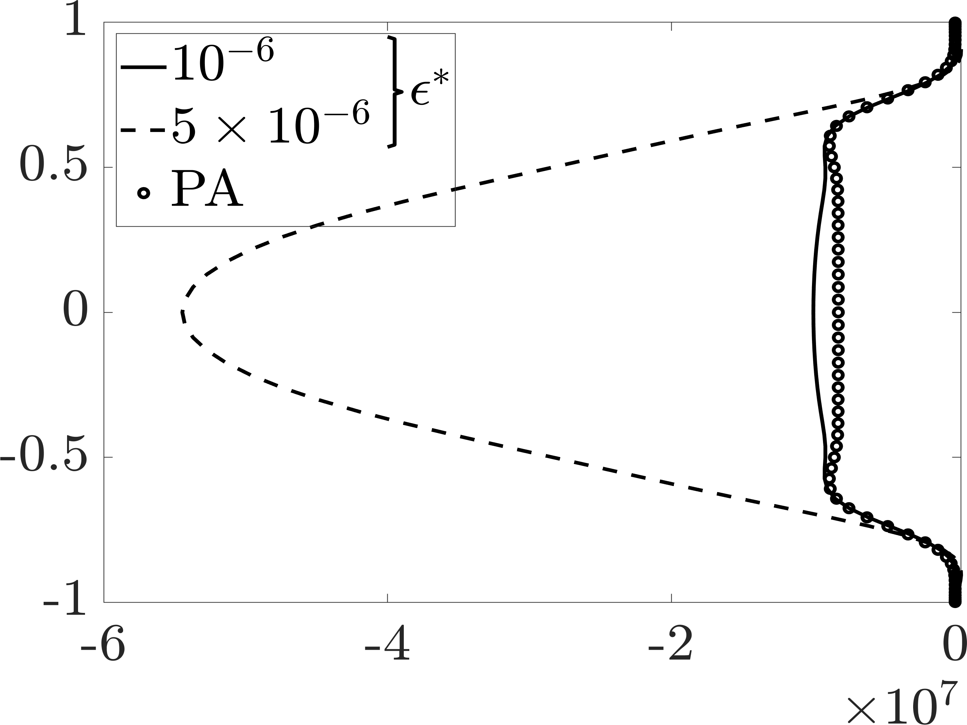}
        \\ [-0.3cm]
        \hspace{0.4cm}{$u_{(0,0)}(y)/\epsilon^{*2}$}
      \end{tabular}
  \end{tabular}
  \vspace{-0.1cm}
  \caption{DNS results for (a) the streak response and (b) the mean-flow deformation, both scaled with $\epsilon^{*2}$, for oblique forcing at $(\bkx, \bkz, \bc) = (0.74, 1.14, 0.396)$. The forcing amplitudes are indicated in the legend, and the results are compared with the $\mathcal{O}(\epsilon^2)$ perturbation-analysis (PA) prediction. Deviations at larger $\epsilon^*$ show the gradual breakdown of the second-order approximation.}
    \label{fig.Oeps2-dns-over}
\end{figure}

As the forcing amplitude $\epsilon^*$ is increased, a gradual breakdown of the second-order approximation is observed; see Fig.~\ref{fig.Oeps2-dns-over}. The earliest deviations arise in the mean-flow modification (Fig.~\ref{fig.Oeps2-scale-mean-over}), followed by an overprediction of the streak amplitude (Fig.~\ref{fig.Oeps2-scale-str-over}). Despite substantial changes in the mean-flow deformation, the streak response remains qualitatively consistent with the second-order prediction. This behavior is in line with earlier weakly nonlinear studies~\cite{PonzianiWNLPois,brandt2002weakly}, which demonstrated that second-order expansions capture qualitative trends but require higher-order corrections for quantitative accuracy. Indeed, these observations motivate a systematic examination of higher-order contributions to energy amplification and their role in precipitating transition, which we undertake in the next section.

	\vspace*{-2ex}
    \section{Higher-order effects on energy amplification and transition}        
\label{sec.higher-order-anal}

	\vspace*{-1ex}
Building on the weakly nonlinear results of the previous section, we now examine how higher-order effects shape the amplification process. Classical studies have examined the influence of large-amplitude streaks on transition by analyzing the stability of the spanwise-periodic base flow they induce. Within this secondary-stability framework, exponentially growing modes emerge once the laminar profile is sufficiently distorted by a finite-amplitude streak~\cite{RSBH1998,brandt2002transition}. If such a modified flow becomes linearly unstable, the resulting fluctuations can grow and ultimately trigger transition. Transient-growth analyses have further shown that spanwise-periodic base flows can exhibit substantial amplification even in the absence of modal instability~\cite{hoebrahen05,schhus02}. 

Together, these observations motivate explicit analysis of higher-order terms in the perturbation expansion of the frequency response about the parabolic base flow.
The structure of these higher-order interactions is summarized in the diagrammatic representation shown in Fig.~\ref{fig.fey-diag}, which recasts the block diagram in Fig.~\ref{fig.bd-eps4} into a form that emphasizes interaction pathways, where resolvent operators associated with the linearized Navier--Stokes equations act as propagators and convective nonlinearities $\cN_1$ and $\cN_2$ serve as quadratic interaction vertices.
This representation reveals an alternating cascade between unsteady oblique waves and steady streamwise streaks across successive perturbation orders.
In this section, we investigate how such higher-order contributions modify the leading-order streak response and alter the amplification pathways governing subcritical transition dynamics.


In Section~\ref{sec.streak-shape}, we show that the spatial structure of higher-order streaks is also captured by the second output singular function of the streamwise-constant resolvent operator. These higher-order contributions are either {\em in phase\/} or {\em out of phase\/} with the second-order response, indicating that they respectively reinforce or attenuate the $\mathcal{O}(\epsilon^2)$ streaks. When reinforcement occurs consistently across perturbation orders, monotonic energy growth and sustained unsteadiness are observed in DNS for forcing amplitudes beyond which the perturbation series diverges; see Section~\ref{sec.eps-critical}. In other cases, nonlinear interactions lead to variations in the orientation of higher-order streaks for specific values of $(k_x,k_z,c)$, producing contributions that include both reinforcing and attenuating effects on the $\mathcal{O}(\epsilon^2)$ response. In this regime, DNS exhibits a quasi-stationary steady state that eventually breaks down via secondary instability; see Sections~\ref{sec.eps-critical} and~\ref{sec.sec-instab} for details.

\begin{figure}
\centering
\input{figures/Figure13}
\caption{Graphical representation of the perturbation expansion of the Navier--Stokes equations, analogous to diagrammatic expansions in statistical physics and quantum field theory. Linearized Navier--Stokes resolvents act as propagators, while convective nonlinearities ${\cN}_1$ and ${\cN}_2$ generate quadratic self- and cross-interactions between modes. Odd orders produce oblique waves with nonzero $(k_x,k_z,\omega)$, whereas even orders yield streamwise streaks at $(0,2k_z,0)$.
This structure leads to systematic reinforcement of streaks through higher-order coupling, consistent with the dominant frequency response shown below.}
    \label{fig.fey-diag}
    \end{figure}

\vspace*{-4ex}
\subsection{Spatial structure and direction of streamwise streaks}
\label{sec.streak-shape}

\begin{figure}[h]
  \centering
  \begin{tabular}{cccr}
  \vspace{-0.1cm}
  \hspace{-0.55cm}
   \subfigure[]{\label{fig.SpatProfP_eps4}}
    & &
    \hspace{-0.5cm}
    \subfigure[]{\label{fig.cos_theta}}
    \vspace{-0.1cm}
    &
   $\cos \theta$ 
    \\[-.19cm]
    \begin{tabular}{c}
      \vspace{0.75cm}
      \normalsize{\rotatebox{90}{$y$}}
    \end{tabular}
    &
    \begin{tabular}{c}
    \hspace{-0.34cm}
        \includegraphics[width=0.3\textwidth]{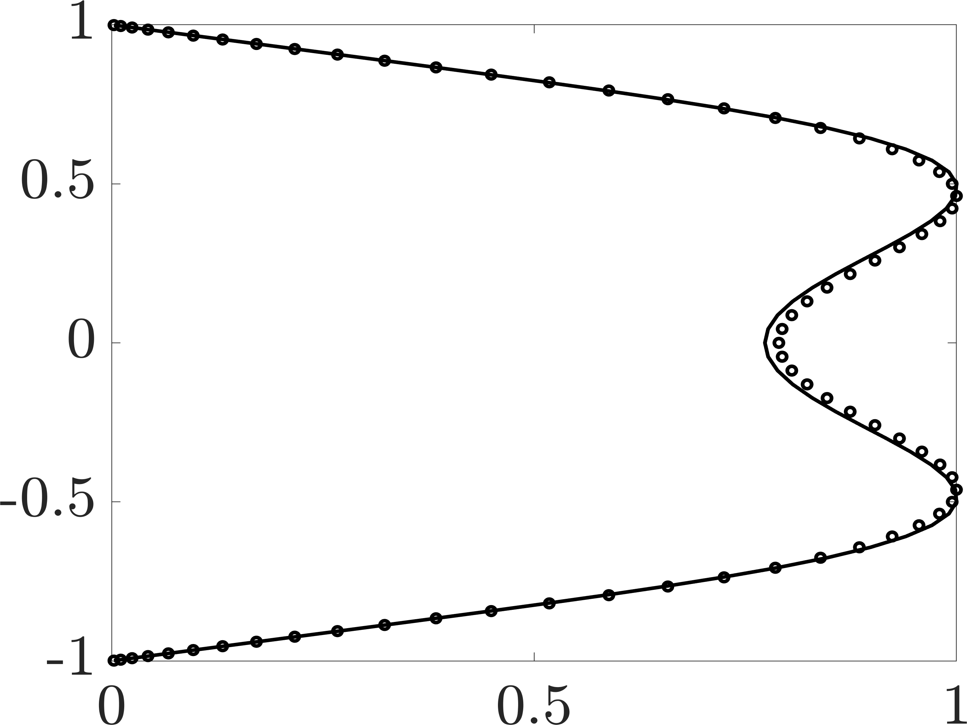}
        \\[-0.051cm] {$u_{\mathrm{s},2.28}^{(2)}(y),u_{\mathrm{s},2.28}^{(4)}(y)$}
    \end{tabular}
    &
    \begin{tabular}{c}
        \vspace{.45cm}
         \hspace{-0.2cm}
        \normalsize{\rotatebox{90}{$\bkx$}}
      \end{tabular}
      &
      \begin{tabular}{c}
      \hspace{-0.35cm}
      \includegraphics[width=0.3\textwidth]{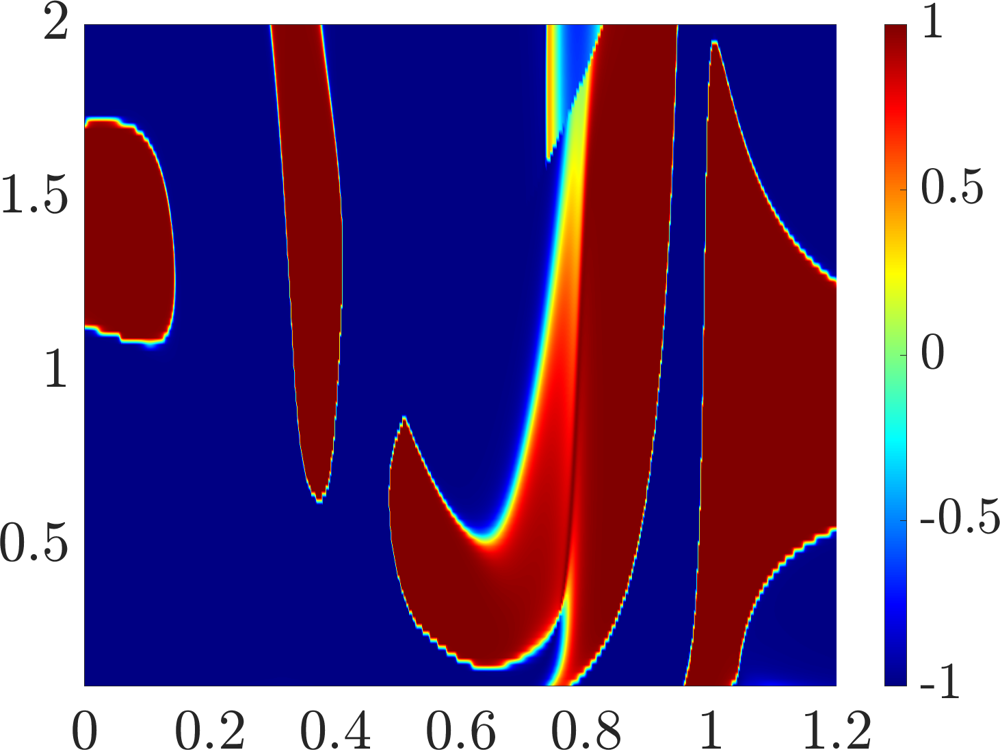}
        \\[-0.051cm] 
        \hspace{-0.1cm}
        {$\bc$}
      \end{tabular} 
  \end{tabular}
  \vspace{-0.1cm}   
   \caption{(a) Wall-normal profiles of the $\mathcal{O}(\epsilon^2)$ (solid) and $\mathcal{O}(\epsilon^4)$ (circles) streaks resulting from $\mathcal{O}(\epsilon)$ oblique-wave forcing with $(k_x, k_z, c) = (0.74, 1.14, 0.396)$. The responses are normalized by their respective peak magnitudes and are accurately captured by the second output singular function of the resolvent operator evaluated at $(k_x, k_z, \omega) = (0, 2.28, 0)$. (b) Cosine of the angle between the $\mathcal{O}(\epsilon^2)$ and $\mathcal{O}(\epsilon^4)$ streaks for $k_{z0} = 2k_z = 2.28$, shown as a function of $k_x$ and $c$, indicating that the two are either in phase (dark-red regions) or out of phase (dark-blue regions).}  
    \label{fig.SpatProfP_all}
\end{figure}

	\vspace*{-1ex}
Extending the perturbation analysis beyond second order reveals a striking persistence in the spatial structure of the resulting streamwise streaks. For the $(k_x, k_z, c)$ values yielding the most energetic $\mathcal{O}(\epsilon^2)$ streaks, Fig.~\ref{fig.SpatProfP_eps4} shows that the $\mathcal{O}(\epsilon^2)$ and $\mathcal{O}(\epsilon^4)$ responses, when normalized by their respective peak amplitudes, exhibit nearly identical wall-normal profiles. In both cases, this structure is determined by the second output singular function of the resolvent operator evaluated at $(k_x, k_z, \omega) = (0, 2.28, 0)$.

To further substantiate these predictions, Appendix~\ref{app.dns-higher-order} applies a reconstruction procedure inspired by~\cite{henlunjoh93} to extract the $\mathcal{O}(\epsilon^4)$ response from DNS. This analysis confirms both the shape and orientation predicted by the perturbation framework, demonstrating consistency with fully nonlinear simulations.

As shown in Fig.~\ref{fig.cos_theta}, this spatial correspondence persists across a broad range of streamwise wavenumbers and temporal frequencies. The cosine of the angle between the $\mathcal{O}(\epsilon^2)$ and $\mathcal{O}(\epsilon^4)$ streaks,
\beq
    \cos \theta 
    \; = \; 
    \frac{\inner{u_{\mathrm{s},2\bkz}^{(2)}}{u_{\mathrm{s},2\bkz}^{(4)}}}
    {\|u_{\mathrm{s},2\bkz}^{(2)}\|_2 \, \|u_{\mathrm{s},2\bkz}^{(4)}\|_2}
\eeq
takes only the values $+1$ or $-1$ throughout the parameter space. Thus, the $\mathcal{O}(\epsilon^4)$ streaks are either {\em in phase\/} (aligned) with the $\mathcal{O}(\epsilon^2)$ streaks or {\em out of phase\/} (anti-aligned), corresponding respectively to $\theta = 0$ and $\theta = \pi$.
	
\begin{figure*}[h]
  \centering
  \begin{tabular}{c@{\hspace{-0.04 cm}}c@{\hspace{-0.01 cm}}l@{\hspace{-0.04 cm}}l@{\hspace{-0.04 cm}}}
    \subfigure[]{\label{fig.Oeps10-maxE}}
    &&
    \subfigure[]{\label{fig.Oeps10-sameDir}}
    &
    \subfigure[]{\label{fig.Oeps10-sameDir-norm}}
    \\[-.03cm]
    \begin{tabular}{c}
      \vspace{.85cm}
      \normalsize{\rotatebox{90}{$y$}}
    \end{tabular}
    &
    \begin{tabular}{c}
      \includegraphics[width=0.3\textwidth]{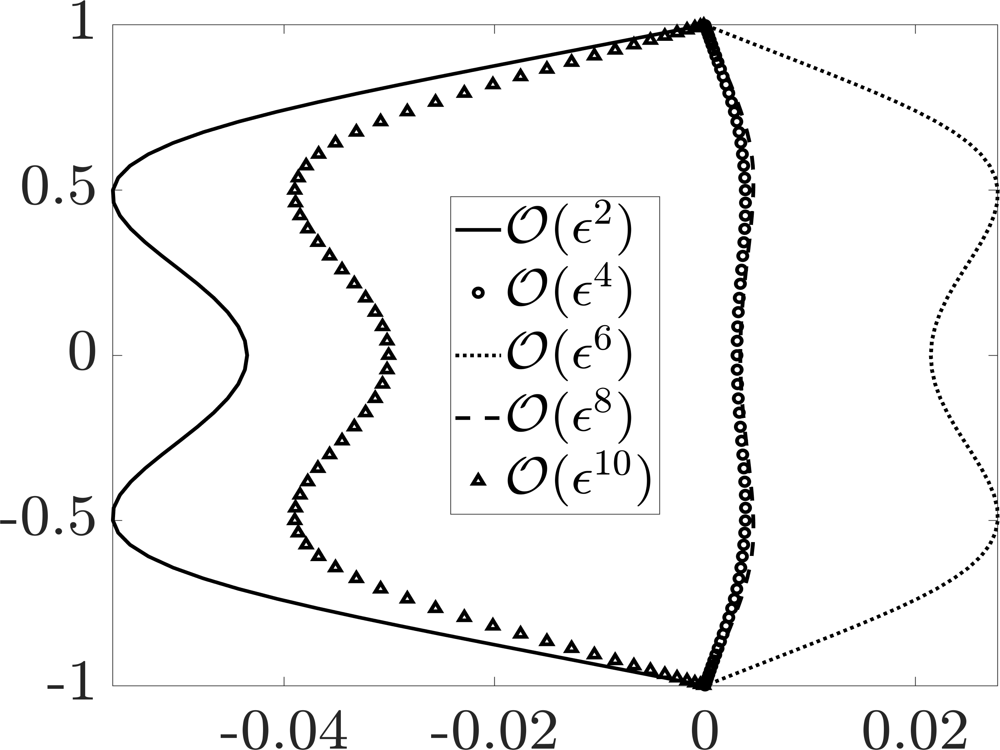}
      \\ 
      \hspace{0.1cm}
      {$u_{\mathrm{s},2k_z}^{(n)}$}
    \end{tabular}
      &
      \begin{tabular}{c}
        \includegraphics[width=0.3\textwidth]{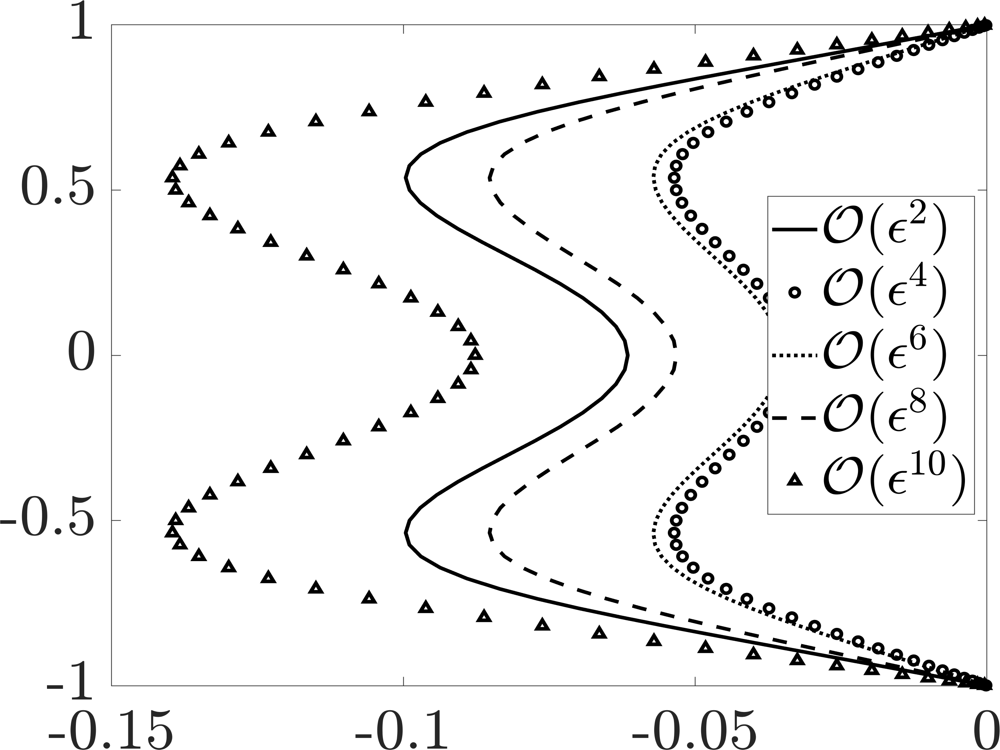}
        \\ 
        {$u_{\mathrm{s},2k_z}^{(n)}$}
      \end{tabular}
      &
      \begin{tabular}{c}
        \includegraphics[width=0.3\textwidth]{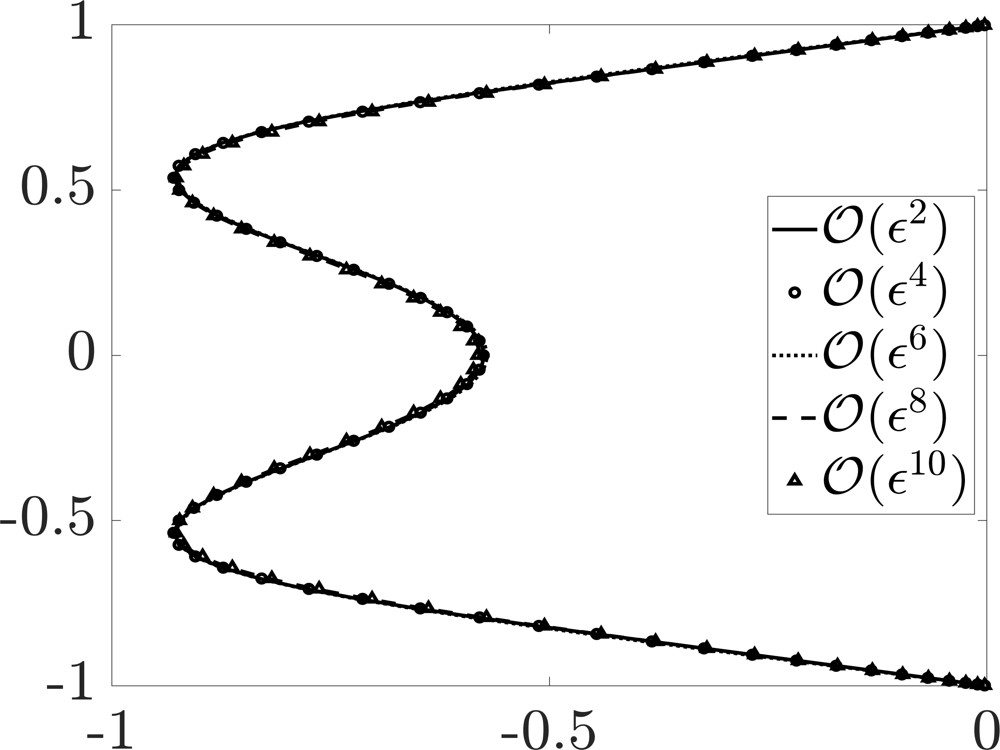}
        \\ 
        {$u_{\mathrm{s},2k_z}^{(n)}$}
      \end{tabular}
  \end{tabular}
  \vspace{-0.1cm}
  \caption{Higher-order streak responses for different oblique-wave forcing parameters. Panel (a) corresponds to $(k_x, k_z, c) = (0.74, 1.14, 0.396)$ with forcing amplitude $\epsilon = 8 \times 10^{-6}$, and panel (b) to $(k_x, k_z, c) = (0.94, 1.48, 0.376)$ with $\epsilon = 2.75 \times 10^{-5}$, both at $Re = 2000$. Panel (c) shows that the responses from (b) collapse on top of each other when normalized to unit energy.}
      \label{fig.higher-order-streaks}
\end{figure*}

Figure~\ref{fig.higher-order-streaks} further confirms that this spatial persistence holds even at higher orders of the perturbation expansion. Accordingly, the second output singular function of the resolvent operator at $k_x = 0$ continues to capture the wall-normal structure of streamwise streaks. For the parameters yielding the most amplified $\mathcal{O}(\epsilon^2)$ streaks, Fig.~\ref{fig.Oeps10-maxE} reveals that higher-order responses exhibit subtle phase variations. In contrast, Fig.~\ref{fig.Oeps10-sameDir} shows that, for certain combinations of $(k_x, k_z, c)$, the higher-order responses remain {\em in phase\/} (aligned) with the $\mathcal{O}(\epsilon^2)$ streaks. As illustrated in Fig.~\ref{fig.Oeps10-sameDir-norm}, the wall-normal profiles of the streaks remain remarkably consistent across perturbation orders.

\begin{figure*}[h]
  \centering
      \begin{tabular}{c@{\hspace{-0.04 cm}}c@{\hspace{-0.01 cm}}l@{\hspace{-0.04 cm}}c@{\hspace{-0.01cm}}}
	{\subfigure[]{\label{fig.E1-Oeps3}}}
	&
	&
	{\subfigure[]{\label{fig.Oeps3-phaseShift}}}
	&
    \\[-.03cm]
    \begin{tabular}{c}
      \vspace{.85cm}
      \normalsize{\rotatebox{90}{$k_x$}}
    \end{tabular}
    &
    \begin{tabular}{c}
      \includegraphics[width=0.3\textwidth]{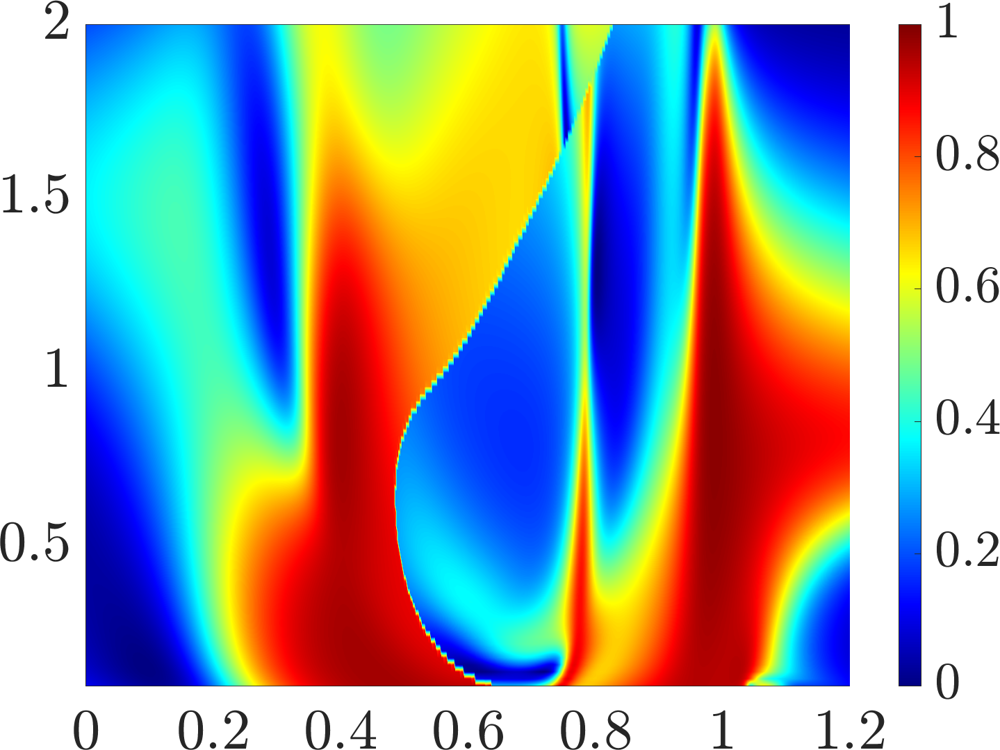}
      \\ 
      \hspace{0.1cm}
      {$c$}
    \end{tabular}
      &&
      \begin{tabular}{c}
        \includegraphics[width=0.3\textwidth]{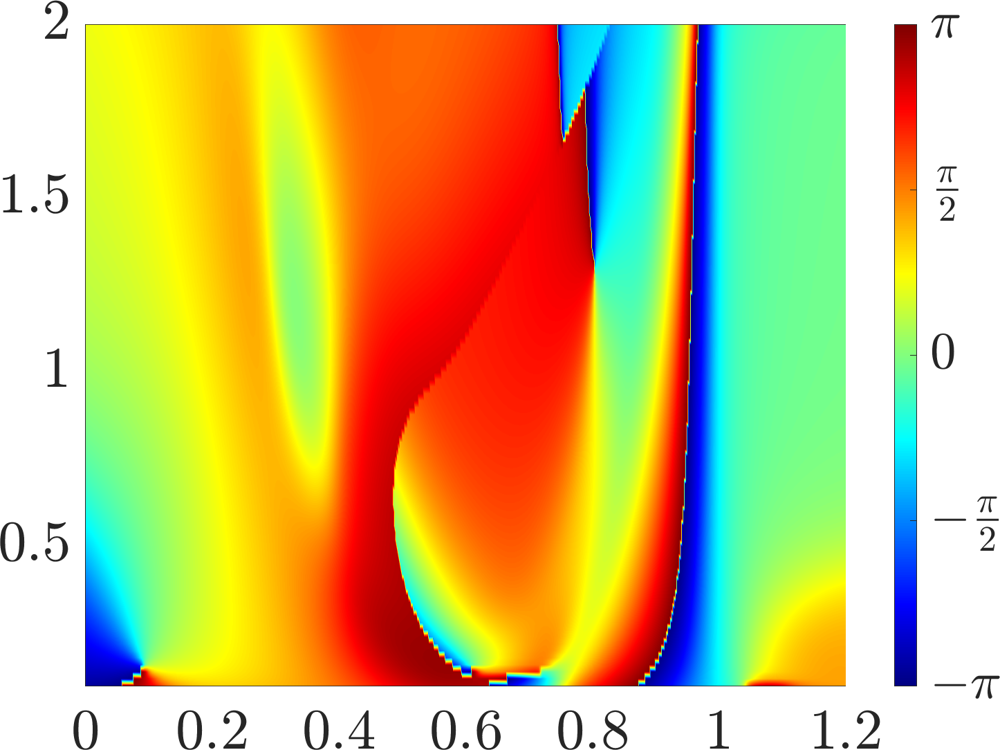}
        \\ {$c$}
      \end{tabular}
  \end{tabular}
  \vspace{-0.1cm}
  \caption{(a) Energy fraction of the principal $\mathcal{O}(\epsilon)$ oblique-wave mode in the $\mathcal{O}(\epsilon^3)$ oblique response as a function of the streamwise wavenumber $k_x$ and phase speed $c$, for fixed spanwise wavenumber $k_z = 1.14$ at $Re = 2000$. (b) Streamwise phase shift $\varphi$ between the principal $\mathcal{O}(\epsilon)$ oblique-wave mode and the corresponding $\mathcal{O}(\epsilon^3)$ response.}
  \label{fig.Normalized_streaks}
\end{figure*}

To gain insight into the observed shape similarity and phase shifts of higher-order streaks, we examine the contribution of oblique waves to the $\mathcal{O}(\epsilon^4)$ response. As illustrated in Fig.~\ref{fig.bd-wn-psi_n}, the unsteady fluctuations that drive the $\mathcal{O}(\epsilon^4)$ streaks arise from nonlinear interactions between the $\mathcal{O}(\epsilon)$ and $\mathcal{O}(\epsilon^3)$ oblique waves. The energy contribution of the principal $\mathcal{O}(\epsilon)$ oblique-wave mode to the $\mathcal{O}(\epsilon^3)$ response, shown in Fig.~\ref{fig.E1-Oeps3}, reveals that the $\mathcal{O}(\epsilon)$ oblique waves responsible for the strongest amplification of the $\mathcal{O}(\epsilon^2)$ streaks also dominate the $\mathcal{O}(\epsilon^3)$ response. This observation motivates the approximation
	\beq\label{eq.u3app}
	\bu_{(\pm k_x,k_z)}^{(3)} (y,c) 
	\; \approx \; 
	\bvartheta_{( \pm k_x,k_z),1}(y,c)\mre^{ \pm \mri \varphi},
	\eeq
where $\varphi$ denotes the phase difference between the $\mathcal{O}(\epsilon)$ and $\mathcal{O}(\epsilon^3)$ oblique-wave responses. Figure~\ref{fig.Oeps3-phaseShift} shows the dependence of $\varphi$ on $k_x$ and $c$. For the parameters that maximize the energy of the $\mathcal{O}(\epsilon^2)$ streaks, the $\mathcal{O}(\epsilon^3)$ response exhibits a phase shift of approximately $1.44$~rad ($82.5^\circ$). This prediction is confirmed in DNS by comparing the reconstructed $\mathcal{O}(\epsilon)$ and $\mathcal{O}(\epsilon^3)$ responses; see Appendix~\ref{app.dns-higher-order}.
	
\begin{figure*}[h]
  \centering
    \begin{tabular}{cr}
    & $\cos \varphi$
    \\
    \begin{tabular}{c}
    \vspace*{.25cm}
        \normalsize{\rotatebox{90}{$k_x$}}
      \end{tabular}
      &
      \begin{tabular}{c}
      \includegraphics[width=0.3\textwidth]{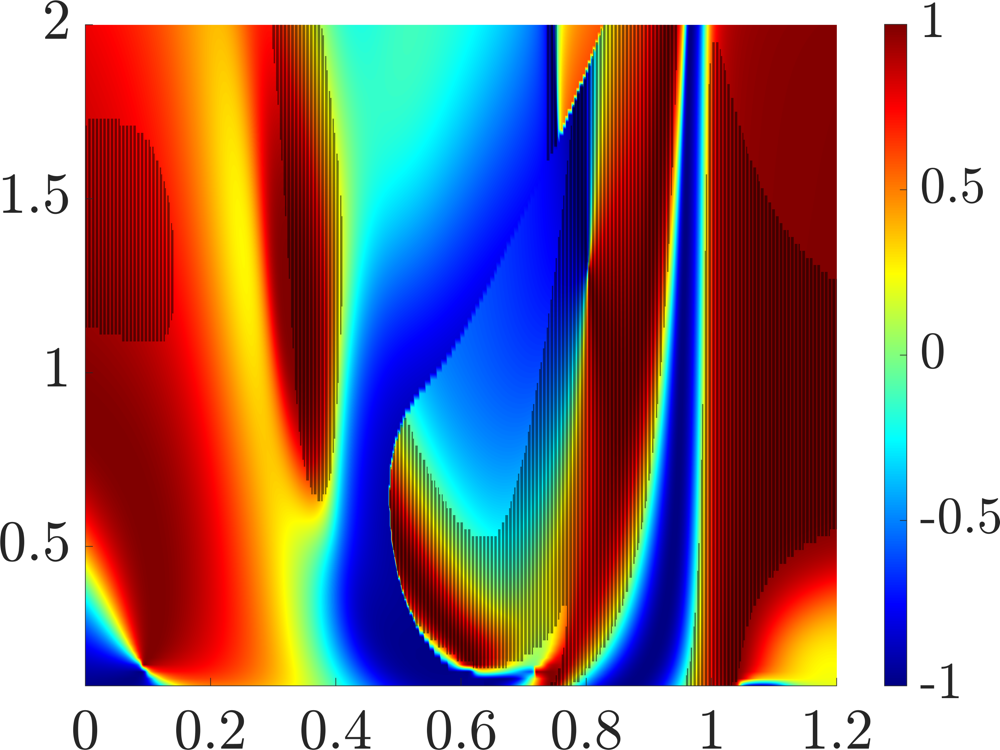}
        \\ [-0.051cm]
        {$c$}
      \end{tabular}
  \end{tabular}
  \vspace{-0.1cm}
  \caption{Cosine of the streamwise phase shift $\varphi$ from Fig.~\ref{fig.Oeps3-phaseShift} (color map), overlaid with regions where $\cos\theta>0$ from Fig.~\ref{fig.cos_theta} (hatched).}
    \label{fig.cos-angles}
\end{figure*}

The nonlinear interactions between these oblique waves then generate the following contribution to the $\mathcal{O}(\epsilon^4)$ streaks:
	\beq\label{eq.d4app}
 	\widetilde{\cN}_2
	\left(\bu_{(k_x,k_z)}^{(1)}, \bu_{(-k_x,k_z)}^{(3)}\right) 
	\,+\, 
	\widetilde{\cN}_2
	\left( \bu_{(-k_x,k_z)}^{(1)}, \bu_{(k_x,k_z)}^{(3)}\right) 
 	\, \approx \; 
	2\widetilde{\cN}_2
	\left( \bu_{(k_x,k_z)}^{(1)}, \bu_{(-k_x,k_z)}^{(1)}\right) \cos \varphi.
	\eeq
Apart from the phase factor $\cos\varphi$, this expression shows that the forcing associated with the $\mathcal{O}(\epsilon^4)$ streaks remains spatially coherent with that of the $\mathcal{O}(\epsilon^2)$ streaks, resulting in strong structural similarity between the corresponding responses. The sign of $\cos\varphi$ directly determines whether the higher-order streaks are in phase (aligned) or out of phase (anti-aligned) relative to the $\mathcal{O}(\epsilon^2)$ streaks---that is, whether they are oriented in the same or opposite directions---as illustrated in Fig.~\ref{fig.cos-angles}; cf.\ Fig.~\ref{fig.cos_theta}. Thus, the streamwise phase shift between the interacting oblique waves governs the direction reversals observed across the parameter space for which approximations~\eqref{eq.u3app}--\eqref{eq.d4app} are valid. Although the approximation is not uniformly accurate across the entire parameter space, it remains reliable within dynamically relevant regions, including those associated with the strongest amplification of the $\mathcal{O}(\epsilon^2)$ streaks (Fig.~\ref{fig.E1-Oeps3}).
	
Higher-order terms generate additional oblique waves and streaks that retain the spatial structure dictated by their respective singular modes. Each successive oblique-wave interaction, however, introduces an additional phase shift, so higher-order phase relationships cannot be inferred from the $\mathcal{O}(\epsilon^4)$ behavior alone; cf.\ Fig.~\ref{fig.Oeps10-maxE}. Consequently, the nonlinear interactions responsible for higher-order streaks lead to variations in their orientation, yielding a perturbation expansion whose terms either reinforce or attenuate the $\mathcal{O}(\epsilon^2)$ response depending on the accumulated phase offset.

When the phase shift between the $\mathcal{O}(\epsilon)$ and $\mathcal{O}(\epsilon^3)$ oblique-wave components is small ($\varphi \approx 0$), all higher-order streaks remain in phase with the leading $\mathcal{O}(\epsilon^2)$ structure; see Fig.~\ref{fig.higher-order-streaks}. This delineates the parameter regimes in which streaks remain coherently aligned and higher-order interactions act constructively. Within these regions, the superposition of higher-order contributions leads to substantial reinforcement of the $\mathcal{O}(\epsilon^2)$ streaks and a pronounced increase in their cumulative energy.

	\vspace*{-4ex}
\subsection{Streak reinforcement and energy amplification}
	\label{sec.eps-critical}
	\vspace*{-2ex}

By extending the perturbation expansion of the frequency response about the parabolic base flow, we have shown that higher-order streaks generated by the original $\cO (\epsilon)$ oblique-wave forcing can reinforce the leading $\mathcal{O}(\epsilon^2)$ response across a broad range of wavenumbers and temporal frequencies. This regime therefore provides a natural starting point for examining how coherent higher-order nonlinear interactions modify energy amplification and set the range of validity of the weakly nonlinear expansion. These nonlinear interactions arise directly from the structure of the governing equations and mirror the self-sustaining mechanism~\cite{wal97} observed in wall-bounded flows, in which aligned streaks amplify one another and perpetuate the turbulence cycle.

Here, we first examine how this nonlinear reinforcement modifies the total perturbation energy and promotes further amplification. We then turn our attention to regimes in which higher-order contributions also include attenuation of the second-order response. In both cases, we employ convergence tests in conjunction with the Shanks transformation to estimate the forcing amplitudes for which the respective perturbation expansions remain valid. These estimates are confirmed by comparing predictions of the perturbation analysis against DNS results for forcing amplitudes within the convergence region as well as near the validity boundary.
 
	\vspace*{1ex}
\noindent {\bf Aligned streak reinforcement.} We first focus on oblique-wave parameters $(k_x, k_z, c)$ for which {\em higher-order streaks are aligned with the $\mathcal{O}(\epsilon^2)$ response\/} and therefore {\em reinforce it.\/} To estimate the forcing amplitude at which higher-order effects become dynamically significant, we define $\epsilon_{\mathrm{cr}}$ as the smallest value of $\epsilon$ for which the expansion no longer converges under the applied convergence tests. In the vicinity of this threshold, higher-order terms exert a pronounced influence on the total response. For Poiseuille flow at $Re = 2000$ with $(k_x, k_z, c) = (0.94, 1.48, 0.376)$, application of the $n$th-root test in conjunction with the Shanks transformation yields $\epsilon_{\mathrm{cr}} \approx 1.9 \times 10^{-5}$. Importantly, $\epsilon_{\mathrm{cr}}$ should not be interpreted as a bifurcation point, but rather as a practical indicator beyond which higher-order interactions dominate and DNS no longer converges to a steady weakly nonlinear state. The estimation procedure is summarized below; see Appendix~\ref{app.nth-root-test} for details.

	\vspace*{-1ex}
\begin{remark}[Procedure for estimating $\epsilon_{\mathrm{cr}}$]
At a fixed Reynolds number, the critical forcing amplitude $\epsilon_{\mathrm{cr}}$ is estimated as follows. Oblique-wave parameters $(k_x,k_z,c)$ are first restricted to cases in which higher-order streaks remain directionally aligned with the second-order response, as indicated by dominance of the $\mathcal{O}(\epsilon)$ principal component in the $\mathcal{O}(\epsilon^3)$ response and a small streamwise phase shift $\varphi$ (Section~\ref{sec.streak-shape}). Higher-order streak responses $u_{\mathrm{s},2k_z}^{(n)}$ are then computed by retaining only the dominant oblique-wave/streak interaction terms associated with the base frequencies. Convergence of the resulting perturbation series is assessed using finite-order approximations of the $n$th-root test,
	$
	\|\epsilon^n u_{\mathrm{s},2k_z}^{(n)}\|_2^{1/n},
	$
with convergence accelerated via the Shanks transformation (Appendix~\ref{app.nth-root-test}). In cases where higher-order contributions also include attenuation of the second-order response, a sliding-window Shanks procedure is employed, and $\epsilon_{\mathrm{cr}}$ is reported as the average of stable accelerated estimates.
	\end{remark}

\begin{figure*}[h]
  \centering
  \begin{tabular}{c@{\hspace{-0.04 cm}}c@{\hspace{-0.01 cm}}c@{\hspace{-0.04 cm}}c@{\hspace{-0.1 cm}}c}
    \subfigure[]{\label{fig.Shank_1em5_streak}}
    &&
    \subfigure[]{\label{fig.Shank_1em5_mean}}&
    \\[-.03cm]
    \begin{tabular}{c}
        \vspace{.8cm}
        \normalsize{\rotatebox{90}{$y$}}
      \end{tabular}
      &
      \begin{tabular}{c}
        \includegraphics[width=0.33\textwidth]{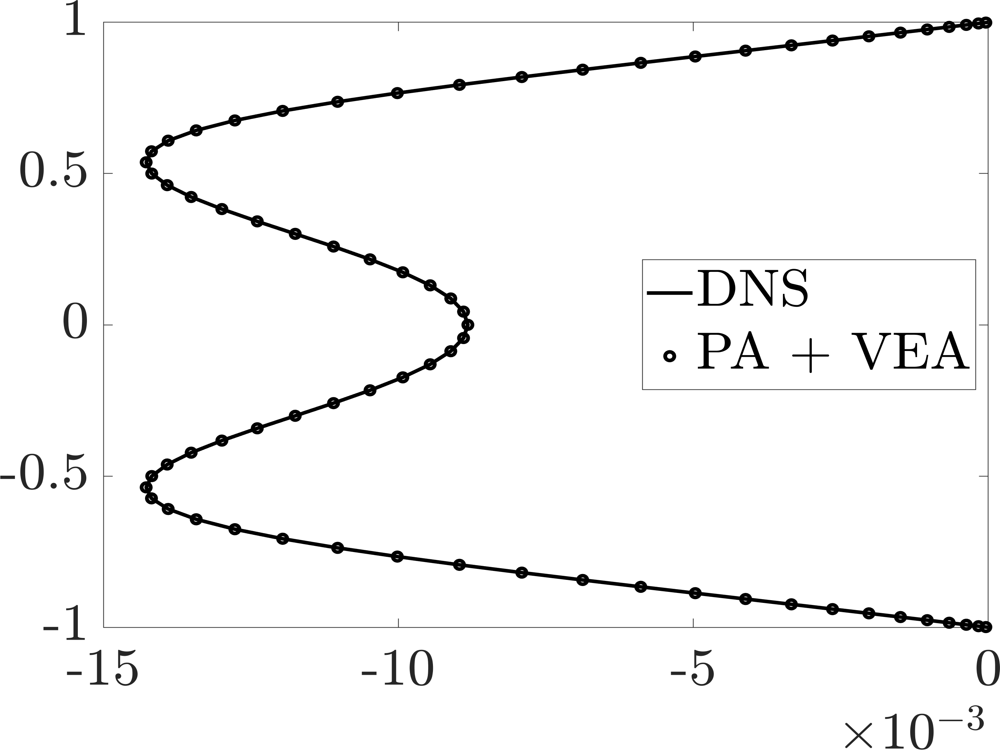}
        \\ [-0.25cm]
        {$u_{\mathrm{s},2.96}(y)$}
      \end{tabular}
      &
    \begin{tabular}{c}
      \vspace{.85cm}
      \normalsize{\rotatebox{90}{$y$}}
    \end{tabular}
    &
    \begin{tabular}{c}
      \includegraphics[width=0.33\textwidth]{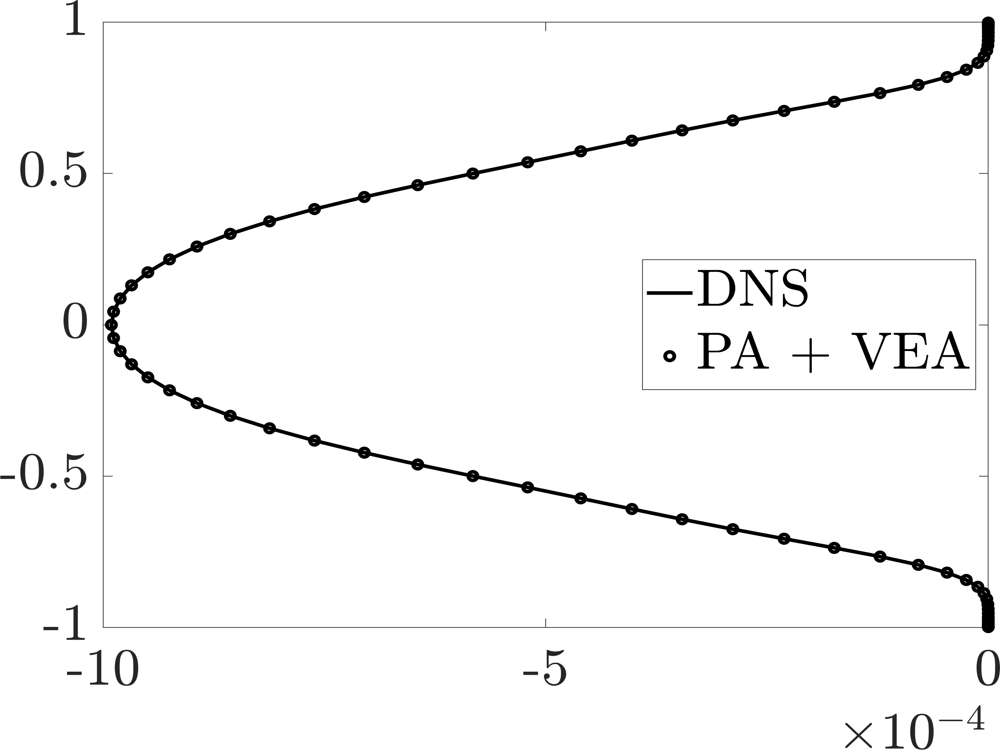}
      \\ [-0.25cm]
      \hspace{0.1cm}
      {$u_{(0,0)}(y)$}
    \end{tabular}
  \end{tabular}
  \vspace{-0.1cm}
  \caption{Comparison between second-order Shanks-transformed perturbation predictions (computed via the vector epsilon algorithm, VEA) and DNS results for Poiseuille flow at $Re = 2000$ with oblique-wave parameters $(k_x, k_z, c) = (0.94, 1.48, 0.376)$ and forcing amplitude $\epsilon^* = 1\times 10^{-5}$. Shown are the steady responses associated with (a) streamwise streaks and (b) mean-flow deformation.}
    \label{fig.Shank_1em5}
\end{figure*}

	\vspace*{-1ex}
Figure~\ref{fig.Shank_1em5} compares the velocity fluctuations predicted by the perturbation framework with DNS results for $\epsilon^* = 10^{-5} < \epsilon_\mathrm{cr} = 1.9 \times 10^{-5}$. These results confirm that a model based on fluctuations about the parabolic base flow remains accurate even in this moderately nonlinear regime. Crucially, inclusion of higher-order terms---beyond $\mathcal{O}(\epsilon^2)$---is required to achieve quantitative agreement. To incorporate these terms efficiently, we employ the Shanks transformation---a classical series convergence-acceleration technique---implemented using the vector epsilon algorithm (VEA)~\cite{gramor00}. Comparison with DNS demonstrates the strong predictive capability of the perturbation framework, even at elevated forcing amplitudes. Moreover, the higher-order expansion accurately captures the modification of the mean-flow deformation relative to the second-order prediction; cf.\ Fig.~\ref{fig.Oeps2-dns-over}. A detailed explanation of this effect based on higher-order contributions is provided in Appendix~\ref{app.mean-higher-order}.

When the amplitude of the oblique-wave forcing slightly exceeds the estimated convergence threshold, $\epsilon^* > \epsilon_\mathrm{cr}$, the perturbation expansion predicts progressively growing amplification of the streak component, indicating loss of validity of the weakly nonlinear description. Each successive term contributes constructively to the leading $\mathcal{O}(\epsilon^2)$ response, producing progressively larger streak modifications. These amplified streaks, in turn, modify the mean flow (cf.\ Appendix~\ref{app.mean-higher-order}), introducing a finite-amplitude correction to the parabolic profile and thereby marking the boundary of validity for the weakly nonlinear model constructed around it. This mean-flow alteration changes the underlying dynamics and steers the system toward a transitional regime.

\begin{figure}[h]
  \centering
  \begin{tabular}{ccl}
  \vspace{-0.1cm}
  \hspace{-0.7cm}
   \subfigure[]{\label{fig.L2norm_fluct_tlin}}
   &
    &
    \subfigure[]{\label{fig.L2norm_components}}
    \\[-.2cm]
    \begin{tabular}{c}
    \hspace{-0.2cm}
      \normalsize{\raisebox{0cm}[0pt][0pt]{\rotatebox{90}{$\| \bu(t)\|^2$}}}
    \end{tabular}
    &
    \begin{tabular}{c}
    \includegraphics[width=0.29\textwidth]{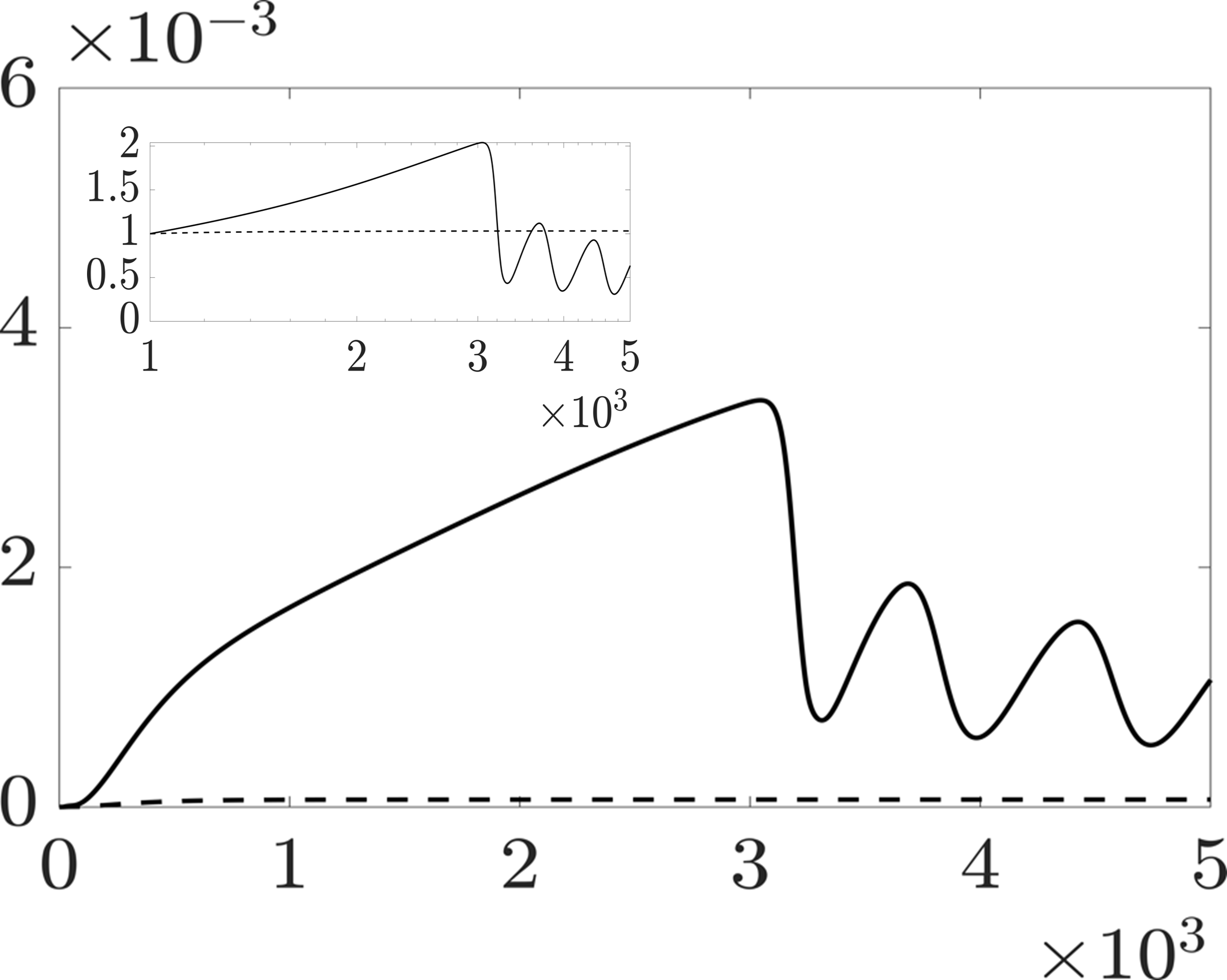}
      \\[-0.4cm] 
      {$t$}
    \end{tabular}
    &
    \begin{tabular}{c}
    \hspace{-0.1cm}
    \vspace{-0.4cm}
    \includegraphics[width=0.3\textwidth]{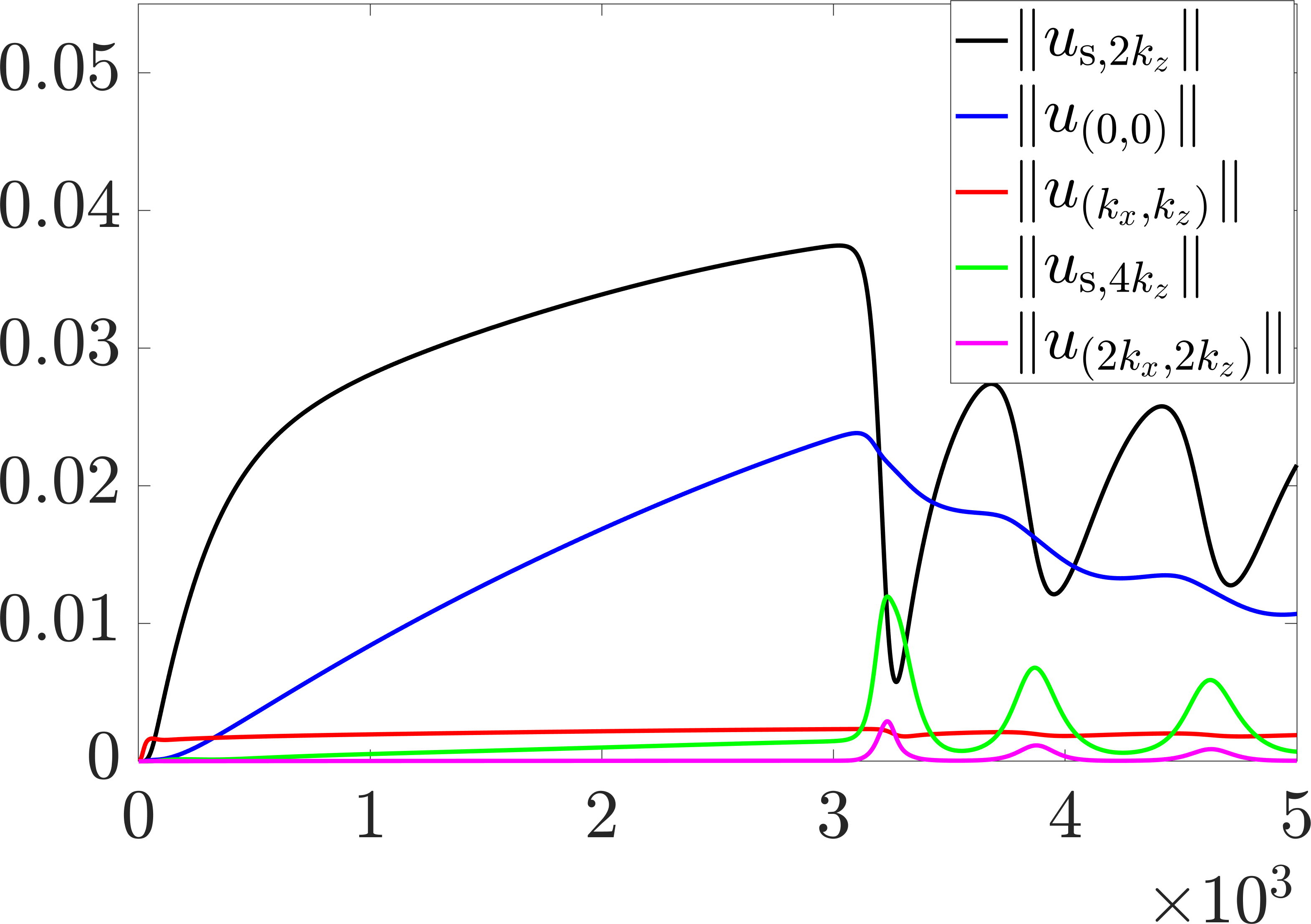}
      \\ [-0.4cm] 
      \hspace{0.1cm}
      \raisebox{-0.35cm}{$t$}
    \end{tabular}  
  \end{tabular}
  \vspace{-0.051cm}  
  \caption{DNS results for Poiseuille flow at $Re = 2000$ with oblique-wave forcing parameters $(k_x, k_z, c) = (0.94, 1.48, 0.376)$.
(a) Temporal evolution of the kinetic energy of velocity fluctuations for forcing amplitudes $\epsilon^* = 10^{-5}$ (dashed) and $\epsilon^* = 2 \times 10^{-5}$ (solid). The inset shows a magnified view for $t \ge 1000$, with energies normalized by their respective initial values $\| \bu(1000) \|_2^2$.
(b) Temporal evolution of the square root of the kinetic energy associated with selected Fourier components of the response for $\epsilon^* = 2 \times 10^{-5}$.}  \label{fig.L2norm_fluct_DNS}
\end{figure}

DNS results corroborate these predictions: for $\epsilon^* = 2\times10^{-5}$, the fluctuation energy no longer saturates but instead exhibits sustained growth, signaling departure from the weakly nonlinear regime and eventual transition to a quasi-periodic state; see Fig.~\ref{fig.L2norm_fluct_tlin}. By contrast, a slightly smaller forcing amplitude, $\epsilon^* < \epsilon_\mathrm{cr}$, maintains laminar behavior over the same time horizon. Figure~\ref{fig.L2norm_components} shows that, once the threshold is exceeded, both the streak and mean-flow fluctuation energies grow until a quasi-periodic regime is established, fully consistent with the behavior predicted by the perturbation expansion.

We observe that the same amplification mechanism persists across a broad range of subcritical Reynolds numbers (not shown). Within the uncertainty of the estimated $\epsilon_{\mathrm{cr}}$, flows forced below this threshold evolve toward a steady state, whereas those forced near it undergo transition marked by pronounced energy growth driven by higher-order streak interactions. For $Re \le 1250$, however, no choice of oblique-wave parameters $(k_x,k_z,c)$ yields higher-order corrections aligned with the $\mathcal{O}(\epsilon^2)$ streaks. This limitation is likely attributable to the parameter-selection criteria, which restrict attention to streak-dominated responses satisfying the small phase-shift condition $\varphi \le 0.1$. Notably, the onset of this amplification mechanism occurs near $Re \approx 1000$, coinciding with the Reynolds numbers at which sustained turbulence first becomes possible~\cite{pathea69,carwidpee82}. As $Re$ increases, the admissible parameter region broadens, enhancing the relevance of this mechanism for realistic flow configurations.

\vspace*{1ex}
\noindent {\bf Coexisting reinforcing and attenuating effects.} The oblique-wave parameters examined thus far were chosen to promote reinforcement of the leading-order response by higher-order streaks. In Poiseuille flow at $Re = 2000$, however, this reinforcement does not hold for the most amplified $\mathcal{O}(\epsilon^2)$ streaks identified in Section~\ref{sec:princWNL}; cf.\ Fig.~\ref{fig.Oeps10-maxE}. In this case, the critical forcing amplitude at which the perturbation series ceases to converge absolutely is estimated using the $n$th-root test (Appendix~\ref{app.nth-root-test}). Because higher-order streak contributions differ in sign, absolute convergence is lost earlier than practical accuracy, and the estimate may slightly underpredict the true threshold. More importantly, the alternating structure of the series reveals that higher-order terms can partially counteract the leading-order response, thereby reducing net energy amplification. As a result, even beyond the formal breakdown of the perturbation expansion, the nonlinear response may remain comparatively weak when higher-order contributions do not reinforce one another coherently.

\begin{figure}[h]
  \centering
  \begin{tabular}{ccccl}
  \vspace{-0.1cm}
  \hspace{-.5cm}
   \subfigure[]{\label{fig.L2norm_maxE}}
    & &
    \hspace{-0.25cm}
    \subfigure[]{\label{fig.Shank_streaks_8em6}}
    \vspace{-0.1cm}
    &
    &
    \subfigure[]{\label{fig.Shank_mean_8em6}}
    \\[.11cm]
    \begin{tabular}{c}
    \hspace{-0.3cm}
      \vspace{0.6cm}
      \normalsize{\rotatebox{90}{$\| \bu(t)\|^2$}}
    \end{tabular}
    &
    \begin{tabular}{c}
       \hspace{-0.1cm}\includegraphics[width=0.31\textwidth]{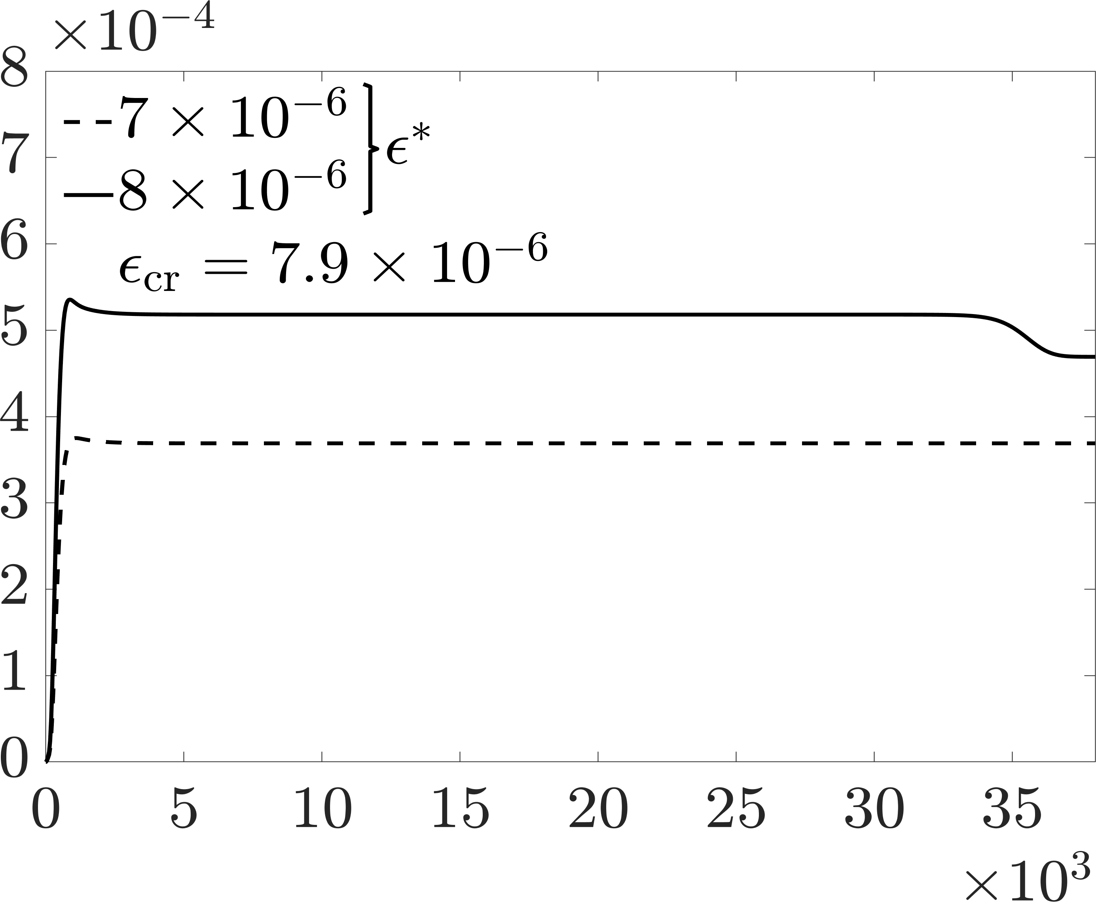}
      \\[-0.1cm] 
      \raisebox{0.3cm}[0pt][0pt]{$t$}
    \end{tabular}
    &
    \begin{tabular}{c}
        \vspace{.55cm}
         \hspace{-0.25cm}
        \normalsize{\rotatebox{90}{$y$}}
      \end{tabular}
      &
      \begin{tabular}{c}
      \hspace{-0.34cm}
        \includegraphics[width=0.3\textwidth]{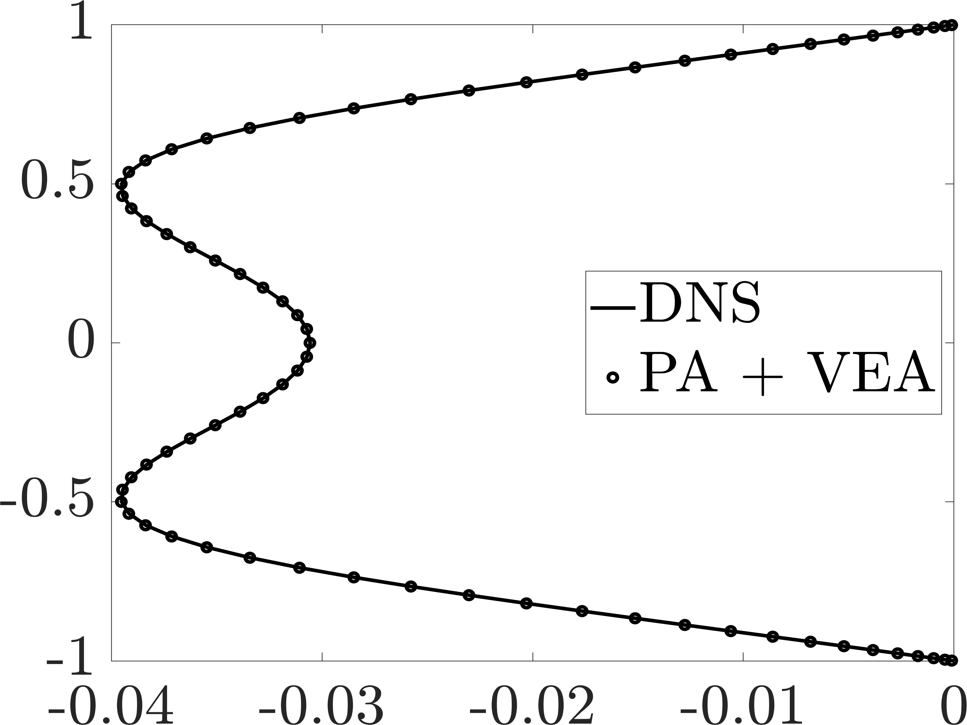}
        \\[-0.1cm] 
        {$u_{\mathrm{s},2.28}(y)$}
      \end{tabular}
    &
    \begin{tabular}{c}
    \hspace{-0.39cm}
      \includegraphics[width=0.33\textwidth]{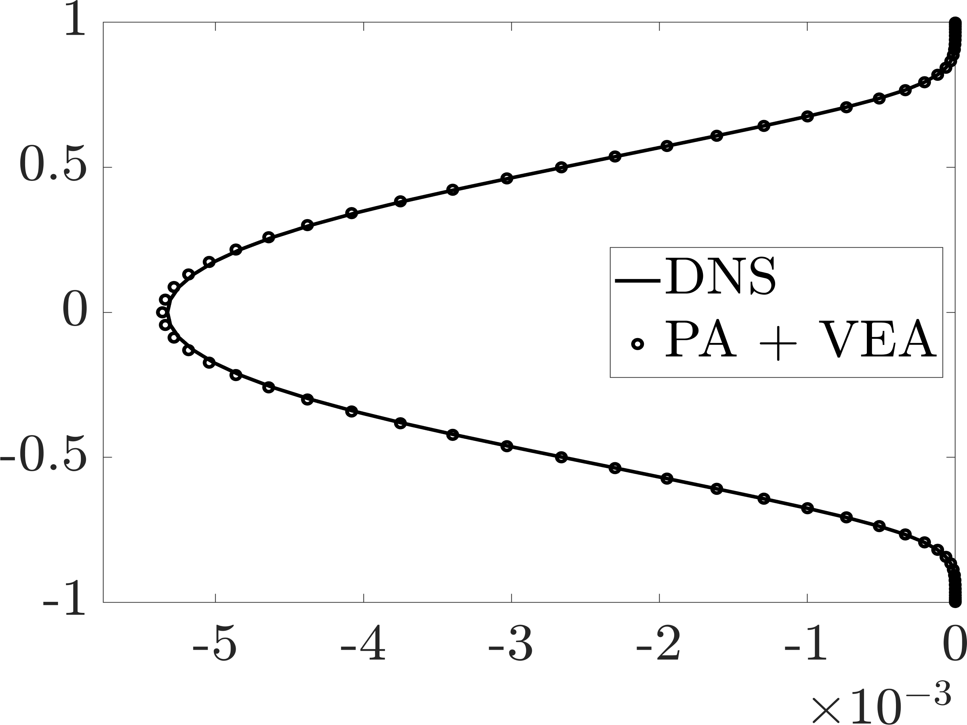}
      \\[-0.44cm]  
      \hspace{0.1cm}
      {$u_{(0,0)}(y)$}
    \end{tabular}  
  \end{tabular}
  \vspace{-0.1cm}  
  \caption{DNS results for Poiseuille flow at $Re = 2000$ with oblique-wave parameters $(k_x, k_z, c) = (0.74, 1.14, 0.396)$. (a) Time evolution of the kinetic energy of velocity fluctuations for $\epsilon^* = 7 \times 10^{-6}$ (dashed) and $\epsilon^* = 8 \times 10^{-6}$ (solid). The estimated convergence threshold is $\epsilon_{\mathrm{cr}} = 7.9 \times 10^{-6}$. (b,c) Comparison between predictions from the sixth-order Shanks-transformed perturbation analysis (via VEA) and the quasi-stationary DNS response at $\epsilon^* = 8 \times 10^{-6}$ and $t = 10{,}000$, showing fluctuations associated with (b) streamwise streaks and (c) mean-flow deformation.}
\label{fig.L2norm_Shank_DNS_maxE}
\end{figure}

This distinction is borne out by the DNS results shown in Fig.~\ref{fig.L2norm_maxE}, which display the temporal evolution of the kinetic energy of velocity fluctuations. The estimated divergence threshold, $\epsilon_{\mathrm{cr}} = 7.9 \times 10^{-6}$, closely matches the forcing amplitude $\epsilon^* = 8 \times 10^{-6}$ at which DNS no longer attains a steady state. Figures~\ref{fig.Shank_streaks_8em6} and~\subref{fig.Shank_mean_8em6} further demonstrate that, even in the quasi-steady regime observed above this threshold, the perturbation expansion continues to provide accurate predictions of the flow fluctuations, albeit requiring a larger number of terms to converge owing to the increased influence of higher-order effects. These results justify the use of the weakly nonlinear framework for analyzing streak instability in Section~\ref{sec.sec-instab}, even when the forcing amplitude modestly \mbox{exceeds the estimated critical value.}

Finally, analogous trends persist across a broad range of subcritical Reynolds numbers (not shown). As in Section~\ref{sec:princWNL}, the oblique-wave parameters are selected to maximize the energy of the $\mathcal{O}(\epsilon^2)$ streaks. The estimated value of $\epsilon_{\mathrm{cr}}$ thus provides a practical indicator of the forcing amplitude beyond which the perturbation series diverges and DNS no longer converges to a steady state. In regimes where higher-order streak contributions alternate in sign, exceeding this threshold leads to weak unsteadiness characterized by slow energy evolution, whereas in regimes dominated by constructive higher-order interactions, the same threshold marks the onset of rapid streak amplification and strong energy growth. Together, these results demonstrate that the nature of higher-order nonlinear interactions---reinforcing versus attenuating---governs both the magnitude and qualitative character of the response near the limits of weakly nonlinear validity, and sets the stage for the secondary instabilities examined in the next section.

	\vspace*{-2ex}
\section{Validation via secondary stability analysis}
	\label{sec.sec-instab}
\vspace*{-1ex}

The classical viewpoint attributes the onset of transition in wall-bounded shear flows to secondary modal instabilities that arise once the laminar base flow is sufficiently distorted by large-amplitude streaks~\cite{RSBH1998,brandt2002transition}. In secondary-stability analysis, an eigenvalue problem is formulated to identify exponentially growing three-dimensional fluctuations about a streamwise-constant, spanwise-periodic base flow,
\beq
\label{eq.streaky-base}
U(y,z)
\;=\;
U(y)
\,+\,
u_{s,k_{z0}}(y)\cos (k_{z0} z)
\eeq
where $U(y)=1-y^2$ denotes the laminar base flow and $u_{s,k_{z0}}(y) \cos  (k_{z0} z)$ represents the streak harmonic with spanwise wavenumber $k_{z0}$ generated by nonlinear interactions. 

We now show that the breakdown of the perturbation-based frequency-response framework---identified without assuming a streak-modified base flow---coincides with the onset of secondary instability, thereby providing a mechanistic bridge between non-modal amplification and classical secondary-stability modal transition theory.

Fluctuations about the spanwise-periodic base flow~\eqref{eq.streaky-base} may be expressed using a Bloch-wave decomposition~\citep{naymoo79},
\begin{equation}
\label{eq.theta-decomp}
\mathbf{q}(x,y,z,t)
\;=\;
\mre^{\lambda t}
\mre^{\mri\alpha x}
\sum_{m \, = \, -M}^{M}
\mathbf{q}_m(y)\,
\mre^{\mathrm{i}(\gamma \, + \, m k_{z0})z}
\end{equation}
where $\lambda$ denotes the temporal growth (or decay) rate, $\alpha$ is the streamwise wavenumber, $\gamma\in[0,k_{z0})$ specifies the phase shift accumulated over one spanwise period, and $M$ determines the number of retained spanwise harmonics. Because no spanwise symmetry is imposed, both sinuous and varicose fluctuations are included in~\eqref{eq.theta-decomp}.

Using the lifting technique~\cite{moajovJFM10,jovARFM21}, the linearized operator governing~\eqref{eq.theta-decomp} can be cast as the sum of a block-diagonal operator (representing uncoupled harmonics indexed by $m$) and a block-Toeplitz operator (representing harmonic coupling induced by the streaks). Since the least stable eigenvalues in wall-bounded shear flows are typically associated with the fundamental Bloch-wave mode~\cite{RSBH1998}, we restrict attention to $\gamma=0$ and select the truncation level $M$ to ensure convergence.

\begin{figure*}[h]
  \centering
  \begin{tabular}{c@{\hspace{-0.04 cm}}c@{\hspace{-0.01 cm}}c@{\hspace{-0.04 cm}}c@{\hspace{-0.1 cm}}c}
    \subfigure[]{\label{fig.evals-maxE}}
    &&
    \subfigure[]{\label{fig.evals-sameDir}}&
    \\[-.03cm]
    \begin{tabular}{c}
        \vspace{.8cm}
        \normalsize{\rotatebox{90}{$\max \Re (\lambda)$}}
      \end{tabular}
      &
      \begin{tabular}{c}
        \includegraphics[width=0.3\textwidth]{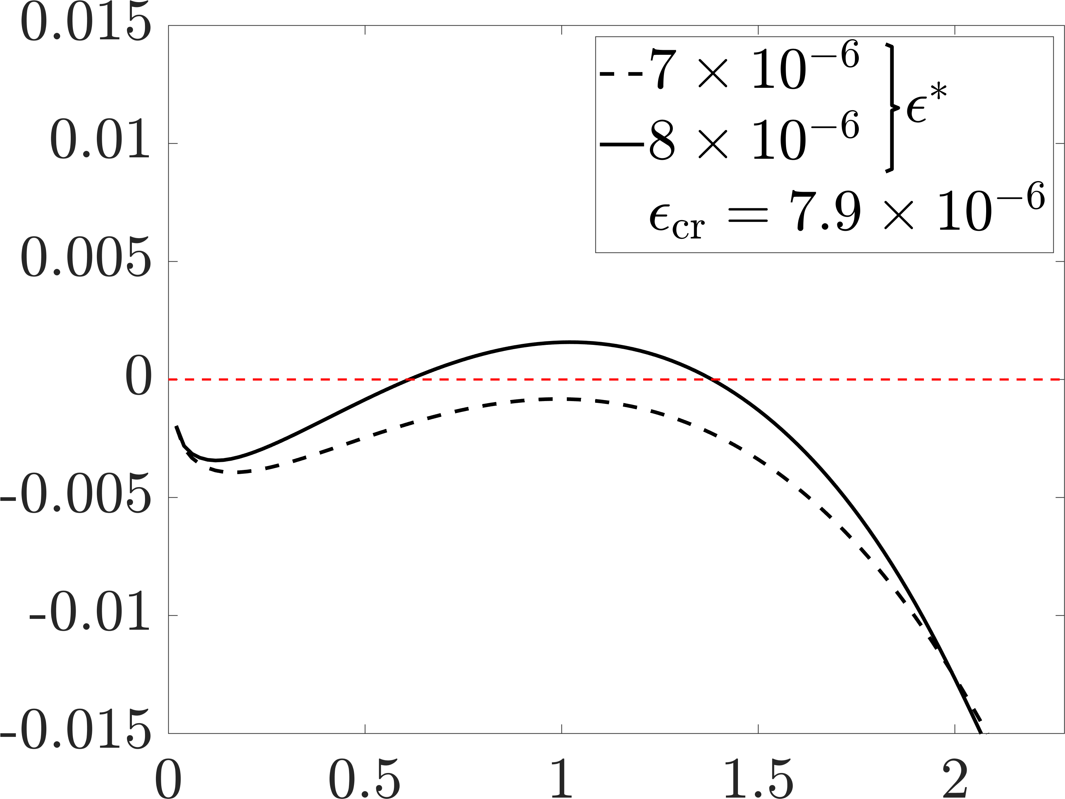}
        \\ [-0.05cm]
        \hspace{0.1cm}
        {$\alpha$}
      \end{tabular}
      &
      \hspace{0.15cm}
      &
    \begin{tabular}{c}
      \includegraphics[width=0.3\textwidth]{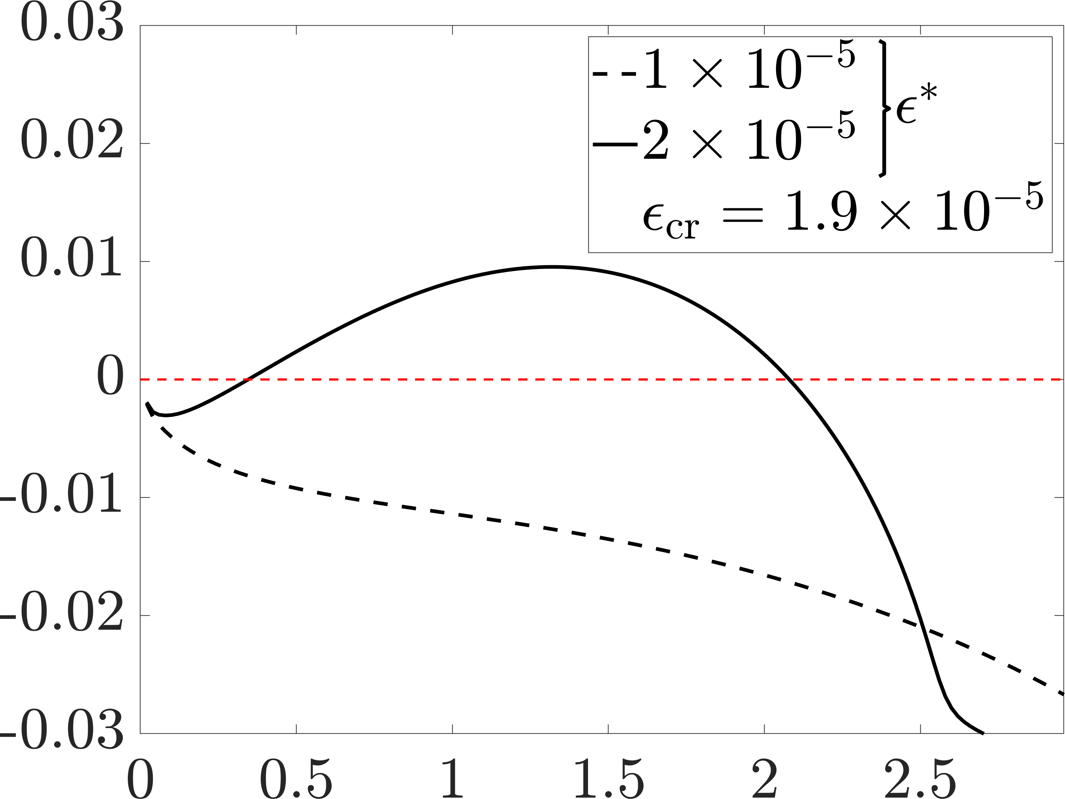}
      \\ [-0.05cm]
      \hspace{0.1cm}
      {$\alpha$}
    \end{tabular}
  \end{tabular}
  \vspace{-0.1cm}
  \caption{(a,b) Largest real part of the eigenspectrum associated with linearization about the spanwise-periodic base flow~\eqref{eq.streaky-base}, shown as a function of the fluctuation streamwise wavenumber $\alpha$, for oblique-wave forcing with (a) $(k_x, k_z, c) = (0.74, 1.14, 0.396)$ and (b) $(k_x, k_z, c) = (0.94, 1.48, 0.376)$, both at $Re = 2000$. In each case, the streak component $u_{\mathrm{s},k_{z0}}(y)$ is obtained from the sixth-order Shanks-transformed perturbation series, and eigenvalue calculations are converged for $M = 8$.}
    \label{fig.secInst-maxE}
\end{figure*}

Figure~\ref{fig.evals-maxE} shows the largest real part of the eigenvalue $\lambda$ for parameter values corresponding to oblique-wave forcing that produces the most amplified $\mathcal{O}(\epsilon^2)$ streak. Results are shown as a function of the streamwise wavenumber $\alpha$ for two forcing amplitudes: $\epsilon^*=7\times10^{-6}$ and $\epsilon^*=8\times10^{-6}$. For $\epsilon^*=7\times10^{-6}$, which lies below the critical value $\epsilon_{\mathrm{cr}}\approx7.9\times10^{-6}$ identified via the perturbation expansion in Section~\ref{sec.eps-critical}, all eigenvalues remain stable. Once the forcing amplitude exceeds this threshold, eigenvalues with positive real parts emerge, signaling the onset of secondary instability. This agreement demonstrates that the loss of convergence of the frequency-response-based perturbation series coincides with the appearance of secondary modal instability in the classical framework.

We next repeat this analysis for oblique-wave forcing configurations in which all higher-order streaks reinforce the leading-order response. In contrast to the previous case, such configurations exhibit monotonic growth of streak energy once $\epsilon_{\mathrm{cr}}$ is exceeded (Section~\ref{sec.higher-order-anal}). Despite this difference in nonlinear saturation behavior, secondary-stability analysis of the Shanks-accelerated streak profile (Fig.~\ref{fig.evals-sameDir}) yields qualitatively similar results: the eigenspectrum remains stable below $\epsilon_{\mathrm{cr}}$ and becomes unstable once this threshold is crossed.
		
\begin{figure}[h]
  \centering
  \begin{tabular}{ccl}
  \vspace{-0.2cm}
  \hspace{-0.7cm}
   \subfigure[]{\label{fig.EpsCr_vs_Eps2nd_maxE}}
   &
    &
    \subfigure[]{\label{fig.EpsCr_vs_Eps2nd_sameDir}}
    \\[.01cm]
    \begin{tabular}{c}
    \hspace{-0.2cm}
      \normalsize{\raisebox{0cm}[0pt][0pt]{\rotatebox{90}{$\epsilon_\mathrm{cr}, \, \epsilon_{2\mathrm{nd}}$}}}
    \end{tabular}
    &
    \begin{tabular}{c}
    \includegraphics[width=0.35\textwidth]{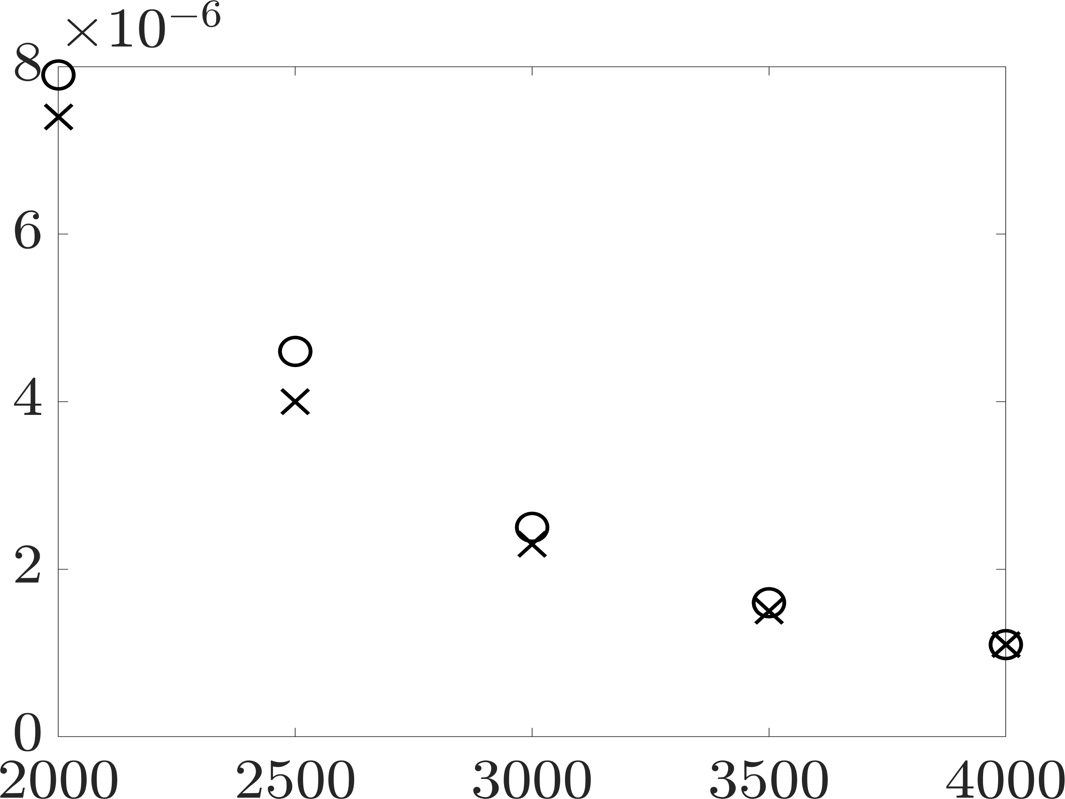}
      \\[-0.1cm] 
      {$Re$}
    \end{tabular}
    &
    \begin{tabular}{c}
    \hspace{-0.1cm}
    \includegraphics[width=0.35\textwidth]{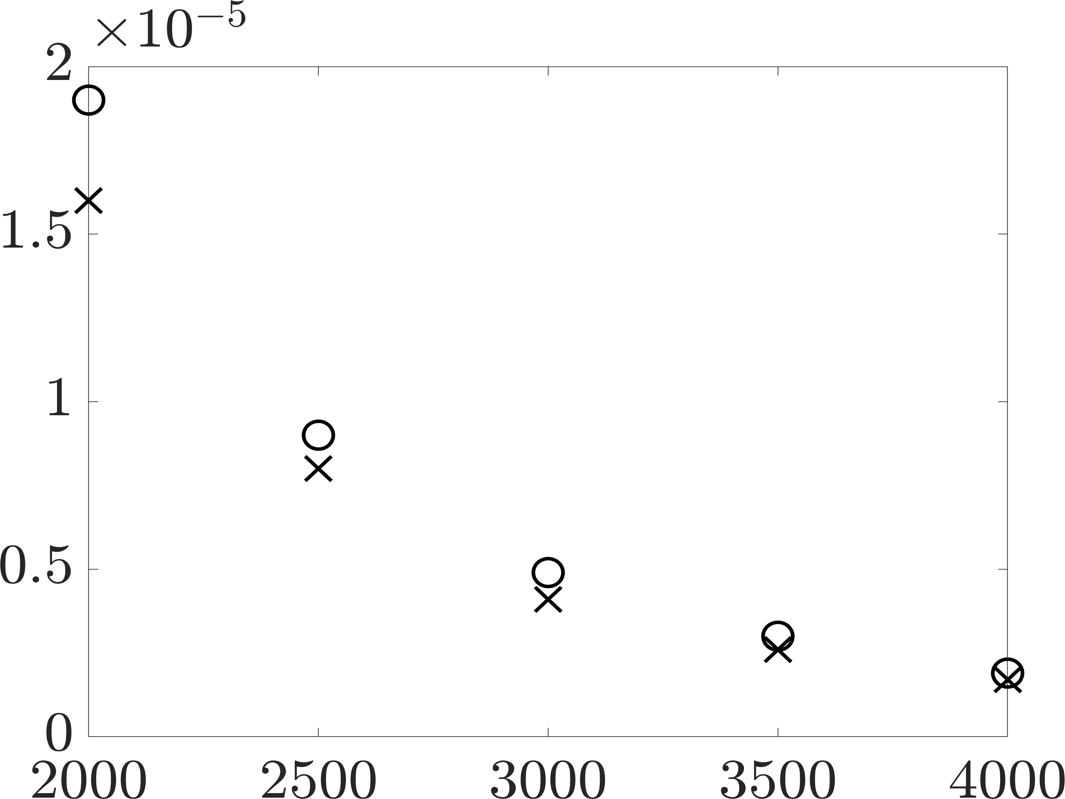}
      \\[-0.1cm]  
      \hspace{0.1cm}
      {$Re$}
    \end{tabular}  
  \end{tabular}
  \vspace{-0.051cm}  
  \caption{(a,b) Estimated critical forcing amplitude $\epsilon_{\mathrm{cr}}$ for divergence of the perturbation series (circles) and the forcing amplitude at which the first unstable mode appears (crosses), shown as functions of the Reynolds number. In panel (a), oblique-wave forcing parameters are chosen to maximize the $\mathcal{O}(\epsilon^2)$ streak energy, whereas in panel (b) parameters are selected such that higher-order streak responses remain aligned with the leading-order streak.}
  \label{fig.EpsCr_vs_Eps2nd}
\end{figure}

	Similar trends are observed across the range of Reynolds numbers considered, for both the most energetic $O(\epsilon^2)$ streak forcing configuration and the case in which higher-order streaks reinforce lower-order streaks; see Fig.~\ref{fig.EpsCr_vs_Eps2nd}. In each case, the forcing amplitude at which the eigenspectrum first becomes unstable, denoted $\epsilon_{2\mathrm{nd}}$, closely matches the critical value $\epsilon_{\mathrm{cr}}$ inferred from the breakdown of the perturbation expansion. This correspondence persists over a broad interval of subcritical Reynolds numbers, indicating that the emergence of secondary instability is tightly linked to the nonlinear amplification mechanisms captured by the frequency-response-based perturbation framework.

\vspace*{-3ex}
\subsection{Secondary stability of streaks resulting from the streamwise-constant model}
\vspace*{-1ex}

Transient-growth analysis of the linearized NS equations identifies streamwise streaks as the most energetic flow structures~\cite{butfar92}. Motivated by this observation, several studies have examined the modal stability of base flows obtained by augmenting the laminar profile with optimal streaks of prescribed amplitude, in both boundary-layer and channel flows~\cite{andersson1999optimal,andersson2001breakdown,RSBH1998}. Building on these results, we now examine the stability of streaks arising from the frequency-response analysis of the linearized NS equations (Section~\ref{sec.FrRes}) and, crucially, of the streaks generated by our perturbation-based frequency-response framework applied to the streamwise-constant nonlinear NS equations.

We focus on streaks obtained by forcing the streamwise-constant nonlinear model with wall-normal and spanwise forcing $d_\psi$. As shown in Section~\ref{sec.FrRes}, the most amplified $\cO (\epsilon)$ streak occurs at $k_{z0}=1.62$. Accordingly, we set $d_\psi$ equal to the principal input singular function $d_\psi^{(1)}$ of the resolvent operator for the linearized NS equations at $(k_x,k_{z0},\omega)=(0,1.62,0)$, and compute the streak response at successive orders using~\eqref{eq.ss-kx0-epsn}. Application of the $n$th-root test shows that the perturbation series for the first streak harmonic $u_{\mathrm{s},k_{z0}}$ begins to lose convergence at $\epsilon_\mathrm{cr}\approx1.9\times10^{-5}$.

\begin{figure}[h]
  \centering
  \begin{tabular}{cccccc}
  \vspace{-0.1cm}
  \hspace{-.5cm}
   \subfigure[]{\label{fig.lin-sec-instab-naive}}
    & &
    \hspace{-0.25cm}
    \subfigure[]{\label{fig.lin-sec-instab-DNS}}
    \vspace{-0.1cm}
    &
    &
    \hspace{-0.5cm}
    \subfigure[]{\label{fig.lin-sec-instab-Ns7}}
    & 
    \\[-.cm]
    \begin{tabular}{c}
    \hspace{-0.2cm}
      \vspace{0.6cm}
      \normalsize{\rotatebox{90}{$\max \Re (\lambda)$}}
    \end{tabular}
    &
    \begin{tabular}{c}
    \hspace{-0.21cm}
       \includegraphics[width=0.3\textwidth]{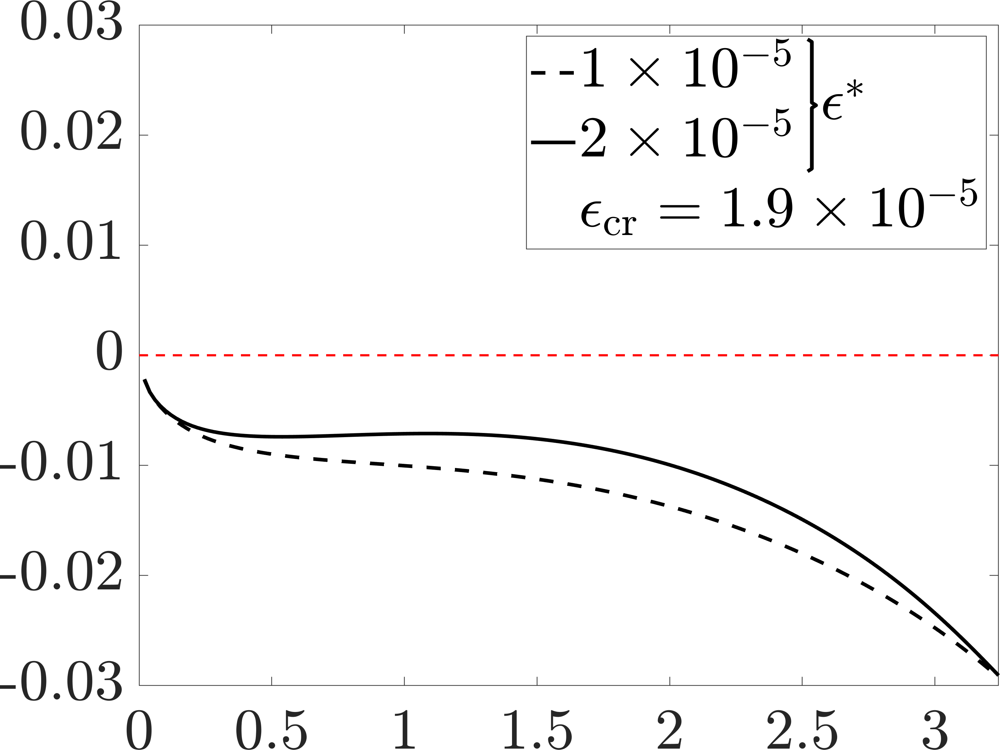}
      \\[-0.1cm] 
      {$\alpha$}
    \end{tabular}
    &
    \begin{tabular}{c}
        \vspace{.55cm}
         \hspace{-0.25cm}
        \normalsize{\rotatebox{90}{$\| \bq_0 \|$}}
      \end{tabular}
      &
      \begin{tabular}{c}
      \hspace{-0.4cm}
        \includegraphics[width=0.3\textwidth]{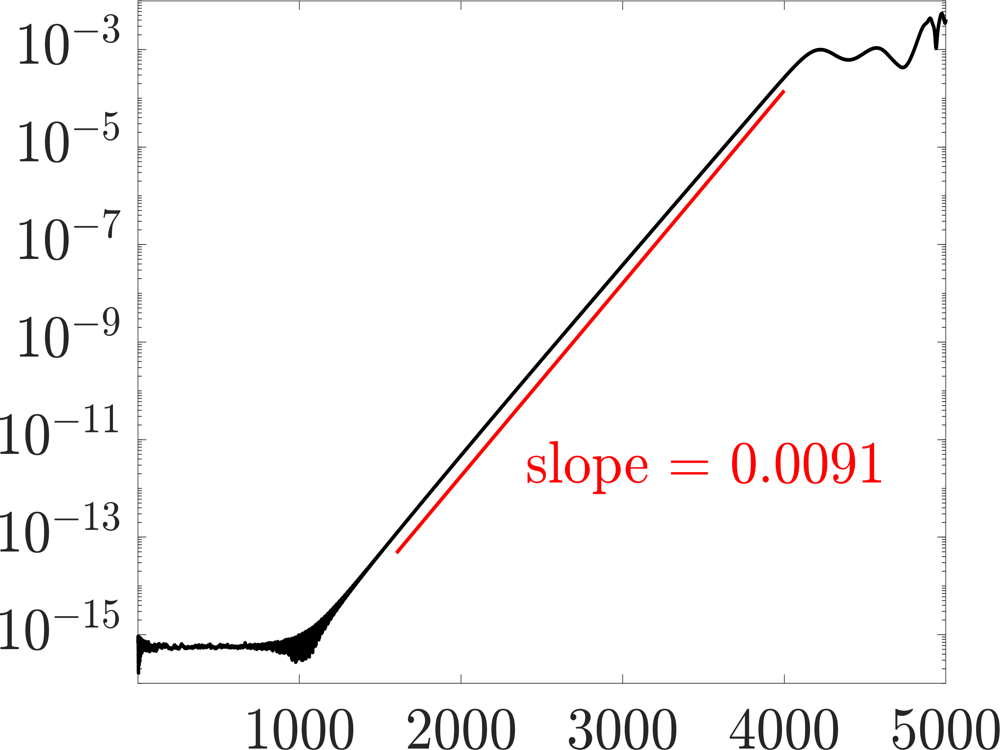}
        \\[-0.1cm] 
        {$t$}
      \end{tabular}
      &
    \begin{tabular}{c}
      \vspace{.7cm}
      \hspace{-0.35cm}
      \normalsize{\rotatebox{90}{$\max \Re (\lambda)$}}
    \end{tabular}
    &
    \begin{tabular}{c}
    \hspace{-0.35cm}
      \includegraphics[width=0.3\textwidth]{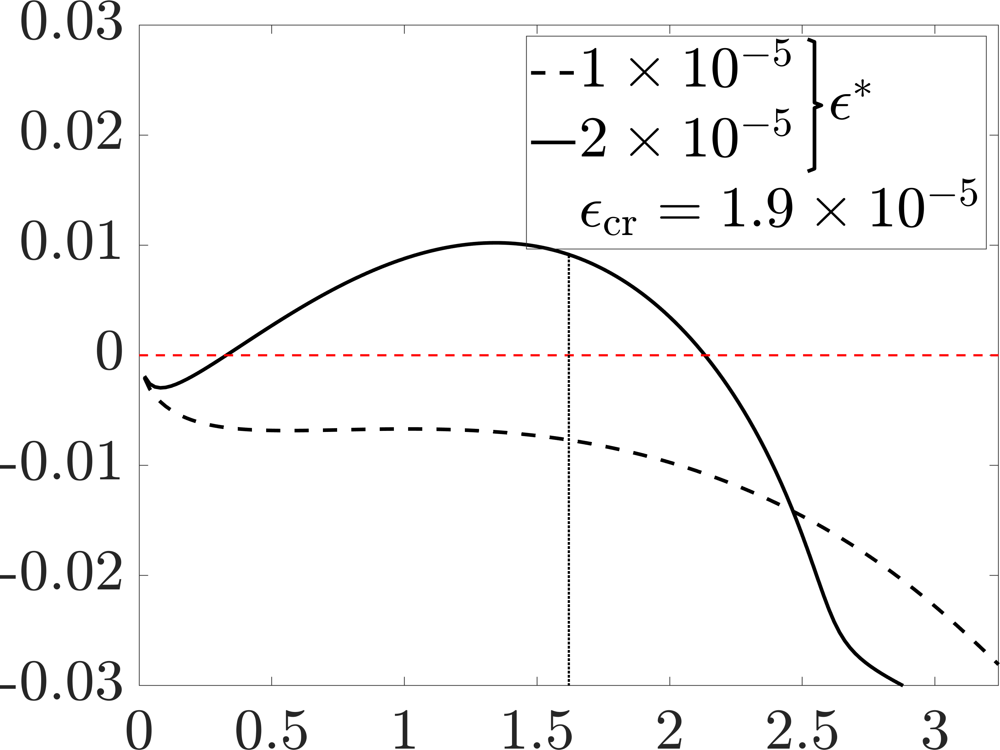}
      \\[-0.1cm]  
      \hspace{0.1cm}
      {$\alpha$}
    \end{tabular}
  \end{tabular}
  \vspace{-0.25cm}  
  \caption{(a) Largest real part of the eigenspectrum obtained by linearizing about the spanwise-periodic base flow~\eqref{eq.streaky-base}, shown as a function of the fluctuation streamwise wavenumber $\alpha$, for linear optimal forcing with $k_{z0}=1.62$ at $Re=2000$. Eigenvalue computations converge with $M=10$ Bloch-wave modes.
(b) Temporal evolution of the square root of the fluctuation energy $\| \bq_0 \|_2$ from DNS for $\alpha=1.62$. The slope corresponding to the growth rate predicted in panel (c) is overlaid for comparison.
(c) Same as in panel (a), but for linearization about the spanwise-periodic base flow~\eqref{eq.app-streaky-base} that incorporates higher streak harmonics. Convergence is achieved with $M=15$ Bloch-wave modes and $N_{\mathrm{s}}=7$ retained streak harmonics.}
   \label{fig.sec-instab-Oeps}
\end{figure}

Eigenvalue analysis of the linearization about the laminar base flow augmented by this single streak harmonic~\eqref{eq.streaky-base}, with fluctuations represented by~\eqref{eq.theta-decomp}, fails to identify secondary instability even for forcing amplitudes $\epsilon^*>\epsilon_\mathrm{cr}$; see Fig.~\ref{fig.lin-sec-instab-naive}. In contrast, DNS initialized with small perturbations about the same streaky flow and streamwise wavenumber $\alpha=k_{z0}$ exhibits clear exponential growth; see Fig.~\ref{fig.lin-sec-instab-DNS}. This discrepancy arises because a single streak harmonic is insufficient to represent the full nonlinear response in this configuration: unlike the oblique-wave-driven case, the $k_{z0}$ component does not overwhelmingly dominate the streak response. As a result, secondary stability analysis based solely on the optimal linear streak provides an incomplete---and potentially misleading---picture of instability onset.

To address this limitation, we perform secondary stability analysis about a spanwise-periodic base flow that incorporates higher streak harmonics,
\begin{equation}
\label{eq.app-streaky-base}
U(y,z)
\;=\;
U(y)
\;+\;
\sum_{r \, = \, 1}^{N_\mathrm{s}} u_{s,rk_{z0}}(y) \cos (rk_{z0}z)
\end{equation}
where $U(y)$ denotes the modified mean flow and $u_{s,rk_{z0}}(y) \cos (rk_{z0}z)$ represents the $r$th streak harmonic generated by nonlinear interactions in the streamwise-constant NS equations. Converged eigenspectra are obtained with $N_\mathrm{s}=7$ retained streak harmonics and $M=15$ Bloch-wave modes in~\eqref{eq.theta-decomp}.

When these higher harmonics are included, the secondary stability analysis reveals modal instability for $\epsilon^*>\epsilon_\mathrm{cr}$; see Fig.~\ref{fig.lin-sec-instab-Ns7}. Notably, although the critical amplitude $\epsilon_\mathrm{cr}$ is estimated using only the base streak harmonic, it nonetheless accurately predicts the onset of instability once the full streak structure is accounted for.

This analysis underscores a key distinction between streaks identified by linear theory and those produced by nonlinear interactions. The failure of the single-harmonic base flow reflects the fact that linearly optimal streaks do not remain the dominant structures once nonlinear effects become significant. By contrast, oblique-wave interactions generate streaks whose harmonics are dynamically coupled and collectively shape the base flow seen by secondary perturbations. The laminar profile augmented by this nonlinear streak response therefore provides the appropriate reference state for capturing instability onset. In this sense, our framework not only retains the essential features of linear amplification mechanisms but also systematically incorporates the leading nonlinear interactions that determine which streaks are physically relevant.

Overall, these results show that classical secondary stability analysis and our perturbation-based frequency-response framework describe successive stages of the same amplification process. Secondary stability analysis detects modal growth only after sizable streak distortions have formed, whereas our expansion identifies the breakdown of the weakly nonlinear regime---and the associated amplification pathways---directly from the governing equations. The close correspondence between divergence of the perturbation series and the emergence of unstable eigenmodes demonstrates that the modal instabilities uncovered by classical analysis originate from the same nonlinear interactions that our framework exposes at finite but small amplitudes. Rather than offering an alternative route to transition, the present approach therefore reveals the physical mechanisms that precede and ultimately give rise to secondary instability, yielding a unified and mechanistically transparent picture of transition onset.

    	\vspace*{-2ex}
\section{Concluding remarks}
	\label{eq.conclusion}
	\vspace*{-2ex} 

We have presented a perturbation-based frequency-response framework that systematically expands the input--output dynamics of the Navier--Stokes equations in the amplitude of external forcing. By embedding linear resolvent analysis within a weakly nonlinear perturbation expansion, this formulation provides a principled description of how non-modal amplification mechanisms evolve into finite-amplitude flow structures, establishing a direct and quantitative connection between linear resolvent analysis and higher-order nonlinear interactions. Fluctuation dynamics at each perturbation order are governed by the resolvent of the laminar base flow, with successive nonlinear interactions among lower-order responses driving the evolution at higher orders. This coupled hierarchy of linear dynamical systems enables identification of dominant flow structures and quantification of kinetic energy redistribution, providing a predictive description of subcritical amplification.

At second order, we showed that quadratic interactions of unsteady oblique-wave fluctuations generate steady streamwise streaks via the lift-up mechanism. The resulting vortical forcing selectively excites the second output singular function of the streamwise-constant resolvent operator, yielding streaks whose wall-normal profile and dominant spanwise length scale differ markedly from predictions based on linear resolvent analysis alone. The regions of strongest amplification in spectral parameter space and the associated wall-normal structure closely match those obtained from gradient-based optimization of the fully nonlinear equations. The accuracy of the second-order weakly nonlinear expansion is further confirmed through direct numerical simulations at small forcing amplitudes.

At higher forcing amplitudes, the dynamics are governed by higher-order interactions between oblique waves and the induced streaks. These interactions generate a streak perturbation series whose wall-normal profile is again captured by the second output singular function and whose relative phase governs reinforcement or attenuation of the leading second-order response. Once the forcing amplitude exceeds a critical value, the perturbation series loses convergence and DNS exhibits a transition to sustained unsteadiness. In parameter regimes where higher-order contributions act constructively, this breakdown is accompanied by pronounced streak amplification and rapid energy growth.

Validation via secondary-stability analysis demonstrates that the loss of convergence of the perturbation expansion coincides with the onset of modal instability of the streak-distorted base flow. Although the expansion is carried out about the laminar base flow, its breakdown accurately predicts the streak amplitudes at which secondary instability emerges once these streaks are superposed onto the parabolic base profile. In this way, the framework provides a mechanistic bridge between resolvent-driven non-modal amplification and the streak-amplitude assumptions traditionally imposed in classical secondary-stability theory, yielding a unified pathway from linear amplification to modal instability. This perspective suggests that secondary instability represents a finite-amplitude manifestation of amplification mechanisms already encoded in the weakly nonlinear frequency-response expansion.

Beyond canonical channel flow, the framework naturally extends to settings with geometric roughness, {time- or spatially-periodic base flows,} compressibility effects, and feedback-control inputs, providing a systematic route for analyzing and influencing subcritical amplification in more realistic configurations. Our recent study~\cite{dwisidjovJFM22} demonstrates the predictive power of a weakly nonlinear frequency-response approach in identifying subcritical transition mechanisms in a separated high-speed boundary-layer flow subject to unsteady oblique disturbances. Even in this markedly different configuration, the spatial structure of streaks is captured by the second output singular function of the steady resolvent operator, evaluated at half the spanwise wavelength of the oblique waves, with predictions in excellent agreement with DNS.

Taken together, these results establish a self-consistent, physics-based framework for analyzing subcritical amplification mechanisms that incorporates nonlinear interactions within a frequency-response paradigm. Despite its reduced complexity, the approach delivers quantitative predictions of transition-relevant dynamics arising from the interplay of non-modal amplification and nonlinear streak--oblique-wave interactions. It provides a computationally efficient alternative to proper orthogonal and dynamic mode decompositions for reduced-order modeling of wall-bounded transitional and turbulent flows~\cite{smimoehol05,row05,sch10,rowmezbagschhen09,jovschnicPOF14} that is tractable for analysis, optimization, and control.

	\vspace*{-3ex}
\section*{Acknowledgements}

	\vspace*{-1ex}	
We thank Bassam Bamieh and Mitul Luhar for insightful discussions and valuable feedback that significantly improved the quality of this paper.

	\appendix
    	\vspace*{-4ex}
\section{Operators in the linearized evolution model}\label{app.Operators}
    	\vspace*{-1ex}

Because the operators $\cA$, $\cB$, and $\cC$ in the linearized NS equations~\eqref{eq.ss-On} are translationally invariant in the streamwise ($x$) and spanwise ($z$) directions, it is convenient to express the dynamics in Fourier space with respect to these coordinates. Upon taking the Fourier transform in $x$ and $z$, the linearized evolution model can be written in terms of the wall-normal coordinate $y$ and the wavenumber vector $\bk = (k_x,k_z)$ as
\begin{align}
\label{eq.NSveta}
\begin{aligned}
\mathcal{A}_\bk 
\,&=\, 
\left[
\begin{array}{cc}
\frac{1}{Re}{\Delta}^{-1} {\Delta}^2 - \mathrm{i}k_x{\Delta}^{-1} U {\Delta} + \mathrm{i}{\Delta}^{-1} U^{\prime \prime} & 0 \\
-\mathrm{i}k_z U^{\prime} & \frac{1}{Re}{\Delta} -\mathrm{i}k_x U 
\end{array}
\right]
\\
\mathcal{B}_\bk 
\,&=\,
\left[
\begin{array}{ccc}
-\mathrm{i}k_x{\Delta}^{-1}\partial_y & -(k_x^2 + k_z^2){\Delta}^{-1} & -\mathrm{i}k_x{\Delta}^{-1}\partial_y\\
\mathrm{i}k_z & 0 & -\mathrm{i}k_x
\end{array}
\right]
\\
\mathcal{C}_\bk 
\,&=\,
\frac{1}{k_x^2+k_z^2}
\left[
\begin{array}{cc}
-\mathrm{i}k_x\partial_y  & -\mathrm{i}k_z   \\
k_x^2+k_z^2 & 0 \\
\mathrm{i}k_z\partial_{y} & \mathrm{i}k_x
\end{array}
\right].
\end{aligned}
\end{align}
Here, $U(y)$ denotes the laminar base flow, with $U' = \mathrm{d}U/\mathrm{d}y$ and $U'' = \mathrm{d}^2U/\mathrm{d}y^2$. The operator
	$
	\Delta = \partial_{yy} - k_x^2 - k_z^2
	$
is the wall-normal Laplacian subject to homogeneous Dirichlet boundary conditions, while
	$
\Delta^2 = \partial_{yyyy} - 2 (k_x^2 + k_z^2)\partial_{yy} + (k_x^2 + k_z^2)^2
	$
denotes the associated biharmonic operator with homogeneous Dirichlet and Neumann boundary conditions. These operators arise naturally in resolvent and Orr--Sommerfeld/Squire formulations governing the wall-normal velocity and vorticity dynamics; see, for example,~\cite{jovbamJFM05,schmid2007nonmodal}.

    	\vspace*{-2ex}
\section{Evaluation of higher order terms in the perturbation expansion}
\label{app.terms-eval}
    	\vspace*{-1ex}

In Section~\ref{sec.response-single-harm}, we showed how single-harmonic response terms can be evaluated efficiently in the perturbation expansion up to $\mathcal{O}(\epsilon^2)$. Here, we demonstrate that higher-order contributions can be computed in an analogous manner by exploiting the structure of the frequency-response operators associated with the linearized NS equations. No additional assumptions beyond those introduced in Section~\ref{sec.response-single-harm} are made in this appendix; the results follow directly from repeated application of the same frequency-response operators to the nonlinear forcing terms generated at successive orders.

At third order, the response consists of a finite set of harmonics generated by quadratic interactions among the $\mathcal{O}(\epsilon)$ and $\mathcal{O}(\epsilon^2)$ components. These responses can be written compactly as
	\beq\label{eq.4th-O-response}
	\ba{rcl}
	\bu_{(\pm\bkx,\bkz)}^{(3)} (y,c)
	& = &
	[
	\cG_{(\pm\bkx,\bkz)} ( c)
	\bd_{(\pm\bkx,\bkz)}^{(3)} ( \, \cdot \, , \bc )
	] (y)
	\\[0.15cm]
        \bu_{(\pm\bkx,3\bkz)}^{(3)} (y,c)
	& = &
	[
	\cG_{(\pm\bkx,3\bkz)} (c)
	\bd_{(\pm\bkx,3\bkz)}^{(3)} ( \, \cdot \, , \bc )
	] (y)
        \\[0.15cm]
	\bu_{(\pm3\bkx,\bkz)}^{(3)} (y,c)
	& = &
	[
	\cG_{(\pm3\bkx,\bkz)} (c)
	\bd_{(\pm3\bkx,\bkz)}^{(3)} ( \, \cdot \, , \bc )
	] (y)
        \\[0.15cm]
        \bu_{(\pm3\bkx,3\bkz)}^{(3)} (y,c)
	& = &
	[
	\cG_{(\pm3\bkx,3\bkz)} (c)
	\bd_{(\pm3\bkx,3\bkz)}^{(3)} ( \, \cdot \, ,  \bc )
	] (y).
	\ea
	\non
	\eeq 
At fourth order, nonlinear interactions among $\bu^{(1)}$, $\bu^{(2)}$, and $\bu^{(3)}$ generate additional harmonics, including streamwise-constant, mean-flow, and higher-wavenumber components. These responses are given by
    \beq
	\ba{rcl}
	\bu_{(0,0)}^{(4)} (y,c)
	& = &
	[
	\cG_{(0,0)} (c)
	\bd_{(0,0)}^{(4)} ( \, \cdot \, , \bc)
	] (y)
	\\[0.15cm]
	\bu_{(0,2\bkz)}^{(4)} (y,c)
	& = &
	[
	\cG_{(0,2\bkz)} (c)
	\bd_{(0,2\bkz)}^{(4)} ( \, \cdot \, , \bc)
	] (y)
	\\[0.15cm]
	\bu_{(2 \bkx ,0)}^{(4)} (y,c)
	& = &
	[
	\cG_{(2 \bkx,0)} (c)
	\bd_{(2 \bkx,0)}^{(4)} ( \, \cdot \, ,  \bc )
	] (y)
	\\[0.15cm]
        \bu_{(0,4\bkz)}^{(4)} (y,c)
	& = &
	[
	\cG_{(0,4\bkz)} (c)
	\bd_{(0,4\bkz)}^{(4)} ( \, \cdot \, , \bc )
	] (y)
	\\[0.15cm]
	\bu_{(4 \bkx ,0)}^{(4)} (y,c)
	& = &
	[
	\cG_{(4 \bkx,0)} (c)
	\bd_{(4 \bkx,0)}^{(4)} ( \, \cdot \, ,  \bc )
	] (y)
	\\[0.15cm]
	\bu_{(\pm2\bkx,2\bkz)}^{(4)} (y,c)
	& = &
	[
	\cG_{(\pm2\bkx,2\bkz)} (c)
	\bd_{(\pm2\bkx,2\bkz)}^{(4)} ( \, \cdot \,, \bc )
	] (y)
    \\[0.15cm]
    \bu_{(\pm2\bkx,4\bkz)}^{(4)} (y,c)
	& = &
	[
	\cG_{(\pm2\bkx,4\bkz)} (c)
	\bd_{(\pm2\bkx,4\bkz)}^{(4)} ( \, \cdot \,, \bc )
	] (y)
    \\[0.15cm]
    \bu_{(\pm4\bkx,2\bkz)}^{(4)} (y,c)
	& = &
	[
	\cG_{(\pm4\bkx,2\bkz)} (c)
	\bd_{(\pm4\bkx,2\bkz)}^{(4)} ( \, \cdot \, ,  \bc )
	] (y)
    \\[0.15cm]
    \bu_{(\pm4\bkx,4\bkz)}^{(4)} (y,c)
	& = &
	[
	\cG_{(\pm4\bkx,4\bkz)} (c)
	\bd_{(\pm4\bkx,4\bkz)}^{(4)} ( \, \cdot \, , \bc )
	] (y).
	\ea
	\eeq
Here, the forcing terms $\bd^{(3)}$ arise from quadratic interactions between $\bu^{(1)}$ and $\bu^{(2)}$, while $\bd^{(4)}$ collects contributions from interactions among $\bu^{(1)}$, $\bu^{(2)}$, and $\bu^{(3)}$; see Fig.~\ref{fig.bd-eps4} for a schematic illustration. Retaining only interaction terms associated with the base frequencies yields a computationally efficient yet accurate representation of higher-order responses of the NS equations to small-amplitude, single-harmonic, three-dimensional forcing.
	
	\vspace*{-3ex}
\section{Formulation of the optimization problem in Section~\ref{sec:princWNL}}
	\label{app.optim}
	
	    	\vspace*{-1ex}
Within the weakly nonlinear framework, the problem of maximizing the energy of the $\cO(\epsilon^2)$ streamwise streaks generated by a unit-energy $\cO(\epsilon)$ oblique-wave disturbance can be posed as a constrained, nonconvex optimization problem. Specifically, we seek the $\cO(\epsilon)$ forcing that maximizes the kinetic energy of the induced second-order streak response, subject to the linear resolvent dynamics governing the oblique-wave response and the quadratic nonlinear interactions that generate the streak forcing,
\begin{equation}
\begin{array}{rl}
\underset{\bd_{\bk}^{(1)}}{\text{maximize}}\quad & \| u_{\mathrm{s},2k_z}^{(2)}(y,c) \|_2^2 \\[1mm]
\text{subject to}\quad 
& 
	u_{\mathrm{s},2k_z}^{(2)}(y,c) 
	\; = \;
	\left[ 
	\cG_{2k_z}\, \bd_{2k_z}^{(2)}(\cdot,c)
	\right] (y)
	\\[0.25cm]
	& 
	\bd_{2k_z}^{(2)}(y,c) 
	\; = \; 
	\left[
	\begin{array}{c} 
	\Delta_{2 k_z}^{-1}
	\!
	\left(
	(\partial_{yy} \, + \, (2 k_z)^2) 
	\Im 
	\big\{
	{v}_{\bk}^{(1)} (y,c)
	{\bar{w}}_{\bk}^{(1)} (y,c)
	\big\}
	\, + \,
	2 k_z \partial_y
	\big(
	| {v}_{\bk}^{(1)} (y,c) |^2 
	\, + \,
	| {w}_{\bk}^{(1)} (y,c) |^2
	\big)
	\right) \\ 
        -\partial_y 
	\Re
	\big\{
	{u}_{\bk}^{(1)} (y,c)
	{\bar{v}}_{\bk}^{(1)} (y,c)
	\big\}
	\, - \,
	2 k_z \Im 
	\big\{
	{u}_{\bk}^{(1)} (y,c)
	{\bar{w}}_{\bk}^{(1)} (y,c)
	\big\}
	\end{array}
	\right]
	\\[0.5cm]
	& 
	\bu_\bk^{(1)}(y,c)  
	\; = \;
	\left[ \cG_\bk(c)\, \bd_{\bk}^{(1)}(\cdot,c) \right] (y) 
	\\[0.25cm]
	& \| \bd_{\bk}^{(1)}(y,c) \|^2 
	\; = \; 
	1.
	\end{array}
	\end{equation}
Here, $\bk = (k_x,k_z)$ denotes the oblique-wave wavenumber vector, $\bd_{\bk}^{(1)}(y,c)$ is the imposed $\cO(\epsilon)$ forcing, and $\bu_\bk^{(1)}(y,c) = [ \, u_\bk^{(1)}\; v_\bk^{(1)}\; w_\bk^{(1)} \, ]^T$ is the corresponding resolvent response. The term $\bd_{2k_z}^{(2)}$ represents the quadratic nonlinear modulation of the $\cO(\epsilon)$ response that generates the $\cO(\epsilon^2)$ streak component $u_{\mathrm{s},2k_z}^{(2)}$. The operators $\cG_\bk(c)=\cH_\bk(-k_x c)$ and $\cG_{2k_z}=\cH_{(0,2k_z)}(0)$ are defined in Section~\ref{sec.freq-NS}.

This formulation makes explicit that the $\cO(\epsilon^2)$ streak response depends quadratically on the $\cO(\epsilon)$ forcing through nonlinear wave--wave interactions between resolvent-amplified oblique-wave responses. As a result, the streak energy $\|u_{\mathrm{s},2k_z}^{(2)}\|_2^2$ depends quartically on the forcing field, leading to a nonconvex dependence on $\bd_{\bk}^{(1)}$. Rather than solving this nonconvex optimization directly, the present work leverages the structure of the perturbation expansion to identify dominant amplification mechanisms without iterative nonlinear optimization.

    	\vspace*{-4ex}
\section{Comparison with optimization procedure based on harmonic-balance approach}\label{app.hbns}
    	\vspace*{-1ex}
 
Here, we compare the streamwise streak response obtained by forcing the NS equations with the principal oblique-wave disturbance identified by our framework to the streaks resulting from a harmonic-balance-based optimization procedure that maximizes the drag coefficient~\cite{rigsipcol21}. The harmonic balance method represents each state and forcing variable as a truncated Fourier expansion in the wall-parallel directions and in time,
\beq
    \bu(x,y,z,t) 
    \; = \; 
    \sum_{m\,=\, -M}^M\sum_{n \,=\,  -N}^N\sum_{h \,=\,-H}^H 
    \hat{\bu}(y)\mre^{\mri(m\bkx x + n\bkz z + h\omega t)}. 
\eeq
Substituting this expansion into the NS equations and collecting terms with identical spatiotemporal frequencies yields a system of $(2M+1)(2N+1)(2H+1)$ coupled nonlinear equations. A locally optimal solution is then obtained via a Newton iteration, while treating the oblique-wave forcing coefficients as optimization variables. Gradient-based ascent is employed to maximize a prescribed objective functional, here taken to be the drag coefficient.

\begin{figure}[h]
  \centering
  \begin{tabular}{cc}
    \begin{tabular}{c}
    \vspace{0.9cm}
        \normalsize{\rotatebox{90}{$y$}}
      \end{tabular}
      &
      \begin{tabular}{c}
    \includegraphics[width=0.3\textwidth]{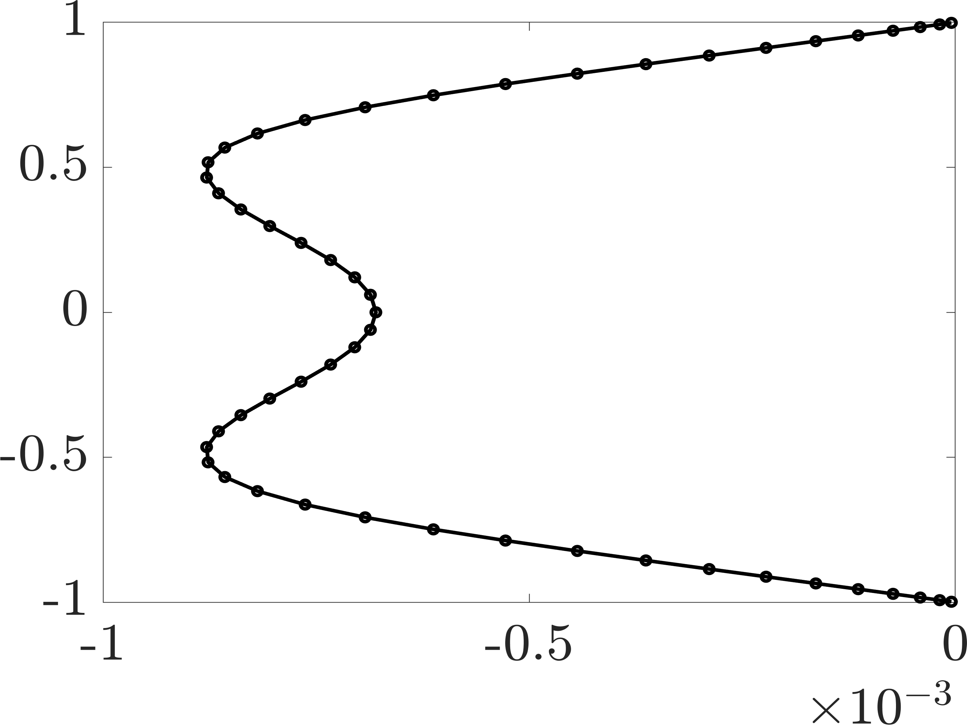}
        \\[-0.21cm] {$u_{s,2k_z}(y)$}
      \end{tabular}
  \end{tabular}
  \caption{Comparison of the steady streamwise streak response obtained using the harmonic balance method (solid lines) and the perturbation-based frequency-response analysis driven by the principal oblique-wave input mode (circles). The forcing amplitude is $\epsilon^* = 10^{-6}$, and the imposed forcing parameters are $(k_x,k_z,\omega) = (0.74, 1.14,-k_x c) = (0.74,1.14,-0.293)$.}
      \label{fig.hbns-comp}
        \vspace{-0.25cm}  
\end{figure}

Figure~\ref{fig.hbns-comp} compares the steady streamwise streak response at $(0,2\bkz,0)$ generated by oblique-wave forcing at $(\pm\bkx,\pm\bkz,\pm\omega)$ and $(\mp\bkx,\pm\bkz,\mp\omega)$ with amplitude $10^{-6}$. The base frequencies are chosen using the optimal wavenumbers identified in Section~\ref{sec.freq-NS}, and the harmonic balance truncation is set to $M=N=H=2$. 

Despite the fact that the harmonic balance formulation retains a substantially larger set of nonlinear interactions and optimizes a different objective functional, the resulting streak responses are nearly indistinguishable. This close agreement demonstrates that the dominant streak-producing mechanisms identified by the perturbation-based frequency-response framework capture the same physics as the fully nonlinear, optimization-based harmonic balance approach, but at a significantly reduced computational cost and with enhanced physical interpretability.

	\vspace*{-2ex}
\section{The Reynolds number dependence of the $\cO(\epsilon^2)$ response}
\label{app.stream-const}
	\vspace*{-1ex}

For the $\cO(\epsilon)$ oblique-wave inputs listed in Table~\ref{tab.ow-ss}, the corresponding first-order velocity field can be written as
\beq
\left[
				\ba{c}
				u^{(1)} (\bar{x},y,z)
				\\[0.1cm]
				v^{(1)} (\bar{x},y,z)
				\\[0.1cm]
				w^{(1)} (\bar{x},y,z)
				\ea
				\right]
				= 
				\left[
				\ba{c}
				\phantom{-} 2 \, \Re 
				\,
				\big\{
				{u}_{(k_x,k_z)}^{(1)} (y,c) \, \mre^{\mri \bkx \bar{x}} 
				\big\}
				\cos k_z z
				\\[0.15cm]
				\phantom{-} 2 \, \Re 
				\,
				\big\{
				{v}_{(k_x,k_z)}^{(1)} (y,c) \, \mre^{\mri \bkx \bar{x}} 
				\big\}
				\cos k_z z
				\\[0.15cm]
				- 2 \, \Im
				\,
				\big\{
				{w}_{(k_x,k_z)}^{(1)} (y,c) \, \mre^{\mri \bkx \bar{x}} 
				\big\}
				\sin k_z z
				\ea
				\right].
\eeq
Substituting these $\cO(\epsilon)$ fields into the quadratic nonlinear terms of the NS equations yields the forcing of the $\cO(\epsilon^2)$ streamwise-constant model~\eqref{eq.ss-kx0-eps2}, which takes the form
	\beq
	\ba{rcl}
	d_{\psi}^{(2)} (y,z)
	& = &
	{d}_{\psi,2 k_z}^{(2)} (y) 
	\sin 2 k_z z
	\\[0.2cm]
	d_{u}^{(2)} (y,z)
	& = &
	{d}_{u, 2 k_z}^{(2)} (y) 
	\cos 2 k_z z
	\ea
	 \label{eq.dpsi-du-eps2}
	\eeq
where
	\beq
	\ba{rcl}
	{d}_{\psi, 2 k_z}^{(2)} (y) 
	& = &
	\Delta_{2 k_z}^{-1}
	\!
	\left(
	(\partial_{yy} \, + \, (2 k_z)^2) 
	\Im 
	\big\{
	{v}_{(k_x,k_z)}^{(1)} (y,c)
	{\bar{w}}_{(k_x,k_z)}^{(1)} (y,c)
	\big\}
	\, + \,
	2 k_z \partial_y
	\big(
	| {v}_{(k_x,k_z)}^{(1)} (y,c) |^2 
	\, + \,
	| {w}_{(k_x,k_z)}^{(1)} (y,c) |^2
	\big)
	\right)
	\\[0.25cm]
	{d}_{u, 2 k_z}^{(2)} (y) 
	& = &
	- \partial_y 
	\Re
	\big\{
	{u}_{(k_x,k_z)}^{(1)} (y,c)
	{\bar{v}}_{(k_x,k_z)}^{(1)} (y,c)
	\big\}
	\, - \,
	2 k_z \Im 
	\big\{
	{u}_{(k_x,k_z)}^{(1)} (y,c)
	{\bar{w}}_{(k_x,k_z)}^{(1)} (y,c)
	\big\}.
	\ea
	\label{eq.dpsi2-du2-2kz}
	\eeq
Here, $\bar{(\cdot)}$ denotes complex conjugation and $|\cdot|$ denotes the modulus, e.g., $|v|^2 = \bar{v}v$. The steady $\cO(\epsilon^2)$ streamwise-constant response is obtained by setting $\partial_t(\cdot)=0$ in~\eqref{eq.ss-kx0-eps2}, yielding
\beq
\ba{rcl}
	\psi^{(2)}
	& = &
	- Re \, \cL^{-1} \, d_{\psi}^{(2)}
	\\[0.2cm]
	u^{(2)}
	& = &
    Re^2 \, \cS^{-1} \, \cC_p \, \cL^{-1}              d_{\psi}^{(2)}
	\; - \;
	Re \, \cS^{-1} \,d_u^{(2)}.
	\ea
    \label{eq.ss-kx0-t0-eps2}
\eeq
Thus, for the input given by~\eqref{eq.dpsi-du-eps2} the $\cO (\epsilon^2)$ steady streamwise constant fluctuations in~\eqref{eq.ss-kx0-t0-eps2} are determined by 
	\beq
	\ba{rcl}
	\psi_{\mathrm{s}}^{(2)} (y,z)
	& = &
	- Re \, \big[ \cL_{2 k_z}^{-1} {d}_{\psi, 2 k_z}^{(2)} (\cdot) \big] (y)
	\sin 2 k_z z
	\\[0.2cm]
	u_{\mathrm{s}}^{(2)} (y,z)
	& = &
	\left(
    Re^2
	 \big[  
	 \cS_{2 k_z}^{-1} \cC_{p, 2k_z} \cL_{2 k_z}^{-1} {d}_{\psi, 2 k_z}^{(2)} (\cdot) 
	 \big] (y)
	\; - \;
	Re \, \big[ \cS_{2 k_z}^{-1} {d}_{u, 2 k_z}^{(2)} (\cdot) \big] (y)
    \right) 
	\cos 2 k_z z.
	\ea
	\label{eq.psi-u-eps2}
	\eeq
This decomposition makes explicit the distinct Reynolds-number scalings of the $\cO(\epsilon^2)$ stream function and streak velocity, and underpins the amplification mechanisms summarized in Theorem~\ref{thm1}.

    	\vspace*{-2ex}
\section{Parity operator and even/odd operators}\label{app.ParityOp}
\vspace*{-1ex}

The parity operator $\Pi$, commonly used in quantum mechanics~\cite{GriSch18}, reverses the sign of all spatial coordinates,
\begin{equation}
\left[ \Pi f \right] (\mathbf{x}) 
\;=\;
f(-\mathbf{x}),
\end{equation}
where $f$ is any function of the spatial coordinates $\mathbf{x}$. The operator $\Pi$ is involutive ($\Pi^{-1}=\Pi$), Hermitian ($\Pi^{\mathrm{ad}}=\Pi$), and therefore unitary, with eigenvalues $\pm 1$. Its eigenfunctions possess definite parity and are either even or odd functions of the spatial coordinates.

An operator $S$ is said to be {\em even\/} if it commutes with the parity operator,
	\begin{equation}\label{eq.ParEven}
	S \Pi \, - \, \Pi S \, = \, 0
	\end{equation}
in which case $S$ and $\Pi$ share the same eigenfunctions. Consequently, the eigenfunctions of an even operator have definite parity. Conversely, an operator $Q$ is said to be {\em odd} if it anti-commutes with $\Pi$,
	\begin{equation}\label{eq.ParOdd}
	Q\Pi \, + \, \Pi Q \, = \, 0
	\end{equation}
For odd operators, no general parity property of the eigenfunctions can be inferred.

The commutation relations~\eqref{eq.ParEven}--\eqref{eq.ParOdd} can equivalently be expressed as
	\begin{equation}
	\begin{aligned}
	S \;&=\; \Pi S \Pi\\
	Q \;&=\; -\Pi Q \Pi.
	\end{aligned}
	\end{equation}
a form that clarifies the algebraic composition rules for even and odd operators. For example, the composition of two odd operators is even:
	\begin{equation}
	\Pi Q_{1}Q_{2}\Pi 
	\,=\, 
	\Pi Q_{1}\Pi^2Q_{2}\Pi 
	\,=\, 
	-Q_{1}(-Q_{2}) 
	\,=\, 
	Q_{1}Q_{2}.
	\end{equation}
More generally, the composition of (i) any number of even operators is even, (ii) an even number of odd operators is even, and (iii) an odd number of odd operators is odd.

In the present setting, we consider families of operators acting on the wall-normal coordinate $y$, parametrized by the wavenumber vector $\bk=(k_x,k_z)$. The streamwise-constant Orr--Sommerfeld and Squire operators are even with respect to $y$, whereas for Poiseuille flow the coupling operator $\cC_{p,2k_z} = -2k_z U'(y)$ is odd. The same parity properties hold for their adjoints. As a result, the composite operator
	\[
	(\cS_{2k_z}^{-1}\cC_{p,2k_z}\cL_{2k_z}^{-1})^{\mathrm{ad}}
	\cS_{2k_z}^{-1}\cC_{p,2k_z}\cL_{2k_z}^{-1}
	\]
is even, and its eigenfunctions---and hence the singular functions of $\cS_{2k_z}^{-1}\cC_{p,2k_z}\cL_{2k_z}^{-1}$---possess definite parity in the wall-normal coordinate $y$. This parity structure underlies the symmetry selection observed in Section~\ref{sec.streak-shape}.

    	\vspace*{-4ex}
\section{Symmetries of principal responses and antisymmetry of the $\cO (\epsilon^2)$ input $d_{\psi,2k_z}^{(2)}$}\label{app.GalilTrEven}
    	\vspace*{-1ex}

In a frame of reference moving with the wave, spatial derivatives remain unchanged, while the time derivative transforms as $\partial_t \mapsto \partial_t - c\,\partial_{\bar{x}}$. Under this transformation, the linearized dynamical generator $\cA_\bk$ takes the form
	\beq
	{\mathcal{A}}_\bk 
	\; = \;
	\left[
	\begin{array}{cc}
	\frac{1}{Re}{\Delta}^{-1} {\Delta}^2
	-
	\mathrm{i}k_x{\Delta}^{-1} (U-c) {\Delta} 
	+ 
	\mathrm{i}{\Delta}^{-1} U^{\prime \prime}  
	& 0 
	\\
	-\mathrm{i}k_z U^{\prime} & \frac{1}{Re} {\Delta} - \mathrm{i}k_x (U-c) 
	\end{array} 
	\right].
	\eeq
The operators $\mathcal{B}_\bk$ and $\mathcal{C}_\bk$ remain unchanged from their definitions in~\eqref{eq.NSveta}. In Poiseuille flow, we observe a block structure consisting of even and odd operators (cf.\ Appendix~\ref{app.ParityOp}). For instance,
	\begin{equation}
	{\mathcal{A}}_\bk
	\,=\,
 	\left[
	\ba{cc}
	S_{11} & 0\\
	Q_{21} & S_{22}
	\end{array}
	\right]
	\end{equation}
where $S_{11}$ and $S_{22}$ are even operators and $Q_{21}$ is an odd operator. Operators $\mathcal{B}_\bk$ and $\mathcal{C}_\bk$, as well as their respective adjoints, can be partitioned in a similar way. The output singular functions of the linearized frequency-response operator are given by the eigenfunctions of $\cG_\bk(\bc)\cG_\bk^{\mathrm{ad}}(\bc)$. Using the aforementioned partitions and the composition rules for operator parity, it can be shown that this operator has the following structure,
\begin{equation}
\cG_\bk(\bc)\cG_\bk^{\mathrm{ad}}(\bc)
\,=\,
\left[
\begin{array}{ccc}
S_{11} & Q_{12} & S_{13}\\
Q_{21} & S_{22} & Q_{23}\\
S_{31} & Q_{32} & S_{33}
\end{array}
\right]
\end{equation}
where $S_{ij}$ and $Q_{ij}$ are even and odd operators, respectively.  This structure reflects the parity-preserving and parity-flipping components inherited from the Orr--Sommerfeld, Squire, and coupling operators.

The eigenfunction problem can now be written in the form:
\begin{equation}
\begin{aligned}
\begin{bmatrix}
S_{11} & Q_{12} & S_{13}\\
Q_{21} & S_{22} & Q_{23}\\
S_{31} & Q_{32} & S_{33}
\end{bmatrix}
\begin{bmatrix}
{u}_1 \\ {v}_1 \\ {w}_1
\end{bmatrix} \,&=\, 
\sigma_1^2 \begin{bmatrix}
{u}_1 \\ {v}_1 \\ {w}_1
\end{bmatrix} \\
\begin{bmatrix}
I & 0 & 0\\
0 & Q_* & 0\\
0 & 0 & I
\end{bmatrix}
\begin{bmatrix}
S_{11} & Q_{12} & S_{13}\\
Q_{21} & S_{22} & Q_{23}\\
S_{31} & Q_{32} & S_{33}
\end{bmatrix}
\begin{bmatrix}
I & 0 & 0\\
0 & Q_*^{-1} & 0\\
0 & 0 & I
\end{bmatrix}
\begin{bmatrix}
I & 0 & 0\\
0 & Q_* & 0\\
0 & 0 & I
\end{bmatrix}
\begin{bmatrix}
{u}_1 \\ {v}_1 \\ {w}_1
\end{bmatrix} \,&=\, 
\sigma_1^2 \begin{bmatrix}
I & 0 & 0\\
0 & Q_* & 0\\
0 & 0 & I
\end{bmatrix}
\begin{bmatrix}
{u}_1 \\ {v}_1 \\ {w}_1
\end{bmatrix} \\
S \begin{bmatrix}
{u}_1 \\ Q_* {v}_1 \\ {w}_1
\end{bmatrix} 
\,&=\,
\sigma_1^2 \begin{bmatrix}
{u}_1 \\ Q_*{v}_1 \\ {w}_1
\end{bmatrix}
\end{aligned}
\end{equation}
where $Q_*$ can be any invertible odd operator of our choice (e.g., a derivative operator), and ${\bvartheta}_{\bk,1} = [  \, {u}_1 \;\, {v}_1 \;\, {w}_1 \, ]^T$ is the principal output singular direction. Since $S$ is an even operator, its eigenfunctions possess definite parity in the wall-normal coordinate. Consequently, the vector $[\, u_1 \;\; Q_* v_1 \;\; w_1 \,]^T$ is either entirely even or entirely odd in $y$. It follows that $u_1$ and $w_1$ share the same wall-normal parity, while $v_1$ necessarily has the opposite parity. Thus, for principal disturbance, the $\cO(\epsilon)$ oblique-wave response $\bu_\bk^{(1)} = \sigma_1\bvartheta_{\bk,1}$ can exhibit only the following two wall-normal parity patterns:
\begin{equation} \label{eq.Oeps-symmetry}
    \ba{llcll}
    \textrm{Case}~1 &  
    & ~~~ & 
    \textrm{Case}~2 & 
    \\[0.1cm]
    u_{\bk}^{(1)}(y,\bc) \! : 
    & 
    \textrm{even}  
    &
    ~~~
    & 
    u_{\bk}^{(1)}(y,\bc) \! :
    & 
    \textrm{odd} 
    \\[0.1cm]
    v_{\bk}^{(1)}(y,\bc) \! : 
    & 
    \textrm{odd}   
    & 
    ~~~
    & 
    v_{\bk}^{(1)}(y,\bc) \! :
    &  
    \textrm{even} 
    \\[0.1cm]
    w_{\bk}^{(1)}(y,\bc) \! : 
    & 
    \textrm{even}  
    & ~~~ & 
    w_{\bk}^{(1)}(y,\bc) \! : 
    & 
    \textrm{odd} 
    \ea
\end{equation}
The convective nonlinearity generates the input to the $\cO (\epsilon^2)$ streak response in the form
\beq
{d}_{\psi, 2 k_z}^{(2)} (y) 
	\,=\, 
	\Delta_{2 k_z}^{-1}
	\!
	\left(
	(\partial_{yy} \, + \, (2 k_z)^2) 
	\Im 
	\big\{
	{v}_{\bk}^{(1)} (y,c)
	{\bar{w}}_{\bk}^{(1)} (y,c)
	\big\}
	\, + \,
	2 k_z \partial_y
	\big(
	| {v}_{\bk}^{(1)} (y,c) |^2 
	\, + \,
	| {w}_{\bk}^{(1)} (y,c) |^2
	\big)
	\right).
\eeq
Using standard rules of even/odd function composition (e.g., the product of an even and an odd function is odd, and the derivative of an even function is odd), it follows that $d_{\psi,2k_z}^{(2)}$ is an odd---i.e., antisymmetric---function of $y$, independent of the specific wall-normal parity pattern of the oblique-wave response.

Higher-order terms can be analyzed in a similar fashion. In particular, for spatio-temporal frequencies where the $\mathcal{O}(\epsilon^2)$ streaks carry an order of magnitude more energy than any other component of the $\mathcal{O}(\epsilon^2)$ response, the $\mathcal{O}(\epsilon^3)$ oblique-wave forcing can be approximated by
\beq \label{eq.d3}
  \bd^{(3)}(\bar{x},y,z) \, \approx \,
  - \left[
  \begin{array}{c}
      u_{\mathrm{s},2k_z}^{(2)}(y,z) \partial_x u^{(1)}(\bar{x},y,z) + v^{(1)}(\bar{x},y,z)\partial_y u_{\mathrm{s},2k_z}^{(2)}(y,z) + w^{(1)}(\bar{x},y,z)\partial_z u_{\mathrm{s},2k_z}^{(2)}(y,z) \\
      u_{\mathrm{s},2k_z}^{(2)}(y,z)\partial_x v^{(1)}(\bar{x},y,z) \\
      u_{\mathrm{s},2k_z}^{(2)}(y,z) \partial_x w^{(1)}(\bar{x},y,z)
  \end{array}
  \right].
\eeq
This expression shows that, when $u_{\mathrm{s},2k_z}^{(2)}$ is an even function of $y$, the $\mathcal{O}(\epsilon^3)$ oblique-wave forcing inherits the same wall-normal symmetry as the $\mathcal{O}(\epsilon)$ response in~\eqref{eq.Oeps-symmetry}. As a result, the $\mathcal{O}(\epsilon^3)$ response admits a nonzero projection onto the principal singular mode, enabling coherent principal-mode dominance to persist at higher orders of the perturbation expansion.

These symmetry considerations explain both the universal antisymmetry of the $\cO(\epsilon^2)$ streak forcing and the robustness of principal-mode alignment observed across successive perturbation orders.

    	\vspace*{-3ex}
\section{Spectral analysis of the streamwise-constant Squire and Orr--Sommerfeld operators}
\label{app.OS-Sq-modes}
    	\vspace*{-1ex}

We next summarize the spectral properties of the streamwise-constant,
Reynolds-number-independent Squire and Orr--Sommerfeld operators.
Specifically, the Squire operator is given by
$\cS_{k_z} \DefinedAs \Delta$, while the Orr--Sommerfeld operator is
$\cL_{k_z} \DefinedAs \Delta^{-1} \Delta^{2}$;
see~\cite{dolph1958spectral} and~\cite[Appendix~B]{jovbamJFM05} for further details.
The Laplacian
	$
	\Delta 
	= 
	\partial_{yy} - k_z^2
	$
satisfies homogeneous Dirichlet boundary conditions in $y$,
and its biharmonic counterpart
	$
	\Delta^2
	=
	\partial_{yyyy} - 2 k_z^2 \partial_{yy} + k_z^4
	$
obeys homogeneous Dirichlet and Neumann boundary conditions.

The eigenvalue decomposition of the Squire operator yields an orthonormal set of eigenfunctions $\{\varphi_m\}$ with corresponding eigenvalues $\{\gamma_m\}$,
    \beq
    \varphi_m (y) 
    \, = \, 
    \sin \left( \frac{m \pi}{2} (y \, + \, 1) \right),
    ~~
    \gamma_{m} (k_z) 
    \, = \, 
    -\left( \frac{m^2\pi^2}{4} \, + \, k_z^2 \right),
    ~~
    m \, \in \, \bbN.
    \non
   \label{eq.eigSquire}
    \eeq 
The eigenfunctions of the Orr--Sommerfeld operator take the form
\beq
	\begin{array}{rcl}
	\phi_m(y,k_z) 
	& = & 
	A_{m} (k_z) 
	\left(
        \cos(p_m (k_z) y) 
        \, - \, 
	\dfrac{\cos(p_m (k_z))}{\cosh(k_z)}\cosh(k_z y) 
	\right) 
	\, +
	\\[0.35cm] 
	& & 
	B_{m} (k_z) 
	\left(
	\sin(p_m (k_z) y) 
	\, - \, 
	\dfrac{\sin(p_m (k_z))}{\sinh(k_z)}\sinh(k_z y)
	\right)
	\end{array}
\label{eq.os-modes}
\eeq
where $p_m (k_z)$ satisfies one of the transcendental equations
	\begin{align}
	\label{eq.pn1}
	p_m (k_z) \tan(p_m (k_z)) &\,=\, -k_z  \tanh(k_z)
	\\
	p_m (k_z) \cot(p_m (k_z)) &\,=\, k_z \coth(k_z).
	\label{eq.pn2}
	\end{align}
The normalization constants $A_m (k_z)$ and $B_m (k_z)$ cannot both be nonzero for the same index $m$. If $p_m (k_z)$ satisfies~\eqref{eq.pn1}, then
	\beq 
	A_m (k_z) 
	\; = \, 
	\left( 
	(p_m^2 (k_z)+ k_z^2)
	\left(
	1+\dfrac{\sin(2p_m (k_z))}{2p_m (k_z)}
	\right) 
	\right)^{-1/2},
	~~
	B_m (k_z) 
	\; = \; 
	0
	\label{eq.Am}
	\eeq
whereas if $p_m (k_z)$ satisfies~\eqref{eq.pn2}, then
	\beq
	A_m (k_z) 
	\; = \; 
	0,
	~~
	B_m (k_z) 
	\; = \, 
	\left( 
	(p_m^2 (k_z)+ k_z^2)
	\left(
	1-\dfrac{\sin(2p_m (k_z))}{2p_m (k_z)}	
	\right) 
	\right)^{-1/2}.
	\label{eq.Bm}
	\eeq
The eigenfunctions are ordered according to increasing magnitude of the corresponding eigenvalues $\lambda_m (k_z)$ of $\cL_{k_z}$,
	\beq
    	\lambda_m (k_z) 
    	\; = \; 
    	-(p_m^2 (k_z) \, + \, k_z^2).
    	\label{eq.os-lam}
	\eeq
The normalization constants $A_m (k_z)$ and $B_m (k_z)$, given by~\eqref{eq.Am} and~\eqref{eq.Bm}, ensure that the eigenfunctions $\{\phi_m\}$ form an orthonormal basis with respect to the inner product defined in~\cite[Appendix~B]{jovbamJFM05}.

	Both the Squire and Orr--Sommerfeld eigenfunctions exhibit symmetry or antisymmetry with respect to the wall-normal coordinate. For the Orr--Sommerfeld eigenfunctions, $A_m \neq 0$ corresponds to even (symmetric) dependence on $y$, while $B_m \neq 0$ corresponds to odd (antisymmetric) dependence. By ordering the eigenvalues according to increasing absolute value, odd and even indices identify, respectively, symmetric and antisymmetric $y$-dependence of the eigenfunctions of both operators; this property is exploited in Appendix~\ref{app.Galerkin}.

\vspace*{-2ex}
\section{Galerkin projection of the streamwise-constant model}
\label{app.Galerkin}
\vspace*{-1ex}

A reduced-order representation of the $\cO(\epsilon^2)$ streak dynamics can be
obtained by applying a Galerkin projection of the streamwise-constant model
\eqref{eq.ss-kx0-eps2} onto the eigenfunctions of the Orr--Sommerfeld and Squire
operators (see Appendix~\ref{app.OS-Sq-modes}),
\beq
\psi_{\mathrm{s},2k_z}^{(2)} (y,t)
\, = \,
{\displaystyle \sum_{m \,=\, 1}^{\infty}} \psi_{m}(t)
\phi_m(y),
~~
u_{\mathrm{s}1,2k_z}^{(2)} (y,t)
\, = \,
{\displaystyle \sum_{m \,=\, 1}^{\infty}} u_{m}(t)
\varphi_m (y)
\label{eq.Galerkin}
\eeq
where $\psi_m(t)$ and $u_m(t)$ denote the spectral coefficients associated with the
$m$th Orr--Sommerfeld and Squire eigenfunctions $\phi_m(y)$ and $\varphi_m(y)$,
respectively.
Substituting~\eqref{eq.Galerkin} into~\eqref{eq.ss-kx0-eps2} with
$d_u^{(2)} = 0$ yields the following coupled system of ordinary differential
equations for the coefficients $\psi_m(t)$ and $u_m(t)$,
\beq
\ba{rcl}
\dot{\psi}_m(t)
& = &
\dfrac{1}{Re} \, \lambda_{m} (2k_z) \psi_m(t)
\, - \,
\inner{\phi_m}{\Delta_{2k_z}d_{\psi,2k_z}^{(2)}}
\\[0.25cm]
\dot{u}_m(t)
& = &
\dfrac{1}{Re} \, \gamma_{m} (2k_z) u_m(t)
\, + \,
\displaystyle\sum_{i\,=\,1}^{\infty}
\inner{\varphi_m}{\cC_{p,2k_z}\phi_{i}}\psi_{i}(t).
\ea
\label{eq.psi-u-os-sq}
\eeq
Here, $\lambda_{m} (2k_z)$ and $\gamma_{m} (2k_z)$ denote the eigenvalues of the
Orr--Sommerfeld and Squire operators corresponding to the $m$th eigenfunctions,
and $\inner{\cdot}{\cdot}$ is the standard $L_2$ inner product.
Since $d_{\psi,2k_z}^{(2)}$ is an antisymmetric function of $y$, it has a
nonzero projection only onto Orr--Sommerfeld modes that are odd in $y$.
Moreover, for Poiseuille flow, the action of the coupling operator
$\cC_{p,2k_z}$ changes the symmetry of the response, yielding an output with
nonzero projection only onto the even (in $y$) Squire modes; see
Appendix~\ref{app.OS-Sq-modes}.

Figure~\ref{fig.1os-2sq-approx} shows that retaining only the second
Orr--Sommerfeld mode together with the first and third Squire modes, i.e.,
\beq
\psi_{\mathrm{s},2k_z}^{(2)} (y,t)
\; = \;
\psi_{2}(t)
\phi_2(y),
~~
u_{\mathrm{s}1,2k_z}^{(2)} (y,t)
\; = \;
u_{1}(t)
\varphi_1 (y) \,+\, u_{3}(t) \varphi_3 (y)
\eeq
provides an accurate approximation of the steady-state solution of
system~\eqref{eq.ss-kx0-eps2} to the $\cO (\epsilon)$ principal oblique disturbance.
This low-order Galerkin projection therefore offers a physically transparent
framework for modeling and interpreting the mechanisms by which
$\cO(\epsilon^2)$ steady streamwise streaks are generated through nonlinear
interactions of $\cO(\epsilon)$ unsteady oblique waves.

\begin{figure}[h]
  \centering
  \begin{tabular}{cc}
    \begin{tabular}{c}
    \vspace{1.15cm}
        \normalsize{\rotatebox{90}{$y$}}
      \end{tabular}
      &
      \begin{tabular}{c}
    \includegraphics[width=0.332\textwidth]{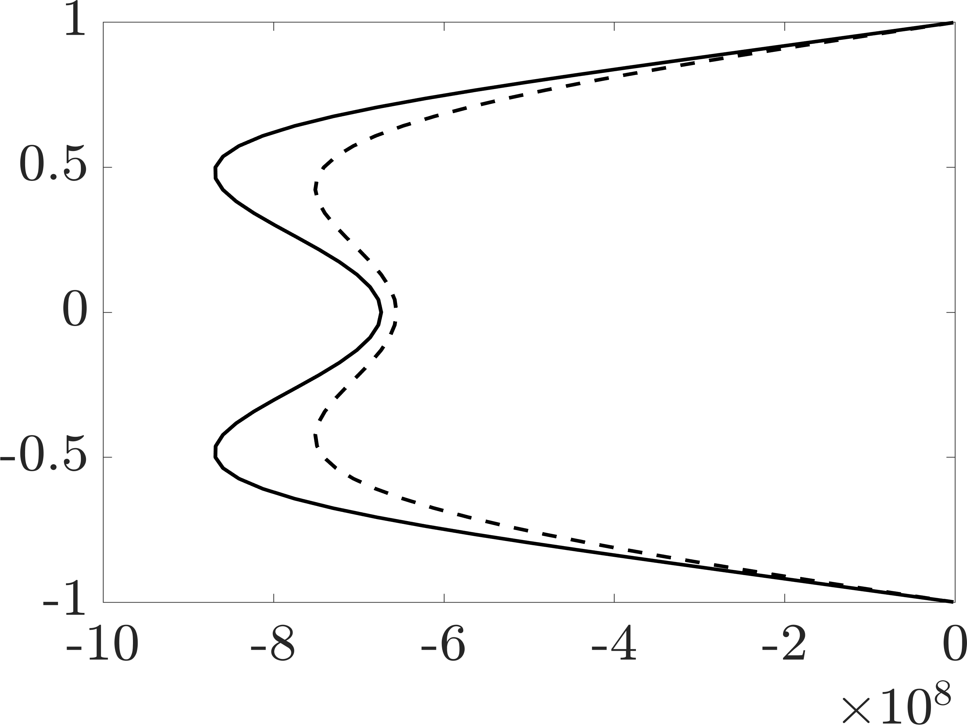}
        \\[-0.21cm]
        {$u_{\mathrm{s}1, 2 k_z}^{(2)} (y)$}
      \end{tabular}
  \end{tabular}
  \caption{The second-order steady streamwise streak response to the
  $\cO (\epsilon)$ principal oblique disturbance.
  The solid curve is obtained from the steady-state solution of
  system~\eqref{eq.ss-kx0-eps2}, while the dashed curve shows the approximation
  obtained using a truncated Galerkin projection retaining only the second
  Orr--Sommerfeld and the first and third Squire modes.}
  \label{fig.1os-2sq-approx}
\end{figure}

The preceding analysis also enables approximation of the spanwise wavenumber
for which the principal input singular direction of the operator
$\cS_{2 k_z}^{-1} \cC_{p, 2k_z} \cL_{2 k_z}^{-1}$ is symmetric, that is, for which
the second input singular direction of the steady streamwise-constant resolvent
operator is antisymmetric.
Owing to the parity-changing action of $\cC_{p, 2k_z}$, the maximal gain
associated with the symmetric input can be approximated by the product of the
gains of the first Orr--Sommerfeld mode and the second Squire mode, namely
$A_{1} (2k_z)/(\lambda_{1} (2k_z) \gamma_{2} (2k_z))$.
Similarly, the maximal amplification associated with the antisymmetric input is
given by $B_{2} (2k_z)/(\lambda_{2} (2k_z)\gamma_{1} (2k_z))$, corresponding to
the product of the second Orr--Sommerfeld mode and the first Squire mode.
Definitions of the normalization constants $A_{1} (2k_z)$ and $B_{2} (2k_z)$,
together with analytical expressions for the eigenvalues
$\lambda_{m} (2k_z)$ and $\gamma_{m} (2k_z)$, are provided in
Appendix~\ref{app.OS-Sq-modes}.
Equating these two gain estimates yields $2k_z \approx 0.9$.
For spanwise wavenumbers larger than this value, the antisymmetric mode
becomes dominant, satisfying
$B_{2} (2k_z)/(\lambda_{2} (2k_z)\gamma_{1}(2k_z))
>
A_{1} (2k_z)/(\lambda_{1} (2k_z)\gamma_{2} (2k_z))$.

Finally, the steady-state solutions $\psi_m$ and $u_m$ of
system~\eqref{eq.psi-u-os-sq} are given by
\beq
\ba{rcl}
\psi_m
& = &
( Re / \lambda_m (2k_z) )
\,
{\inner{\phi_m ( \, \cdot \, ,2k_z)}
{\Delta_{2k_z} d_{\psi,2k_z}^{(2)}(\cdot)}}
\\[0.25cm]
u_m
& = &
- ( Re / \gamma_{m} (2k_z) )
\displaystyle
\sum_{i\,=\,1}^{\infty}
\inner{\varphi_m ( \, \cdot \, ,2k_z)}
{\cC_{p,2k_z}\phi_{i} ( \, \cdot \, ,2k_z)}\psi_{i},
\ea
\label{eq.psi_m-u_m}
\eeq
so that computation of $\psi_m$ requires projection of the input
$d_{\psi,2k_z}^{(2)}$ onto the $m$th Orr--Sommerfeld mode.
Since $d_{\psi,2k_z}^{(2)}$ is antisymmetric in $y$
(Appendix~\ref{app.GalilTrEven}), $\psi_m$ vanishes for odd $m$.
Likewise, the steady value of $u_m$ in~\eqref{eq.psi_m-u_m} can be \mbox{written as}
\beq
\ba{lcl}
    u_m
    & = &
    -Re\,{\displaystyle \sum_{m'=1}^{\infty}}\psi_{2m'}\frac{8k_z}{16 k_z^2+m^2 \pi^2} \Bigg(\frac{4B_{2m'}(2k_z)\sin(p_{2m'}(2k_z))}{\left(16k_z^2+m^2 \pi^2\right)^2}
\Big(-m \pi  (-1+\cos(m \pi)) \left(16 k_z^2+m^2 \pi^2 - 16 k_z \coth(2 k_z)\right)  \\
&&
	+~
	2 \left(-16 k_z^2+m^2 \pi^2+2 k_z (16 k_z^2+m^2 \pi^2\right)
	\coth(2 k_z) ) \sin(m \pi)\Big)+ \frac{1}{\left(-4 p_{2m'}^2(2k_z)
	+m^2 \pi^2\right)^2}\\
&& \times\Big(4 B_{2m'}(2k_z) m \pi  (-1+\cos(m \pi ))\left(-8 p_{2m'}(2k_z) \cos(p_{2m'}(2k_z))+\left(-4 p_{2m'}^2(2k_z)+m^2 \pi ^2\right)
\sin(p_{2m'}(2k_z))\right) \\
&&
	-~
	8 B_{2m'}(2k_z) ((-4 p_{2m'}^3(2k_z)+m^2 p_{2m'}(2k_z) \pi ^2 ) \cos(p_{2m'}(2k_z)) \\
&&
	+~
	(4 p_{2m'}^2(2k_z)+m^2 \pi^2 ) \sin(p_{2m'}(2k_z)) ) \sin(m \pi )\Big)\Bigg),
\ea
\eeq
where $u_m = 0$ for even $m$.

    	\vspace*{-2ex}
\section{Higher-order perturbation series responses extracted from DNS}
\label{app.dns-higher-order}
\vspace*{-1ex}

For larger forcing amplitudes, DNS data can be used to approximate higher-order
corrections to the frequency response of the NS equations.
Let $u_1^{\textrm{DNS}}$ and $u_2^{\textrm{DNS}}$ denote the steady streak responses
obtained from DNS subject to oblique-wave forcing of amplitudes
$\epsilon_1$ and $\epsilon_2$, respectively.
Inspired by the procedure described in~\cite[Section~6.2]{henlunjoh93},
the $\mathcal{O}(\epsilon^2)$ and $\mathcal{O}(\epsilon^4)$ contributions to the
DNS streak response can be extracted by solving the following system of linear equations:
\beq
\left[
\ba{cc}
\epsilon_1^2 & \epsilon_1^4
\\
\epsilon_2^2 & \epsilon_2^4
\ea
\right]
\left[
\ba{c}
u^{(2)}
\\
u^{(4)}
\ea
\right]
=
\left[
\ba{c}
u_1^{\textrm{DNS}}
\\
u_2^{\textrm{DNS}}
\ea
\right]
~
\Rightarrow
~
\left[
\ba{c}
u^{(2)}
\\
u^{(4)}
\ea
\right]
=
\dfrac{1}{\epsilon_1^2 \epsilon_2^2(\epsilon_2^2 - \epsilon_1^2)}
\left[
\ba{c}
\epsilon_2^4 u_1^{\textrm{DNS}}
-
\epsilon_1^4 u_2^{\textrm{DNS}}
\\
\epsilon_1^2 u_2^{\textrm{DNS}}
-
\epsilon_2^2 u_1^{\textrm{DNS}}
\ea
\right].
\label{eq.dns-Opes4}
\eeq
The same procedure can be employed to compute the
$\mathcal{O}(\epsilon^2)$ and $\mathcal{O}(\epsilon^4)$ corrections to the laminar base flow.

\begin{figure}[t]
  \centering
  \begin{tabular}{c@{\hspace{-0.04 cm}}c@{\hspace{-0.1 cm}}l@{\hspace{-0.04 cm}}c@{\hspace{-0.1 cm}}c}
    \subfigure[]{\label{fig.Oeps4-scale-str}}
    &&
    \subfigure[]{\label{fig.Oeps4-scale-mean}}&
    \\[-.13cm]
    \begin{tabular}{c}
      \vspace{1cm}
      \normalsize{\rotatebox{90}{$y$}}
    \end{tabular}
    &
    \begin{tabular}{c}
      \includegraphics[width=0.332\textwidth]{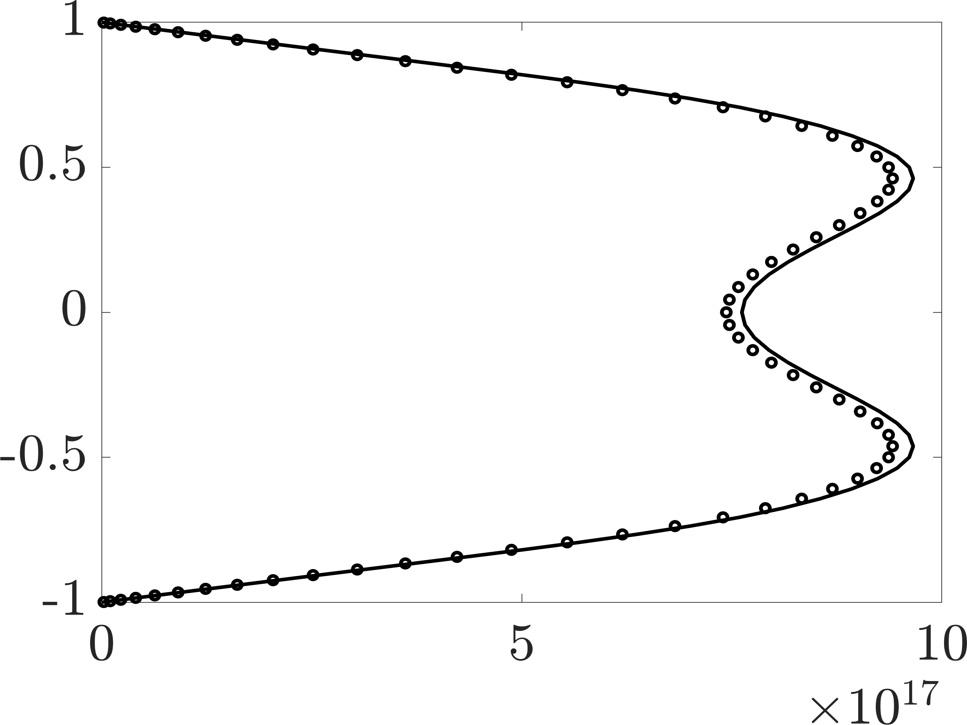}
      \\ [-0.3cm]
      \hspace{0.4cm}{$u_{\mathrm{s},2.28}^{(4)}(y)$}
    \end{tabular}
    &
      \begin{tabular}{c}
        \includegraphics[width=0.332\textwidth]{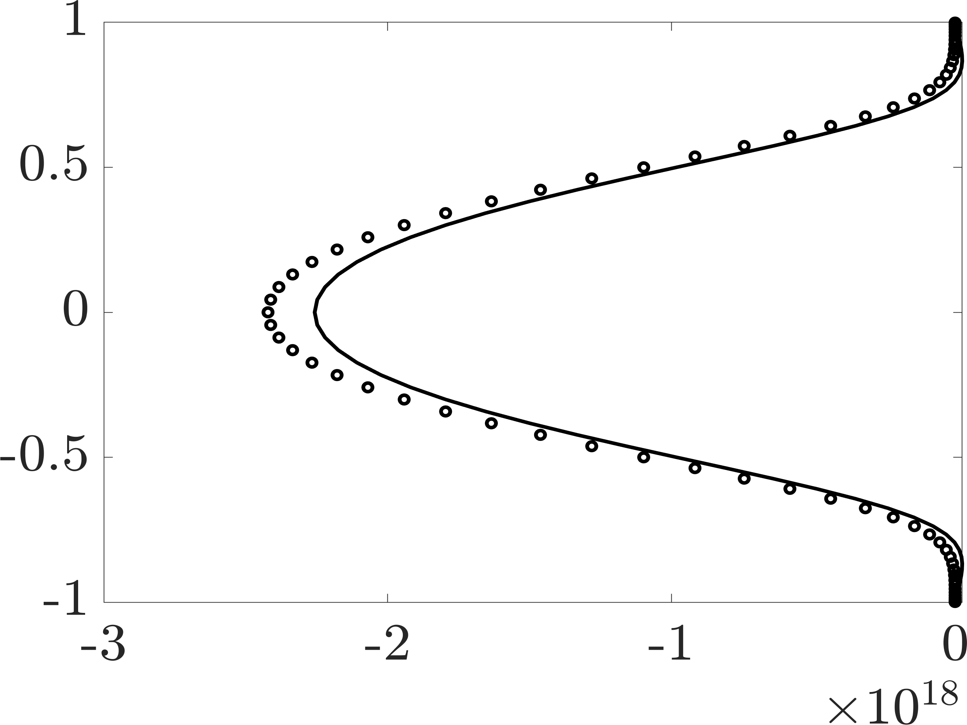}
        \\ [-0.3cm]
        \hspace{0.4cm}{$u_{(0,0)}^{(4)}(y)$}
      \end{tabular}
  \end{tabular}
  \vspace{-0.1cm}
  \caption{Approximated fourth-order (a) streak and (b) mean-flow fluctuation responses from DNS (solid lines) for forcing amplitudes $\epsilon_1^* = 1 \times 10^{-7}$ and $\epsilon_2^* = 5 \times 10^{-7}$, compared with predictions from the perturbation analysis (circles).}
    \label{fig.Oeps4-scale}
\end{figure}

Figure~\ref{fig.Oeps4-scale} demonstrates close agreement between the
$\mathcal{O}(\epsilon^4)$ correction to the DNS streak response obtained from~\eqref{eq.dns-Opes4} and the perturbation-analysis prediction based on~\eqref{eq.4th-O-response}.
Apart from a reversal in sign, the wall-normal profile of the
$\mathcal{O}(\epsilon^4)$ streak response is qualitatively similar to the
second-order profile shown in Fig.~\ref{fig.Oeps2-scale}.
Depending on the values of $k_x$, $k_z$, and $c$, the
$\mathcal{O}(\epsilon^4)$ correction to steady streamwise streaks either reinforces
or attenuates the $\mathcal{O}(\epsilon^2)$ contribution; see
Section~\ref{sec.streak-shape} for further discussion.

\begin{figure}[h]
  \centering
  \begin{tabular}{ll}
  \vspace{-0.1cm}
   \subfigure[]{\label{fig.DNS-OW-Oeps3-kx074}}
    & 
    \hspace{-0.5cm}
    \subfigure[]{\label{fig.DNS-OW-Oeps3-kx074-shift}}
    \\[-.09cm]
    \begin{tabular}{c}
      \includegraphics[width=0.3\textwidth]{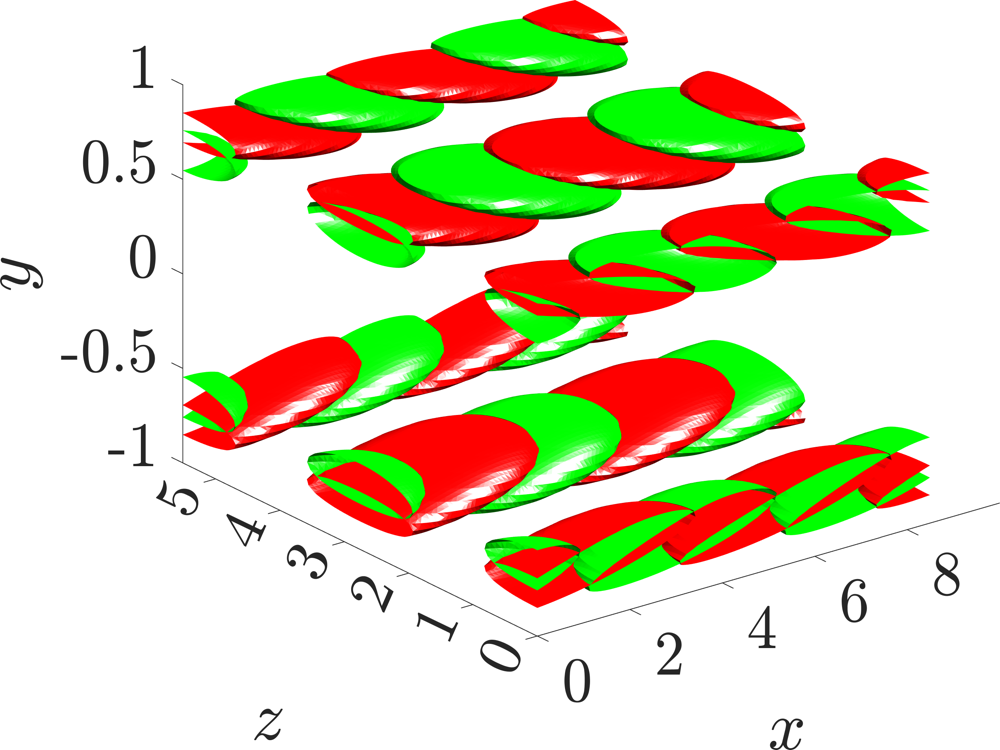}
    \end{tabular}
      &
      \begin{tabular}{c}
        \includegraphics[width=0.3\textwidth]{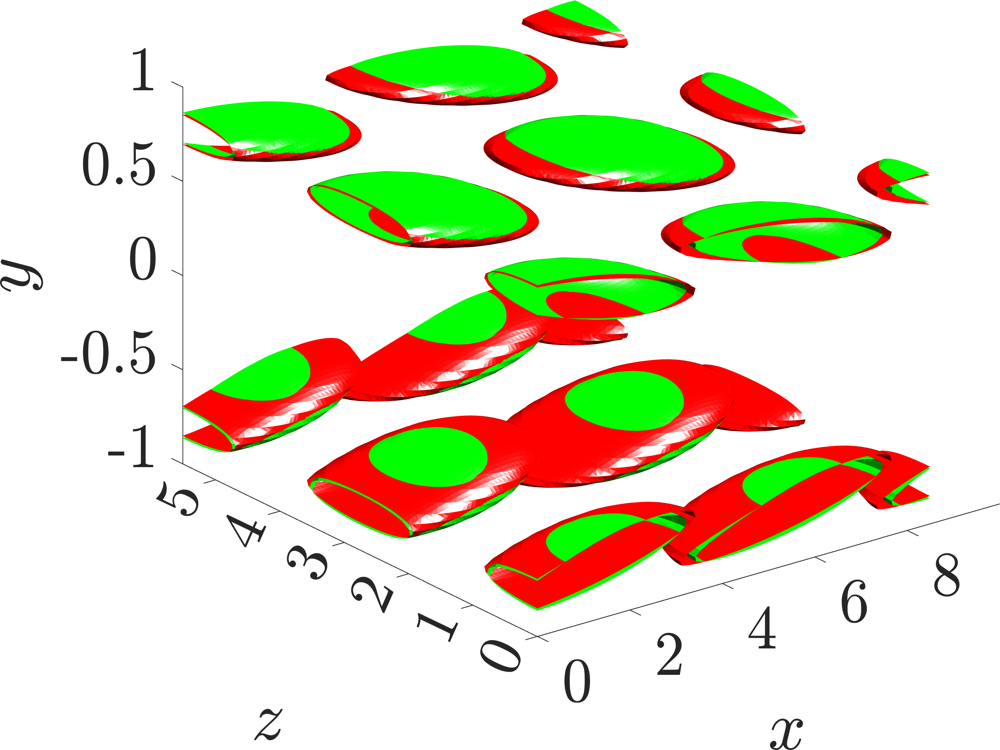}
      \end{tabular}
  \end{tabular}
  \vspace{-0.1cm}   
  \caption{Isosurfaces of the oblique-wave velocity components $|u_{(k_x,k_z)}^{(1)}|$ (green) and $|u_{(k_x,k_z)}^{(3)}|$ (red) extracted from DNS at $50\%$ of the maximum velocity-fluctuation magnitude. The $\mathcal{O}(\epsilon)$ response is obtained for $\epsilon^* = 10^{-8}$, while the $\mathcal{O}(\epsilon^3)$ response is extracted at $\epsilon^* = 10^{-6}$. Simulations are performed at $Re = 2000$ with $(k_x,k_z,c)=(0.74,1.14,0.396)$ in panel (a), and with the $\mathcal{O}(\epsilon)$ component additionally shifted by $1.44$~rad in the streamwise direction in panel (b).}
  \label{fig.DNS-OW-Oeps3}
\end{figure}

The system employed in~\cite[Section~6.2]{henlunjoh93} cannot be used to extract the
$\mathcal{O}(\epsilon^3)$ oblique-wave response while preserving a possible
streamwise phase shift.
To address this limitation, we assume that the oblique-wave responses
$u_1^{\mathrm{DNS}}$ and $u_2^{\mathrm{DNS}}$ obtained from DNS at sufficiently small
and moderate forcing amplitudes $\epsilon_1$ and $\epsilon_2$ are accurately
described by the linearized dynamics together with the leading odd-order correction,
that is,
\beq
\ba{rcl}
u_1^{\mathrm{DNS}}
& = &
\epsilon_1 u^{(1)}
\\
u_2^{\mathrm{DNS}}
& = &
\epsilon_2 u^{(1)}
\, + \,
\epsilon_2^3 u^{(3)}.
\ea
\eeq
Solving this system yields the third-order correction $u^{(3)}$ to the oblique-wave response,
\beq
u^{(3)}
\, = \, 
\frac{u_2^{\mathrm{DNS}}}{\epsilon_2^3}
\, - \,
\frac{u_1^{\mathrm{DNS}}}{\epsilon_1 \epsilon_2^2}.
\eeq
Apart from a streamwise phase shift of approximately $1.44$ radians, the
$\mathcal{O}(\epsilon^3)$ oblique-wave response is spatially similar to the
$\mathcal{O}(\epsilon)$ response
$u^{(1)} = u_1^{\mathrm{DNS}} / \epsilon_1$;
see Fig.~\ref{fig.DNS-OW-Oeps3}.
This observation is consistent with the perturbation-analysis predictions
presented in Section~\ref{sec.higher-order-anal}.

	\vspace*{-3ex}
\section{Procedure for estimating $\epsilon_{\mathrm{cr}}$}
\label{app.nth-root-test}
\vspace*{-1ex}

We begin by identifying, at a fixed Reynolds number, the oblique-wave forcing parameters $(k_x, k_z, c)$ that generate streak responses which remain spatially aligned across successive perturbation orders. As discussed in Section~\ref{sec.streak-shape}, such alignment occurs when both the $\mathcal{O}(\epsilon)$ and $\mathcal{O}(\epsilon^3)$ oblique-wave responses project predominantly onto the principal singular direction and when the associated streamwise phase shift $\varphi$ is small. Under these conditions, higher-order streak contributions act coherently, enabling assessment of absolute convergence of the perturbation expansion. The $\mathcal{O}(\epsilon^3)$ response used in this selection is obtained from the approximate forcing defined in~\eqref{eq.d3}.

Following the procedure of Section~\ref{sec:princWNL}, we perform a parameter sweep subject to two approximate criteria: (i) the $\mathcal{O}(\epsilon^3)$ response contains at least $50\%$ of its energy in the principal direction, and (ii) the phase shift satisfies $\varphi \leq 0.1$. The first condition ensures that the response is dominated by the principal mode, while the second guarantees that, even under the conservative assumption that each successive perturbation order accumulates an additional phase shift of magnitude $\varphi$, the streak response remains directionally aligned up to high order (e.g., $\cos(15\varphi) > 0$ at $\mathcal{O}(\epsilon^{30})$).

When the streak responses at all orders remain aligned in the same direction, convergence of the perturbation series is governed by absolute convergence. Since the streak expansion is a power series in the forcing amplitude $\epsilon$, we assess convergence using the $n$th-root test,
\begin{equation}
\label{eq.nthroot}
\lim_{n \, \to \, \infty} \sqrt[n]{\| \epsilon^n u_{\mathrm{s},2k_z}^{(n)} \|_2}
\end{equation}
where $\|\cdot\|_2$ denotes an $L_2$ norm in the wall-normal coordinate $y$. If the limit in~\eqref{eq.nthroot} exceeds unity, the series diverges. We denote by $\epsilon_{\mathrm{cr}}$ the forcing amplitude at which this condition is first satisfied, which provides an estimate of the breakdown of the weakly nonlinear expansion.

\begin{figure}[t]
  \centering
  \begin{tabular}{c@{\hspace{-0.04 cm}}c@{\hspace{-0.1 cm}}l@{\hspace{-0.04 cm}}c@{\hspace{-0.05 cm}}c}
    \subfigure[]{\label{fig.finite-epsCr-sameDir}}
    &&
    \subfigure[]{\label{fig.finite-epsCr-maxE}}&
    \\[-.13cm]
    \begin{tabular}{c}
      \vspace{0.51cm}
      \normalsize{\rotatebox{90}{$\epsilon_\mathrm{cr}$}}
    \end{tabular}
    &
    \begin{tabular}{c}
      \includegraphics[width=0.3\textwidth]{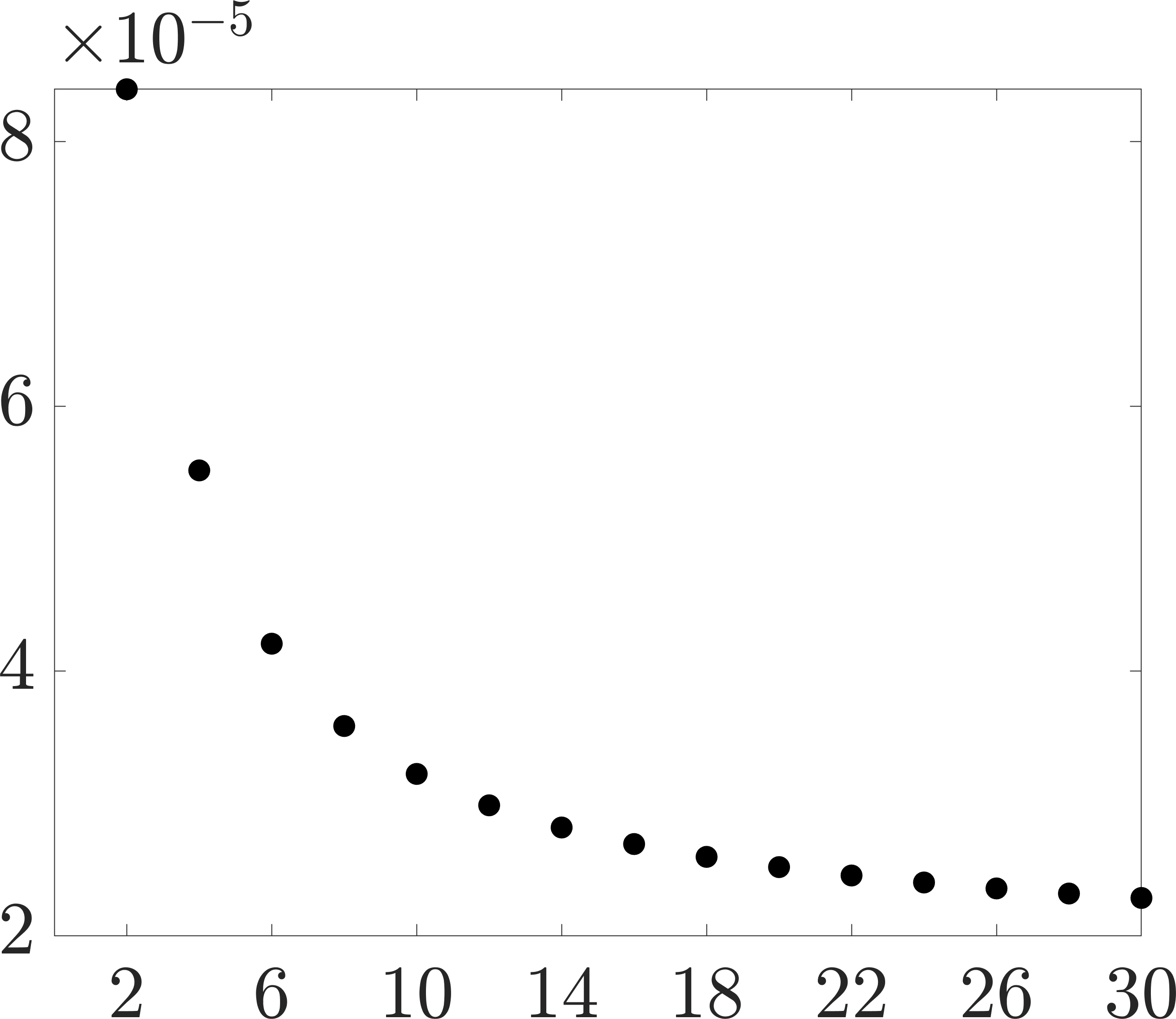}
      \\ [-0.1cm]
      \hspace{0.4cm}{$n$}
    \end{tabular}
    &
      \begin{tabular}{c}
    \includegraphics[width=0.3\textwidth]{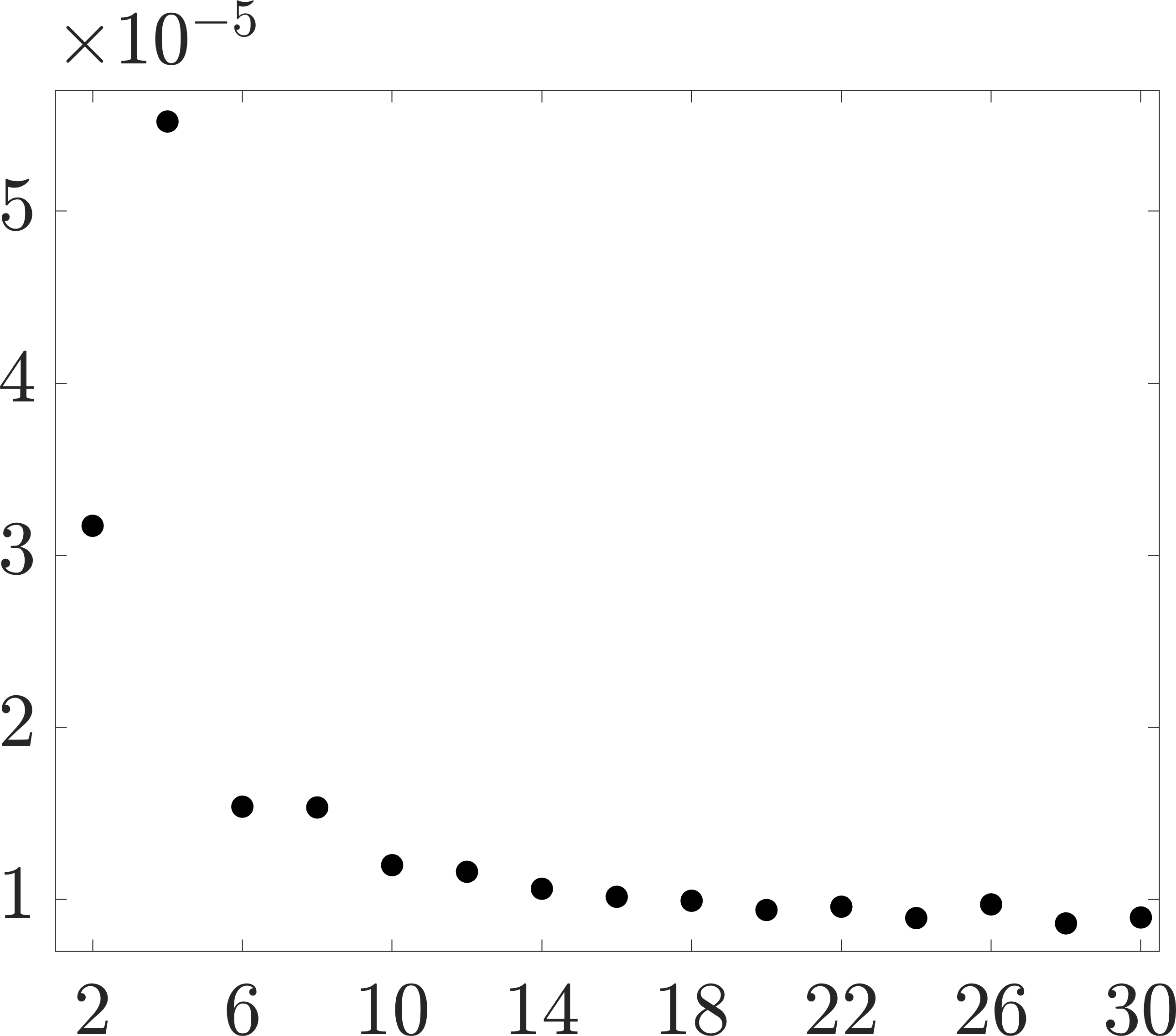}
        \\ [-0.1cm]
        \hspace{0.4cm}{$n$}
      \end{tabular}
  \end{tabular}
  \vspace{-0.1cm}
  \caption{Estimated critical forcing amplitude obtained by approximating the limit in~\eqref{eq.nthroot} using finite truncation orders $n$, for oblique-wave forcing with (a) $(k_x,k_z,c) = (0.94,1.48,0.376)$ and (b) $(k_x,k_z,c) = (0.74,1.14,0.396)$; both at $Re = 2000$.}
   \vspace{-0.25cm}
    \label{fig.eps-cr}
\end{figure}

To compute high-order streak responses efficiently, we retain only the oblique-wave/streak interaction terms associated with the base frequencies. Finite-order approximations of the limit in~\eqref{eq.nthroot} yield a sequence of estimates for $\epsilon_{\mathrm{cr}}$, shown in Fig.~\ref{fig.finite-epsCr-sameDir}. These estimates themselves display a convergent, series-like behavior. To accelerate convergence, we apply the Shanks transformation to the sequence of $\epsilon_{\mathrm{cr}}$ values.

\begin{figure}[h]
  \centering
  \begin{tabular}{c@{\hspace{-0.04 cm}}c@{\hspace{-0.1 cm}}l@{\hspace{-0.04 cm}}c@{\hspace{-0.05 cm}}c}
    \subfigure[]{\label{fig.epsCr-sameDir-sensitivity}}
    &
    \raisebox{-0.18cm}{$\epsilon_\mathrm{cr}$}
    &
    \subfigure[]{\label{fig.kx-sameDir-sensitivity}}
        \raisebox{-0.18cm}
    {\hspace{2.5cm}$k_x$}
    &
    \\[-.23cm]
    \begin{tabular}{c}
      \vspace{0.81cm}
      \normalsize{\rotatebox{90}{Energy threshold}}
    \end{tabular}
    &
    \raisebox{0.15cm}{
    \begin{tabular}{c}
      \includegraphics[width=0.32\textwidth]{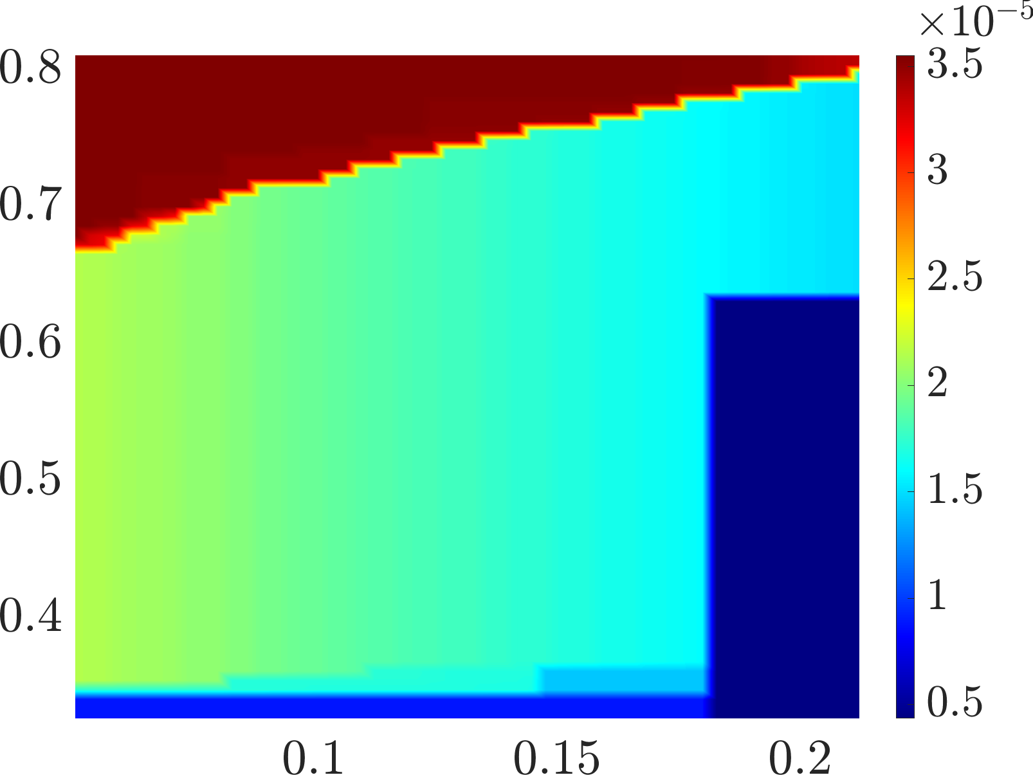}
      \\ [-0.1cm]
      \hspace{0.34cm}{Phase threshold}
    \end{tabular}}
    &
      \begin{tabular}{c}
    \includegraphics[width=0.3\textwidth]{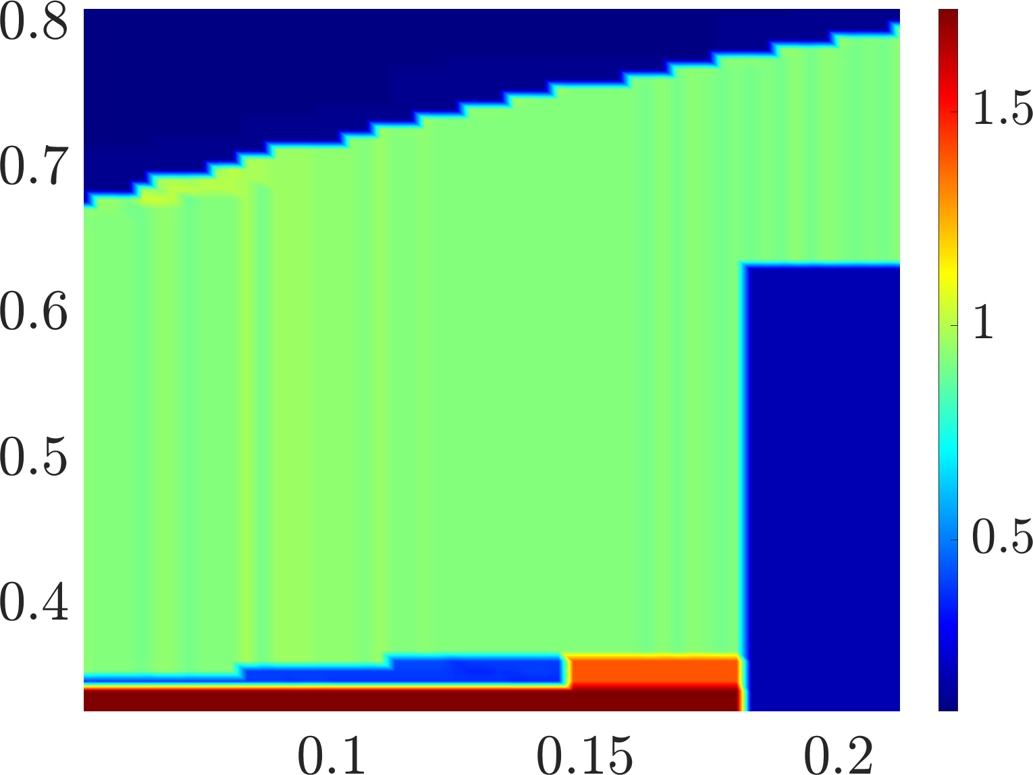}
        \\ [-0.1cm]
        \hspace{0.34cm}{Phase threshold}
      \end{tabular}
  \end{tabular}
  \vspace{-0.1cm}
 \caption{(a) Estimated minimal critical forcing amplitude $\epsilon_{\mathrm{cr}}$ as a function of phase-shift and energy-projection thresholds at $Re = 2000$. (b) Streamwise wavenumber $k_x$ corresponding to $\epsilon_{\mathrm{cr}}$ in (a). Regions with similar $\epsilon_{\mathrm{cr}}$ exhibit nearly constant $k_x$.}  
 \label{fig.eps-cr-sensitivity}
\end{figure}

For Poiseuille flow at $Re = 2000$, this procedure identifies the forcing parameters $(k_x, k_z, c) = (0.94, 1.48, 0.376)$ as yielding the smallest Shanks-accelerated estimate, $\epsilon_{\mathrm{cr}} \approx 1.9 \times 10^{-5}$. {The minimum estimated $\epsilon_{\mathrm{cr}}$ is robust to moderate variations in the energy-fraction and phase-shift thresholds (Fig.~\ref{fig.eps-cr-sensitivity}). Within a broad admissible range, $\epsilon_{\mathrm{cr}} \approx 2 \times 10^{-5}$, with oblique-wave parameters clustered around $k_x \in [0.9,1]$ (Fig.~\ref{fig.eps-cr-sensitivity}(b)), $k_z \in [1.4,1.5]$, and $c \in [0.37,0.38]$. For overly permissive thresholds (energy $\lesssim 0.35$, phase $\gtrsim 0.18$), higher-order streaks lose coherence with the leading-order mode, indicating breakdown of the aligned-streak assumption. Conversely, for overly restrictive energy thresholds ($\gtrsim 0.75$), no oblique-wave satisfies the selection criterion; in this case, $k_x \approx 0$ (Fig.~\ref{fig.kx-sameDir-sensitivity}), indicating that only $\mathcal{O}(\epsilon)$ streaks persist.} The parameters used for other Reynolds numbers are selected using the same methodology.

We next consider parameter regimes in which some higher-order streak responses are oriented opposite to the leading-order $\mathcal{O}(\epsilon^2)$ streak, as occurs for the oblique-wave parameters that maximize the energy of the $\mathcal{O}(\epsilon^2)$ response. In this case, the perturbation series alternates in sign, and the absolute-convergence criterion provided by the $n$th-root test yields only a lower bound on the true radius of convergence. Moreover, the finite-order $n$th-root estimates exhibit pronounced oscillations rather than monotonic convergence, as shown in Fig.~\ref{fig.finite-epsCr-maxE}. This behavior complicates the application of standard convergence-acceleration techniques.

To mitigate this sensitivity, we employ a sliding-window Shanks transformation. Specifically, for window sizes of $9$, $11$, $13$, and $15$ terms, we compute Shanks-accelerated limits over windows that are shifted one index at a time along the sequence of $\epsilon_{\mathrm{cr}}$ estimates. For each window size, we identify the first Shanks limit for which successive window shifts change the estimate by less than $5\%$. The final reported value of $\epsilon_{\mathrm{cr}}$ is then taken as the average of the stable limits obtained across the four window sizes.

\begin{figure}[h]
  \centering
  \begin{tabular}{cc}
    \begin{tabular}{c}
    \vspace*{0.65cm}
        \normalsize{\rotatebox{90}{$y$}}
      \end{tabular}
      &
      \begin{tabular}{c}
      \includegraphics[width=0.315\textwidth]
      {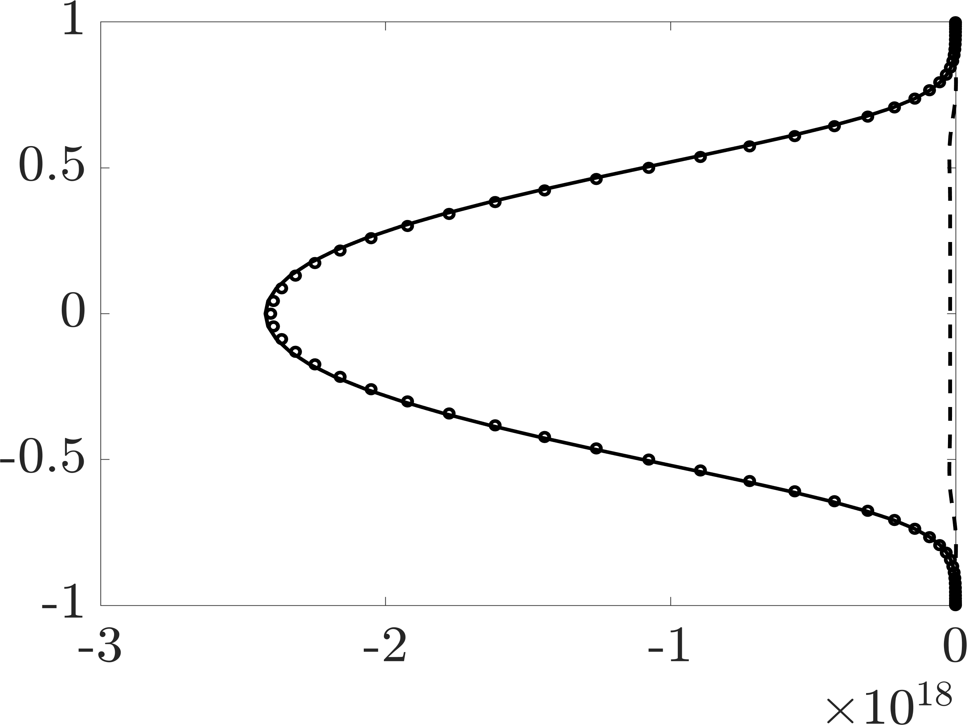}
        \\[-0.4cm] { $u_{(0,0)}^{(4)}(y)$}
        \end{tabular}
    \end{tabular}
  \vspace{-0.1cm}
  \caption{Perturbation-analysis predictions of the dominant contributions to the $\mathcal{O}(\epsilon^4)$ mean-flow response: full response (solid line), contribution from streamwise-constant terms (circles), and contribution from streamwise-fluctuating terms (dashed). Results are shown for Poiseuille flow at $Re = 2000$ with oblique-wave forcing $(k_x, k_z, c) = (0.74, 1.14, 0.396)$.}
  \label{fig.Oeps4-split}
\end{figure}

    	\vspace*{-5ex}
\section{Higher-order corrections to the laminar base flow}
\label{app.mean-higher-order}
\vspace*{-1ex}

In contrast to the steady streaks, the $\mathcal{O}(\epsilon^2)$ and
$\mathcal{O}(\epsilon^4)$ corrections to the laminar base flow differ
substantially.
The $\mathcal{O}(\epsilon^2)$ mean-flow deformation is driven entirely by
quadratic interactions among the streamwise-varying components associated
with the $\mathcal{O}(\epsilon)$ oblique waves, namely,
\begin{equation}
\bu_{(0,0)}^{(2)}(y,0)
 \, = \, 
\frac{1}{4}
\left(
\cG_{(0,0)}(0)
\Big(
\widetilde{\cN}_2 \big(\bu_{(\bkx,\bkz)}^{(1)}(\cdot,c),
\bar{\bu}_{(\bkx,\bkz)}^{(1)}(\cdot,c)\big)
 \, + \,
\widetilde{\cN}_2 \big(\bu_{(-\bkx,\bkz)}^{(1)}(\cdot,c),
\bar{\bu}_{(-\bkx,\bkz)}^{(1)}(\cdot,c)\big)
\Big)
\right)(y).
\end{equation}
Here, $\cG_{(0,0)}(0)$ denotes the frequency-response operator,
$\bu_{\bk}^{(1)}$ is the $\mathcal{O}(\epsilon)$ oblique-wave response with
wavenumber vector $\bk$, and $\bar{(\cdot)}$ denotes complex conjugation.
The operator $\widetilde{\cN}_2$ is the Fourier-transformed cross-interaction nonlinear operator defined in~\eqref{eq.NM},
\[
\widetilde{\cN}_2(\bu_{\bk},\bu_{\bk'})
\, \DefinedAs \,
-(\bu_{\bk}\!\cdot\!\nabla_{\bk'})\bu_{\bk'}
-(\bu_{\bk'}\!\cdot\!\nabla_{\bk})\bu_{\bk},
~~
\nabla_{\bk} \DefinedAs [\, \mri k_x \;\; \partial_y \;\; \mri k_z \,]^T .
\]

By contrast, the $\mathcal{O}(\epsilon^4)$ correction to the base flow receives
contributions from both steady streaks and streamwise-varying fluctuations,
\begin{equation}
\label{eq.o4mean}
\begin{aligned}
\bu^{(4)}_{(0,0)}
=\,&
\frac{1}{4}\,
\cG_{(0,0)}
\Big(
\overbrace{
\widetilde{\cN}_2\Big(\bu_{(0,2\bkz)}^{(2)},\bar{\bu}_{(0,2\bkz)}^{(2)}\Big)
}^{\ba{c} \mbox{\bf streamwise-constant terms} \ea}
+
\\[0.15cm]
&\underbrace{
\widetilde{\cN}_2\Big(\bu_{(2\bkx,0)}^{(2)},\bar{\bu}_{(2\bkx,0)}^{(2)}\Big)
+
\widetilde{\cN}_2\Big(\bu_{(2\bkx,2\bkz)}^{(2)},\bar{\bu}_{(2\bkx,2\bkz)}^{(2)}\Big)
+
\widetilde{\cN}_2\Big(\bu_{(-2\bkx,2\bkz)}^{(2)},\bar{\bu}_{(-2\bkx,2\bkz)}^{(2)}\Big)
}_{\ba{c} \mbox{\bf streamwise-varying terms} \ea}
+
\\[0.cm]
&\overbrace{
\widetilde{\cN}_2\Big(\bu_{(\bkx,\bkz)}^{(1)},\bar{\bu}_{(\bkx,\bkz)}^{(3)}\Big)
+
\widetilde{\cN}_2\Big(\bar{\bu}_{(-\bkx,\bkz)}^{(1)},\bu_{(-\bkx,\bkz)}^{(3)}\Big)
+
\widetilde{\cN}_2\Big(\bu_{(\bkx,\bkz)}^{(3)},\bar{\bu}_{(\bkx,\bkz)}^{(1)}\Big)
+
\widetilde{\cN}_2\Big(\bar{\bu}_{(-\bkx,\bkz)}^{(3)},\bu_{(-\bkx,\bkz)}^{(1)}\Big)
}
\Big).
\end{aligned}
\end{equation}
Here, $\bu_{\bk}^{(2)}$ and $\bu_{\bk}^{(3)}$ denote the
$\mathcal{O}(\epsilon^2)$ and $\mathcal{O}(\epsilon^3)$ velocity fluctuations
with wavenumber vector $\bk$, respectively; the dependence on $y$ and $c$ is
suppressed for brevity.
Equation~\eqref{eq.o4mean} shows that, in addition to unsteady oblique-wave
interactions, steady streamwise streaks contribute directly to higher-order
modifications of the laminar base flow.
Figure~\ref{fig.Oeps4-split} indicates that the contribution arising from the
$\mathcal{O}(\epsilon^2)$ steady streaks is substantially larger than that of
the streamwise-varying terms, whose effect is comparatively weak.

Because the leading-order streak shape is largely preserved at higher orders
in the perturbation expansion (see Fig.~\ref{fig.higher-order-streaks}), the
associated mean-flow deformation inherits the same spatial structure.
This explains why the mean flow obtained from DNS transitions from agreement
with the $\mathcal{O}(\epsilon^2)$ prediction at small forcing amplitudes to
closely following the $\mathcal{O}(\epsilon^4)$ shape as the amplitude
increases.
Consequently, inclusion of $\mathcal{O}(\epsilon^4)$ (and higher-order) terms
is essential for accurate prediction of the mean flow.
This observation is of potential relevance for improving modeling approaches
based on mean-flow perturbations; see, for example,~\cite{prabotche15}.

	\newpage
\providecommand{\noopsort}[1]{}\providecommand{\singleletter}[1]{#1}%

\end{document}

%% file: commands_ozaslan.tex
\input{command_widebar}

\newcommand{\cN}{\mathcal{N}}

\newcommand{\inner}[2]{\langle #1,#2\rangle}    

\newcommand{\enma}[1]{\ensuremath{#1}}

\newcommand{\non}{\nonumber}

\newcommand{\beq}{\begin{equation}}
\newcommand{\eeq}{\end{equation}}
\newcommand{\ba}{\begin{array}}
\newcommand{\ea}{\end{array}}
\newcommand{\bseq}{\begin{subequations}}
\newcommand{\eseq}{\end{subequations}}

\newcommand{\DefinedAs}[0]{\mathrel{\mathop:}=}



%% file: command_widebar.tex
\makeatletter
\newcommand*\rel@kern[1]{\kern#1\dimexpr\macc@kerna}
\newcommand*\widebar[1]{%
	\begingroup
	\def\mathaccent##1##2{%
		\rel@kern{0.8}%
		\overline{\rel@kern{-0.8}\macc@nucleus\rel@kern{0.2}}%
		\rel@kern{-0.2}%
	}%
	\macc@depth\@ne
	\let\math@bgroup\@empty \let\math@egroup\macc@set@skewchar
	\mathsurround\z@ \frozen@everymath{\mathgroup\macc@group\relax}%
	\macc@set@skewchar\relax
	\let\mathaccentV\macc@nested@a
	\macc@nested@a\relax111{#1}%
	\endgroup
}

%% file: notation_ozaslan.tex
\newcommand{\cL}{\enma{\mathcal L}}

\newcommand{\cB}{\enma{\mathcal B}}
\newcommand{\cH}{\enma{\mathcal H}}

\newcommand{\cC}{\enma{\mathcal C}}

\newcommand{\cA}{\enma{\mathcal A}}

%% file: figures/Figure2a.tex
{
	\setlength{\unitlength}{1cm}
	\thicklines
	\begin{tikzpicture}[auto, node distance=3.5cm, >=latex']
	    \node [bigblock, name=system] at (0,0)
	    {\tc{black}{$ \ba{c}\mbox{\sc LNSE} \\[0.05cm]
	    \ba{rcl}
	    \partial_t \bq^{(1)} & \! = & \tc{blue}{\cA} \, \bq^{(1)} \, + \; \tc{red}{\cB \, \bd^{(1)}} \\[0.1cm]
	    \bu^{(1)} & \! = & \cC \, \bq^{(1)}
	    \ea \ea$}};
	    
	    \node [input, name=input] [right of=system] {};
	    \node [coordinate] (end) [left of=system] {};
	    
	    \path[very thick, ->] (input) edge node[above] {
	    $
	    \ba{l}
	    \mbox{\tc{black}{$\ba{c}\mbox{\tc{red}{$\bd^{(1)}$}} \ea$}}
	    \ea
	    $
	    } (system);
	    
	    \draw[very thick, ->] (system) edge node[above] {
	    $
	    \ba{l}
	    \mbox{\tc{black}{$\ba{c}\mbox{$\bu^{(1)}$} \ea$}}
	    \ea
	    $
	    } (end);
	\end{tikzpicture}
	}

%% file: figures/Figure2b.tex
{
	\setlength{\unitlength}{1cm}
	\thicklines
	\begin{tikzpicture}[auto, node distance=6.8cm, >=latex']
	    \node [bigblock, name=system] at (0,0)
	    {\tc{black}{$ \ba{c}\mbox{\sc LNSE} \\[0.05cm]
	    \ba{rcl}
	    \partial_t \bq^{(n)} & \! = & \tc{blue}{\cA} \, \bq^{(n)} \, + \; \tc{red}{\cB \, \bd^{(n)}} \\[0.1cm]
	    \bu^{(n)} & \! = & \cC \, \bq^{(n)}
	    \ea \ea$}};
	    
	    \node [input, name=input] [right of=system] {};
	    \node [coordinate] (end) at ($(system.west)+(-1.25,0)$) {};
	    
	    \path[very thick, ->] (input) edge node[above] {
	    $
	    \ba{l}
	    \mbox{\tc{black}{$\ba{c}\mbox{\tc{red}{$\bd^{(n)} \DefinedAs -\sum_{r = 1}^{n-1} ( \bu^{(r)} \! \cdot \nabla ) \bu^{(n-r)}$}} \ea$}}
	    \ea
	    $
	    } (system);
	    
	    \draw[very thick, ->] (system) edge node[above] {
	    $
	    \ba{l}
	    \mbox{\tc{black}{$\ba{c}\mbox{$\bu^{(n)}$} \ea$}}
	    \ea
	    $
	    } (end);
	\end{tikzpicture}
	}

%% file: figures/Figure2c.tex
\begin{tikzpicture}[>=latex', node distance=1cm and 1cm]
    \node[block, minimum width=3em, minimum height=2em, label={[yshift=-2pt]above:$\cO(\epsilon^4)$}] (L4) at (0,0) {LNSE};
    \node[block, minimum width=3em, minimum height=2em, fill=customRed] (M1) [right=2cm of L4] {$\cN_2( \, \cdot \, , \, \cdot \, )$};
    \node[block, minimum width=3em, minimum height=2em, label={[yshift=-2pt]above:$\cO(\epsilon^3)$}] (L3) [right=1cm of M1] {LNSE};
    \node[block, minimum width=3em, minimum height=2em,fill=customRed] (M2) [right=1cm of L3] {$\cN_2 ( \, \cdot \, , \, \cdot \, )$};
    \node[block, minimum width=3em, minimum height=2em, label={[yshift=-2pt]above:$\cO(\epsilon^2)$}] (L2) [right=1cm of M2] {LNSE};
    \node[block, minimum width=3em, minimum height=2em,fill=customRed] (N_main) [right=1cm of L2] {$\cN_1( \, \cdot \, )$};
    \node[block, minimum width=3em, minimum height=2em, label={[yshift=-2pt]above:$\cO(\epsilon)$}] (L1) [right=1.25cm of N_main] {LNSE};
    
    \draw[->, very thick] (L1.west) -- (N_main.east) node[midway, above, xshift = 6.5pt] {$\bu^{(1)}$};
    \draw[->, very thick] (L2.west) -- (M2.east) node[midway, above, xshift = 3.5pt] {$\bu^{(2)}$};
    \draw[->, very thick] (L3.west) -- (M1.east) node[midway, above, xshift = 3.5pt] {$\bu^{(3)}$};
    \draw[->, very thick] (L4.west) -- ++(-1,0) node[midway, above, xshift = 3.5pt] {$\bu^{(4)}$};
    
    \draw[->, very thick] ($(L1.east)+(1,0)$) -- node[midway, above,xshift=3.5pt] {$\bd^{(1)}$} (L1.east);
    \draw[->, very thick] ($(L2.east)+(1,0)$) -- node[midway, above, xshift = 3.5pt] {$\bd^{(2)}$} (L2.east);
    \draw[->, very thick] ($(L3.east)+(1,0)$) -- node[midway, above, xshift = 3.5pt] {$\bd^{(3)}$} (L3.east);

    \coordinate (u1start) at ($(N_main.east)+(0.5,0)$);
    \filldraw (u1start) circle (2pt);
    \draw[very thick, -] (u1start) -- ++(0,1) coordinate (u1up);
    \draw[->, very thick] (u1up) -| (M2.north);
    \draw[->, very thick] (u1up) -| (M1.north);
    
    \coordinate (u2branch) at ($(M2.east)+(0.5,0)$);
    \filldraw (u2branch) circle (2pt);
    \node[block, minimum width=3em, minimum height=2em,fill=customRed] (N_extra) at ($ (L4)!0.33!(L2) + (0,-1.2)$) {$\cN_1( \, \cdot \, )$};
    \draw[->, very thick] (u2branch) -- ++(0,-1.2) coordinate (A) -- (A -| N_extra.east);

    \coordinate (joinCircle) at ($(L4.east)+(1.1,0)$);
    \draw[very thick, fill=none] (joinCircle) circle (4.5pt) node {+};
    \draw[->, very thick, shorten >=4.5pt] (N_extra.west) -- ($(joinCircle)+(0,-1.2)$) -- ++(0,1) -- (joinCircle);

    \draw[->, very thick] ($(joinCircle)-(4.5pt,0)$) -- (L4.east) node[midway, above, xshift = 3.5pt] {$\bd^{(4)}$};

    \draw[->, very thick, shorten >=4.5pt] (M1.west) -- (joinCircle);
    
\end{tikzpicture}

%% file: figures/Figure4a.tex
      \tikzset{every picture/.style={line width=0.75pt}} 

\begin{tikzpicture}[x=0.75pt,y=0.75pt,yscale=-0.8,xscale=0.8]

\draw    (161.74,53.31) -- (161.2,305.73) ;
\draw    (301.56,191.89) -- (-0.13,191.89) ;
\draw    (206.8,188.19) -- (206.8,195.41) ;
\draw    (164.82,147.32) -- (158.37,147.32) ;
\draw    (26.29,188.19) -- (26.29,195.41) ;
\draw    (116.46,188.01) -- (116.46,195.24) ;
\draw    (164.82,236.11) -- (158.37,236.11) ;
\draw  [fill={rgb, 255:red, 0; green, 0; blue, 0 }  ,fill opacity=1 ] (302.75,191.8) -- (296.93,194.38) -- (296.93,189.22) -- cycle ;
\draw  [fill={rgb, 255:red, 0; green, 0; blue, 0 }  ,fill opacity=1 ] (161.74,53.31) -- (164.36,59.03) -- (159.11,59.03) -- cycle ;
\draw  [fill={rgb, 255:red, 0; green, 0; blue, 0 }  ,fill opacity=1 ] (157.76,103.05) .. controls (157.76,100.95) and (159.49,99.24) .. (161.63,99.24) .. controls (163.77,99.24) and (165.5,100.95) .. (165.5,103.05) .. controls (165.5,105.15) and (163.77,106.85) .. (161.63,106.85) .. controls (159.49,106.85) and (157.76,105.15) .. (157.76,103.05) -- cycle ;
\draw  [fill={rgb, 255:red, 0; green, 0; blue, 0 }  ,fill opacity=1 ] (67.41,191.59) .. controls (67.41,189.49) and (69.14,187.79) .. (71.28,187.79) .. controls (73.42,187.79) and (75.15,189.49) .. (75.15,191.59) .. controls (75.15,193.7) and (73.42,195.4) .. (71.28,195.4) .. controls (69.14,195.4) and (67.41,193.7) .. (67.41,191.59) -- cycle ;
\draw  [fill={rgb, 255:red, 0; green, 0; blue, 0 }  ,fill opacity=1 ] (67.41,103.12) .. controls (67.41,101.02) and (69.14,99.31) .. (71.28,99.31) .. controls (73.42,99.31) and (75.15,101.02) .. (75.15,103.12) .. controls (75.15,105.22) and (73.42,106.92) .. (71.28,106.92) .. controls (69.14,106.92) and (67.41,105.22) .. (67.41,103.12) -- cycle ;
\draw  [fill={rgb, 255:red, 0; green, 0; blue, 0 }  ,fill opacity=1 ] (157.81,191.81) .. controls (157.81,189.7) and (159.54,188) .. (161.68,188) .. controls (163.82,188) and (165.55,189.7) .. (165.55,191.81) .. controls (165.55,193.91) and (163.82,195.61) .. (161.68,195.61) .. controls (159.54,195.61) and (157.81,193.91) .. (157.81,191.81) -- cycle ;
\draw  [fill={rgb, 255:red, 0; green, 0; blue, 0 }  ,fill opacity=1 ] (247.78,191.86) .. controls (247.78,189.76) and (249.51,188.05) .. (251.65,188.05) .. controls (253.79,188.05) and (255.52,189.76) .. (255.52,191.86) .. controls (255.52,193.96) and (253.79,195.66) .. (251.65,195.66) .. controls (249.51,195.66) and (247.78,193.96) .. (247.78,191.86) -- cycle ;
\draw  [fill={rgb, 255:red, 0; green, 0; blue, 0 }  ,fill opacity=1 ] (247.83,103.01) .. controls (247.83,100.91) and (249.57,99.21) .. (251.7,99.21) .. controls (253.84,99.21) and (255.58,100.91) .. (255.58,103.01) .. controls (255.58,105.11) and (253.84,106.82) .. (251.7,106.82) .. controls (249.57,106.82) and (247.83,105.11) .. (247.83,103.01) -- cycle ;
\draw  [fill={rgb, 255:red, 0; green, 0; blue, 0 }  ,fill opacity=1 ] (248.1,280.39) .. controls (248.1,278.29) and (249.83,276.58) .. (251.97,276.58) .. controls (254.11,276.58) and (255.85,278.29) .. (255.85,280.39) .. controls (255.85,282.49) and (254.11,284.19) .. (251.97,284.19) .. controls (249.83,284.19) and (248.1,282.49) .. (248.1,280.39) -- cycle ;
\draw  [fill={rgb, 255:red, 0; green, 0; blue, 0 }  ,fill opacity=1 ] (157.76,280.39) .. controls (157.76,278.29) and (159.49,276.58) .. (161.63,276.58) .. controls (163.77,276.58) and (165.5,278.29) .. (165.5,280.39) .. controls (165.5,282.49) and (163.77,284.19) .. (161.63,284.19) .. controls (159.49,284.19) and (157.76,282.49) .. (157.76,280.39) -- cycle ;
\draw  [fill={rgb, 255:red, 0; green, 0; blue, 0 }  ,fill opacity=1 ] (67.41,280.39) .. controls (67.41,278.29) and (69.14,276.58) .. (71.28,276.58) .. controls (73.42,276.58) and (75.15,278.29) .. (75.15,280.39) .. controls (75.15,282.49) and (73.42,284.19) .. (71.28,284.19) .. controls (69.14,284.19) and (67.41,282.49) .. (67.41,280.39) -- cycle ;
\draw  [line width=1.5]  (110.97,230.76) -- (121.94,230.76) -- (121.94,241.54) -- (110.97,241.54) -- cycle ;
\draw  [line width=1.5]  (110.97,141.65) -- (121.94,141.65) -- (121.94,152.43) -- (110.97,152.43) -- cycle ;
\draw  [line width=1.5]  (201.32,141.96) -- (212.29,141.96) -- (212.29,152.75) -- (201.32,152.75) -- cycle ;
\draw  [line width=1.5]  (201.32,230.76) -- (212.29,230.76) -- (212.29,241.54) -- (201.32,241.54) -- cycle ;
\draw  [fill={rgb, 255:red, 0; green, 0; blue, 0 }  ,fill opacity=1 ] (120.24,143.32) -- (120.24,143.32) -- (116.61,146.89) -- (120.24,150.46) -- (120.08,150.6) -- (116.46,147.04) -- (112.83,150.6) -- (112.83,150.6) -- (116.46,147.04) -- (112.83,143.47) -- (112.98,143.32) -- (116.61,146.89) -- cycle ;
\draw  [fill={rgb, 255:red, 0; green, 0; blue, 0 }  ,fill opacity=1 ] (210.43,143.79) -- (210.43,143.79) -- (206.8,147.35) -- (210.43,150.92) -- (210.28,151.07) -- (206.65,147.5) -- (203.02,151.07) -- (203.02,151.07) -- (206.65,147.5) -- (203.02,143.94) -- (203.17,143.79) -- (206.8,147.35) -- cycle ;
\draw  [fill={rgb, 255:red, 0; green, 0; blue, 0 }  ,fill opacity=1 ] (210.43,232.58) -- (210.43,232.58) -- (206.8,236.15) -- (210.43,239.72) -- (210.28,239.86) -- (206.65,236.3) -- (203.02,239.86) -- (203.02,239.86) -- (206.65,236.3) -- (203.02,232.73) -- (203.17,232.58) -- (206.8,236.15) -- cycle ;
\draw  [fill={rgb, 255:red, 0; green, 0; blue, 0 }  ,fill opacity=1 ] (120.24,232.43) -- (120.24,232.43) -- (116.61,236) -- (120.24,239.57) -- (120.08,239.71) -- (116.46,236.15) -- (112.83,239.71) -- (112.83,239.71) -- (116.46,236.15) -- (112.83,232.58) -- (112.98,232.43) -- (116.61,236) -- cycle ;

\begin{scope}[shift={(250,80)}, scale=0.75] 
    
    %
    \draw  [fill={rgb, 255:red, 0; green, 0; blue, 0 }  ,fill opacity=1 ] (2,-49) -- (2,-49) -- (-2,-45) -- (2,-41) -- (2,-41) -- (-2,-45) -- (-6,-41) -- (-6,-41) -- (-2,-45) -- (-6,-49) -- (-6,-49) -- (-2,-45)
     -- cycle ;
    \node[right] at (8,-46) {\small $\cO(\epsilon)$ input};
    
    \draw[line width=1.5] (-6,-32) rectangle (6,-20);
    \node[right] at (8,-25) {\small $\cO(\epsilon)$ output};
    
    \draw[fill=black] (0,-2) circle (4.5); 
    \node[right] at (8,-2) {\small $\cO(\epsilon^2)$ output};
\end{scope}

\draw (295.64,194.89) node [anchor=north west][inner sep=0.75pt]   [align=left] {$\displaystyle n k_{x}$};
\draw (120,48) node [anchor=north west][inner sep=0.75pt]   [align=left] {$\displaystyle m k_{z}$};
\draw (239.86,197.17) node [anchor=north west][inner sep=0.75pt]   [align=left] {$\displaystyle 2k_{x}$};
\draw (199.99,197.32) node [anchor=north west][inner sep=0.75pt]   [align=left] {$\displaystyle k_{x}$};
\draw (6.98,197.85) node [anchor=north west][inner sep=0.75pt]   [align=left] {$\displaystyle -3k_{x}$};
\draw (51.62,197.98) node [anchor=north west][inner sep=0.75pt]   [align=left] {$\displaystyle -2k_{x}$};
\draw (102.96,197.26) node [anchor=north west][inner sep=0.75pt]   [align=left] {$\displaystyle -k_{x}$};
\draw (118,271.25) node [anchor=north west][inner sep=0.75pt]   [align=left] {$\displaystyle -2k_{z}$};
\draw (126,227.58) node [anchor=north west][inner sep=0.75pt]   [align=left] {$\displaystyle -k_{z}$};
\draw (140.61,137.06) node [anchor=north west][inner sep=0.75pt]   [align=left] {$\displaystyle k_{z}$};
\draw (127,95) node [anchor=north west][inner sep=0.75pt]   [align=left] {$\displaystyle 2k_{z}$};
\end{tikzpicture}

%% file: figures/Figure4b.tex
 { \setlength{\unitlength}{0.8cm}
        \thicklines
       \begin{tikzpicture}[auto, node distance=3.8cm, >=latex']
  \node[draw, block, minimum width=2cm, minimum height=0.8cm, fill=blue!20,label={[yshift=-2pt]above:$\cO(\epsilon^2)$}] (H0)
    {$\cG_{(0,2\bkz)}(\bc)$};
  
  \node[draw, block, minimum width=1cm, minimum height=0.8cm, fill=customRed, right of=H0] (M)
    {$\widetilde{\cN}_2(\,\cdot\,,\,\cdot\,)$};
  \node[draw, block, minimum width=2cm, minimum height=0.8cm, fill=blue!20, right of=M,xshift=0.5cm, yshift=0.8cm,label={[yshift=-2pt]above:$\cO(\epsilon)$}] (Hkx)
    {$\cG_{(\bkx,\bkz)}(c)$};
  \node[draw, block, minimum width=2cm, minimum height=0.8cm, fill=blue!20, right of=M,xshift=0.5cm, yshift=-0.8cm] (Hmx)
    {$\cG_{(-\bkx,\bkz)}(c)$};
  
  \draw[->, very thick] (M.west) -- (H0.east)
    node[midway, above] {$\bd_{(0,2\bkz)}^{(2)}$};
  
  \coordinate (uH0) at ([xshift=-2cm]H0.west);
  \draw[->, very thick] (H0.west) -- (uH0)
    node[midway, above] {$\ba{c} 
    \mbox{steady} \\ \mbox{streaks} \\ \mbox{$\bu_{(0,2\bkz)}^{(2)}$} \ea$};
  
  \coordinate (dHkx) at ([xshift=2cm]Hkx.east);
  \draw[->, thick] (dHkx) -- (Hkx.east)
    node[midway, above] {$\ba{c} \mbox{unsteady}\\ \mbox{3D input}\\ \mbox{$\bd_{(\bkx,\bkz)}^{(1)}$} \ea$};
  
  \coordinate (dHmx) at ([xshift=2cm]Hmx.east);
  \draw[->, very thick] (dHmx) -- (Hmx.east)
    node[midway, above] {$\bd_{(-\bkx,\bkz)}^{(1)}$};
  
  \coordinate (Utemp1) at ($(Hkx.west)+(-1.8,0)$);
  \coordinate (Utemp2) at ($(Utemp1 |- M.east)+(0,0.2)$);
  \draw[-, very thick] (Hkx.west) -- (Utemp1) node[midway, above] {$\ba{c} \mbox{oblique} \\ \mbox{waves} \\ \mbox{$\bu_{(\bkx,\bkz)}^{(1)}$} \ea$};
  \draw[-, very thick] (Utemp1) -- (Utemp2);
  \draw[->, very thick] (Utemp2) -- ($(M.east)+(0,0.2)$);
  
  \coordinate (Ltemp1) at ($(Hmx.west)+(-1.8,0)$);
  \coordinate (Ltemp2) at ($(Ltemp1 |- M.east)+(0,-0.2)$);
  \draw[-, very thick] (Hmx.west) -- (Ltemp1) node[midway, above] {$\bu_{(-\bkx,\bkz)}^{(1)}$};
  \draw[-, very thick] (Ltemp1) -- (Ltemp2);
  \draw[->, very thick] (Ltemp2) -- ($(M.east)+(0,-0.2)$);
\end{tikzpicture} 
               }

%% file: tables/table_eps2.tex
	\begin{table}
		\centering
		\small
			\begin{tabular}{ |{c}|{c}| } 
				\hline 
				& 
				\\[-.36cm]
				{\sc Order} & {\sc Input/Response} 
				\\[0.15cm]
				\toprule
				$\cO(\epsilon)$
				& 
				$
				\ba{ccc}
				\mbox{\bf oblique wave input}
				&
				&
				\mbox{\bf oblique wave output}
				\\[0.15cm]
				\left[
				\ba{c}
				d_u^{(1)} (\bar{x},y,z)
				\\[0.1cm]
				d_v^{(1)} (\bar{x},y,z)
				\\[0.1cm]
				d_w^{(1)} (\bar{x},y,z)
				\ea
				\right]
				=
				\left[
				\ba{c}
				\phantom{-} 2 \, \Re 
				\big\{
				{d}_{u,\bk}^{(1)} (y,c) \, \mre^{\mri \bkx \bar{x}} 
				\big\}
				\cos k_z z
				\\[0.15cm]
				\phantom{-} 2 \, \Re 
				\big\{
				{d}_{v,\bk}^{(1)} (y,c) \, \mre^{\mri \bkx \bar{x}} 
				\big\}
				\cos k_z z
				\\[0.15cm]
				- 2 \, \Im
				\big\{
				{d}_{w,\bk}^{(1)} (y,c) \, \mre^{\mri \bkx \bar{x}} 
				\big\}
				\sin k_z z
				\ea
				\right]
				&
				\Rightarrow
				&
				\left[
				\ba{c}
				u^{(1)} (\bar{x},y,z)
				\\[0.1cm]
				v^{(1)} (\bar{x},y,z)
				\\[0.1cm]
				w^{(1)} (\bar{x},y,z)
				\ea
				\right]
				= 
				\left[
				\ba{c}
				\phantom{-} 2 \, \Re 
				\big\{
				{u}_{\bk}^{(1)} (y,c) \, \mre^{\mri \bkx \bar{x}} 
				\big\}
				\cos k_z z
				\\[0.15cm]
				\phantom{-} 2 \, \Re 
				\big\{
				{v}_{\bk}^{(1)} (y,c) \, \mre^{\mri \bkx \bar{x}} 
				\big\}
				\cos k_z z
				\\[0.15cm]
				- 2 \, \Im
				\big\{
				{w}_{\bk}^{(1)} (y,c) \, \mre^{\mri \bkx \bar{x}} 
				\big\}
				\sin k_z z
				\ea
				\right]
				\ea
				$
				\\[-0.25cm]
				\\
				\midrule
				$\cO (\epsilon^2)$ 				
				&
				$
				\ba{ccc}
				\mbox{\bf  streak input}
				&
				&
				\mbox{\bf streak output}
				\\[0.15cm] 
				\left[
				\ba{c}
				d_{\psi}^{(2)} (y,z)
				\\[0.15cm]
				d_{u}^{(2)} (y,z)
				\ea
				\right]
				=
				\left[
				\ba{c}
				d_{\psi, 2 k_z}^{(2)} (y) \sin 2 k_z z
				\\[0.15cm]
				d_{u, 2 k_z}^{(2)} (y) \cos 2 k_z z
				\ea
				\right]
				&
				\Rightarrow
				&
				\left[
				\ba{c}
				u_{\mathrm{s}}^{(2)} (y,z)
				\\[0.15cm]
				v_{\mathrm{s}}^{(2)} (y,z)
				\\[0.15cm]
				w_{\mathrm{s}}^{(2)} (y,z)
				\ea
				\right]
				=
				\left[
				\ba{c}
				u_{\mathrm{s}, 2 k_z}^{(2)} (y) \cos 2 k_z z
				\\[0.15cm]
				v_{\mathrm{s}, 2 k_z}^{(2)} (y) \cos 2 k_z z
				\\[0.15cm]
				w_{\mathrm{s}, 2 k_z}^{(2)} (y) \sin 2 k_z z
				\ea
				\right]
				\ea
				$  
				\\[-0.25cm]
				\\
				\bottomrule
			\end{tabular}		
			\caption{Disturbances and corresponding responses arising from oblique-wave--induced streak interactions, accounting for the streamwise and spanwise symmetries of the channel. Here, $\Re\{\cdot\}$ and $\Im\{\cdot\}$ denote the real and imaginary parts, respectively.}			
			\label{tab.ow-ss}
		\end{table}

%% file: figures/Figure10.tex
\begin{tikzpicture}[>=latex', node distance=0.5cm and 0.5cm]
    \node[block, draw, minimum width=4em, minimum height=3em, label={[yshift=1pt]above:$\tc{RoyalBlue}{\ba{c}\mbox{\bf viscous} \\ \mbox{\bf dissipation}\ea}$}] (ReS1) at (0.5,1.5) {$- \tc{red}{Re} \, \cS_{2 k_z}^{-1}$};

    \draw[->, very thick] (ReS1.west) -- ++(-1.4,0) node[midway, above] 
    {$u_{\mathrm{s}, 2 k_z}^{(2)}$};

    \node[block, draw, minimum width=4em, minimum height=3em, top color=red!20, bottom color=red!20, label={[yshift=1pt]above:$\tc{red}{\ba{c} \mbox{\bf vortex}\\ \mbox{\bf tilting} \ea}$}] (Cp) at (4.5,1.5) {$\cC_{p, 2 k_z}$};
    \node[block, draw, minimum width=4em, minimum height=3em, label={[yshift=1pt]above:$\tc{RoyalBlue}{\ba{c} \mbox{\bf generalized}\\ \mbox{\bf diffusion} \ea}$}] (ReL) at (7.5,1.5) {$-\tc{red}{Re} \, \cL_{2 k_z}^{-1}$};

    \coordinate (plus) at ($(Cp.west)!0.5!(ReS1.east)$);
    \draw[thick] (plus) circle (0.2) node {$+$};

    \draw[->, very thick, shorten <=0.2cm] (plus) -- (ReS1.east);
    
    \draw[->, very thick, shorten >=0.2cm] (Cp.west) -- (plus);
    
    \coordinate (plusAbove) at ($(plus)+(0,1.4)$);
    \draw[->, very thick, shorten >=0.2cm] (plusAbove) -- (plus) node[midway, right, yshift=0.4cm] {${d}_{u, 2 k_z}^{(2)}$};

    \draw[->, very thick] (ReL.west) -- (Cp.east) node[midway, above] {$\psi_{\mathrm{s}, 2 k_z}^{(2)}$};

    \coordinate (dpsi_in) at ($(ReL.east)+(1.4,0)$);
    \draw[->, very thick] (dpsi_in) -- (ReL.east) node[midway, above] {${d}_{\psi, 2 k_z}^{(2)}$};
\end{tikzpicture}

%% file: figures/Figure13.tex
%
%
\usetikzlibrary{decorations.pathmorphing,arrows.meta,positioning,calc,backgrounds}

\definecolor{cOblique}{RGB}{ 30, 80,160}   
\definecolor{cStreak} {RGB}{200,100, 0}   
\definecolor{cWaveIn} {RGB}{ 0,120,60}   
\definecolor{cOpFill} {RGB}{255,225,225}   
\definecolor{cOpFill1} {RGB}{255,225,225}   
\definecolor{cOpFill2} {RGB}{255,210,225}   
\definecolor{cOpEdge} {RGB}{180, 45, 45}   
\definecolor{cSoftGray}{RGB}{235,238,245}   


\tikzset{
  every node/.style={font=\small}
}

\begin{tikzpicture}[
    x=1cm, y=1cm,
    >={Stealth[length=2.4mm,width=1.8mm]},
    bigOblique/.style={circle,fill=cOblique,draw=cOblique,
                      minimum size=6.0mm,inner sep=0pt},
    bigStreak/.style={circle,fill=cStreak ,draw=cStreak ,
                      minimum size=6.0mm,inner sep=0pt},
    op/.style={circle,fill=cOpFill,draw=cOpEdge,line width=0.65pt,
                      minimum size=6.0mm,inner sep=0pt,
                      text=cOpEdge,font=\normalsize\itshape},
     op1/.style={circle,fill=cOpFill1,draw=cOpEdge,line width=0.65pt,
                      minimum size=6.0mm,inner sep=0pt,
                      text=cOpEdge,font=\normalsize\itshape},  
     op2/.style={circle,fill=cOpFill2,draw=cOpEdge,line width=0.65pt,
                      minimum size=6.0mm,inner sep=0pt,
                      text=cOpEdge,font=\normalsize\itshape},                                  
    obliqueArr/.style={->,draw=cOblique,line width=1.15pt},
    streakArr/.style={->,draw=cStreak ,line width=1.15pt},
    wavyGreen/.style={draw=cWaveIn,line width=1.15pt,
                      decorate,decoration={snake,amplitude=0.8mm,
                                           segment length=2mm,
                                           pre length=0.6mm,post length=1.4mm}},
    colLabelB/.style={color=cOblique, align=center},
    colLabelO/.style={color=cStreak, align=center},
    divider/.style={dashed,gray!85,line width=0.35pt},
    wlArr/.style={<->,>={Stealth[length=2.4mm,width=1.8mm]},
                  draw=black,line width=0.7pt},
]



=========================================================
\foreach \x in {3.4,7.0,10.6}
    \draw[divider] (\x,0.3) -- (\x,4.9);

\node[colLabelB] at (1.5 ,4.15) {$\mathcal{O}(\epsilon)$ \\ oblique waves\\[-1pt] {\small $(k_x,k_z,\omega)$}};
\node[colLabelO] at (5.2 ,4.15) {$\mathcal{O}(\epsilon^{2})$ \\ streamwise streaks\\[-1pt] {\small $(0,2k_z,0)$}};
\node[colLabelB] at (8.8 ,4.15) {$\mathcal{O}(\epsilon^{3})$ \\ oblique waves\\[-1pt] {\small $(k_x,k_z,\omega)$}};
\node[colLabelO] at (12.9,4.15) {$\mathcal{O}(\epsilon^{4})$ \\ streamwise streaks\\[-1pt] {\small $(0,2k_z,0)$}};

\node[bigOblique] (u1) at (1.8 ,2.9) {};
\node[op1]         (N1) at (4.3 ,1.9) {$\cN_1$};
\node[bigStreak]  (u2) at (5.8 ,1.35){};
\node[op2]         (M1) at (8.0 ,2.1) {$\cN_2$};
\node[bigOblique,minimum size=6.0mm] (u3) at (9.5 ,2.1) {};
\node[op1]         (N2) at (11.6 ,0.9) {$\cN_1$};
\node[op2]         (M2) at (12.0,2.1) {$\cN_2$};
\node[bigStreak]  (u4) at (13.2,0.9) {};

\node[cOblique, anchor=south] at (2.95,3.00) {$\mathbf{u}^{(1)}$};
\node[cOblique,font=\small,anchor=north] at (2.95,2.82) {$(k_x,k_z,\omega)$};
\node[cStreak , anchor=south] at (6.70,1.45) {$\mathbf{u}^{(2)}$};
\node[cStreak ,font=\small,anchor=north] at (6.70,1.27) {$(0,2k_z,0)$};
\node[cOblique, anchor=south] at (10.75,2.22) {$\mathbf{u}^{(3)}$};
\node[cOblique,font=\small,anchor=north] at (10.75,2.04) {$(k_x,k_z,\omega)$};
\node[cStreak , anchor=south] at (14.20,1.02) {$\mathbf{u}^{(4)}$};
\node[cStreak ,font=\small,anchor=north] at (14.20,0.84) {$(0,2k_z,0)$};

\draw[wavyGreen,->] (-0.3,2.9) -- (u1.west);
\node[color=cWaveIn, anchor=south] at (0.6,3.00) {$\mathbf{d}^{(1)}$};
\node[color=cWaveIn,font=\small,anchor=north] at (0.6,2.78) {$(k_x,k_z,\omega)$};

\draw[obliqueArr] (u1.east) -- (11.6,2.9) to[out=0,in=95,looseness=0.9] (M2.north);
\draw[obliqueArr] (4.00,2.9) to[out=0,in=90,looseness=0.8] (N1.north);
\draw[obliqueArr] (7.75,2.9) to[out=0,in=90,looseness=0.9] (M1.north);

\draw[streakArr] (N1.south) .. controls +(0,-0.45) and +(-0.6,0.1) .. (u2.west);

\draw[streakArr] (u2.east) -- (6.95,1.35) to[out=0,in=180,looseness=0.9] (N2.west);
\draw[streakArr] (6.95,1.35) to[out=0,in=-90,looseness=0.9] (M1.south);

\draw[obliqueArr] (M1.east) -- (u3.west);
\draw[obliqueArr] (u3.east) -- (M2.west);
\draw[streakArr] (N2.east) -- (u4.west);
\draw[streakArr] (M2.south) .. controls +(0,-0.4) and +(0,0.55) .. (u4.north);
\draw[streakArr] (u4.east) -- (15,0.9);

\node at ($(7.5,-1.1)$) { 
\begin{tikzpicture}[
    every node/.style={font=\small}
]

\node[bigOblique,minimum size=6.0mm,anchor=west] (c1a) at (0,0.4) {};
\node[anchor=west,align=left]
      at (c1a.east) {LNSE propagator: resolvent \\ (oblique waves)};

\node[bigStreak,minimum size=6.0mm,anchor=west] (c1b) at (0,-0.6) {};
\node[anchor=west,align=left]
      at (c1b.east) {LNSE propagator: resolvent \\ (streamwise streaks)};

\node[op1,minimum size=6.0mm,anchor=west] (c2a) at (5.5,0.4) {$\cN_1$};
\node[anchor=west,align=left]
      at (c2a.east) {quadratic \\ self-interaction};

\node[op2,minimum size=6.0mm,anchor=west] (c2b) at (5.5,-0.6) {$\cN_2$};
\node[anchor=west,align=left]
      at (c2b.east) {quadratic \\ cross-interaction};

\draw[wavyGreen,->] (9.8,0.5) -- ++(0.8,0);
\node[anchor=west] at (10.7,0.5) {external forcing};

\draw[obliqueArr] (9.8,-0.3) -- ++(0.8,0);
\node[anchor=west] at (10.7,-0.3) {oblique waves};

\draw[streakArr] (9.8,-1.1) -- ++(0.8,0);
\node[anchor=west] at (10.7,-1.1) {streamwise streaks};

\end{tikzpicture}
};

\end{tikzpicture}
